\documentstyle[aps,eqsecnum,preprint,floats,epsf,epsfig]{revtex}
\begin{document}
\def\be{\begin{eqnarray}}
\def\en{\end{eqnarray}}
\def\non{\nonumber}
\def\la{\langle}
\def\ra{\rangle}
\def\nc{N_c^{\rm eff}}
\def\vp{\varepsilon}
\def\vma{{_{V-A}}}
\def\vpa{{_{V+A}}}
\def\m{\hat{m}}
\def\ov{\overline}
\def\etapp{{\eta^{(')}}}
\def\fp{{f_{\eta'}^{(\bar cc)}}}
\def\half{{{1\over 2}}}
\def\pr{{\sl Phys. Rev.}~}
\def\prl{{\sl Phys. Rev. Lett.}~}
\def\pl{{\sl Phys. Lett.}~}
\def\np{{\sl Nucl. Phys.}~}
\def\zp{{\sl Z. Phys.}~}
\def\lsim{ {\ \lower-1.2pt\vbox{\hbox{\rlap{$<$}\lower5pt\vbox{\hbox{$\sim$}
}}}\ } }
\def\gsim{ {\ \lower-1.2pt\vbox{\hbox{\rlap{$>$}\lower5pt\vbox{\hbox{$\sim$}
}}}\ } }

\font\el=cmbx10 scaled \magstep2 {\obeylines \hfill IP-ASTP-01-99
\hfill June, 1999}

\vskip 1.5 cm

\centerline{\large\bf Charmless Hadronic Two-body Decays of $B_u$ and $B_d$
Mesons}
\medskip
\bigskip
\medskip
\centerline{\bf Yaw-Hwang Chen}
\medskip
\centerline{Department of Physics, National Cheng-Kung University}
\centerline{Tainan, Taiwan 700, Republic of China}
\bigskip
\centerline{\bf Hai-Yang Cheng,~ B. Tseng,~ Kwei-Chou Yang}
\medskip
\centerline{Institute of Physics, Academia Sinica}
\centerline{Taipei, Taiwan 115, Republic of China}
\bigskip
\bigskip
\bigskip
\centerline{\bf Abstract}
\bigskip
{\small Two-body charmless nonleptonic decays of $B_u$ and $B_d$
mesons are studied within the framework of generalized
factorization in which the effective Wilson coefficients $c^{\rm
eff}_i$ are renormalization-scale and -scheme independent while
factorization is applied to the tree-level hadronic matrix
elements. Contrary to previous studies, our $c_i^{\rm eff}$ do not
suffer from gauge and infrared problems. Nonfactorizable effects
are parametrized in terms of $\nc(LL)$ and $\nc(LR)$, the
effective numbers of colors arising from $(V-A)(V-A)$ and
$(V-A)(V+A)$ four-quark operators, respectively. Tree and penguin
transitions are classified into six different classes. The data of
$B^-\to\rho^0\pi^-$ and $B^-\to\phi K^-$ clearly indicate that
$\nc(LR)\neq \nc(LL)$: The first measurement of the $b\to u$ mode
$B^-\to\rho^0\pi^-$ and the experimental information on the
tree-dominated mode $B^-\to\omega\pi^-$ all imply that $\nc(LL)$
is less than 3, whereas the CLEO measurement of $B^-\to\phi K^-$
shows $\nc(LR)>3$. For given input parameters, the prediction of
${\cal B}(B\to\eta' K)$ is largely improved by setting
$\nc(LL)\sim 2 $ and $\nc(LR)>\nc(LL)$; in particular, the charm
content of the $\eta'$ contributes in the right direction. The
decay rate of $B\to\phi K^*$ is very sensitive to the form-factor
ratio $A_2/A_1$; the absence of $B\to\phi K$ events does not
necessarily invalidate the factorization approach. If the
branching ratio of $B^-\to\omega K^-$ is experimentally found to
be significantly larger than that of $B^-\to\rho^0 K^-$, we argue
that inelastic final-state rescattering may account for the
disparity between $\omega K^-$ and $\rho^0 K^-$. By contrast, if
${\cal B}(B^-\to\rho^0 K^-)\sim {\cal B}(B^-\to\omega K^-)$ is
observed, then $W$-annihilation and/or spacelike penguins could
play a prominent role. The decay modes $\ov B^0_d\to
\phi\pi^0,\,\phi\eta,\,\phi\eta',\,\phi\rho^0,\,\phi\omega,~B
^-\to\phi\pi^-,\,\phi\rho^-$ involving a vector meson $\phi$ are
dominated by electroweak penguins. We show that a unitarity angle
$\gamma$  larger than $90^\circ$ is helpful for explaining the
$\pi^+\pi^-$, $\pi K$ and $\eta' K$ data. The relative magnitudes
of tree, QCD penguin and electroweak penguin amplitudes are
tabulated for all charmless $B\to PP,VP,VV$ decays. Our favored
predictions for branching ratios are those for $\nc(LL)\approx 2$
and $\nc(LR)\sim 5$.

}

\pagebreak

%%%%%%%%%%%%%%%%%%%%%%%%%%%%%%%%%%%%%%%%%%%%%%%%%%%%%%%%%
%%%%%%%%%%%%%%%%%%%%%%%%%%%%%%%%%%%%%%%%%%%%%%%%%%%%%%%%%%%%%%%

\section{Introduction}
The study of exclusive nonleptonic weak decays of $B$ mesons is of
great interest for several reasons: many of rare hadronic $B$
decay modes are dominated by the gluonic penguin mechanism and
large direct $CP$ asymmetries are expected in many charged $B$
decays. Hence the analysis and measurement of charmless hadronic
$B$ decays will enable us to understand the QCD and electroweak
penguin effects in the Standard Model (SM) and provide a powerful
tool of seeing physics beyond the SM. The sizable direct $CP$
violation expected in exclusive rare decay modes of $B$ mesons
will allow the determination of the Cabibbo-Kobayashi-Maskawa
(CKM) unitarity angles.

In past years we have witnessed remarkable progress in the study
of exclusive charmless $B$ decays. Experimentally, CLEO
\cite{CLEO} has discovered many new two-body decay modes
\be
B\to\eta' K^\pm,~\eta' K^0,~\pi^\pm K^0,~\pi^\pm K^\mp,~\pi^0
K^\pm,~\rho^0\pi^\pm,~\omega K^\pm,
\en
and found a possible evidence for $B\to \phi K^*$. Moreover, CLEO
has provided new improved upper limits for many other decay modes.
While all the channels that have been measured so far are penguin
dominated, the most recently observed $\rho^0\pi^-$ mode is
dominated by the tree diagram. In the meantime, updates and new
results of many $B\to PV$ decays with $P=\eta,\eta',\pi,K$ and
$V=\omega,\phi,\rho, K^*$ as well as $B\to PP$ decays will be
available soon. With the $B$ factories Babar and Belle starting to
collect data, many exciting and harvest years in the arena of $B$
physics and $CP$ violation are expected to come.

Some of the CLEO data are surprising from the theoretical point of
view: The measured branching ratios for $B^\pm\to\eta' K^\pm$ and
$B^\pm\to\omega K^\pm$ are about several times larger than the
naive theoretical estimate. Since then the theoretical interest in
hadronic charmless $B$ decays is surged and recent literature is
rife with all kinds of interesting interpretations of data, both
within and beyond the SM.

An earlier systematic study of exclusive nonleptonic two-body
decays of $B$ mesons was made in \cite{Chau1}. Two different
approaches were employed in this reference: the effective
Hamiltonian approach in conjunction with the factorization
hypothesis for hadronic matrix elements and a model-independent
analysis based on the quark-diagram approach developed by Chau and
one of us (H.Y.C.) \cite{CC}. Many significant improvements and
developments have been achieved over past years. For example, a
next-to-leading order effective Hamiltonian for current-current
operators and QCD as well as electroweak penguin operators becomes
available. The renormalization scheme and scale problems with the
factorization approach for matrix elements can be circumvented by
employing scale- and scheme-independent effective Wilson
coefficients. Heavy-to-light form factors have been computed using
QCD sum rules, lattice QCD and potential models. A great interest
in the flavor-SU(3) quark diagram approach was also revived in
recent years. In particular, this method has been widely utilized
as a model-independent extraction of the CKM unitary triangle.

We will present in this paper an updated and vigorous analysis of
hadronic two-body charmless decays of $B_u$ and $B_d$ mesons (for
$B_s$ mesons, see \cite{CCT}). We will pay special attention to
two important issues: the gauge and infrared problems with the
effective Wilson coefficients, and the nonfactorized effect
characterized by the parameter $\nc$, the effective number of
colors.

One of the principal difficulties with naive factorization is that
the hadronic matrix element under the factorization approximation
is renormalization scale $\mu$ independent as the vector or
axial-vector current is partially conserved. Consequently, the
amplitude $c_i(\mu) \la O\ra_{\rm fact}$ is not physical as the
scale dependence of Wilson coefficients does not get compensation
from the matrix elements. A plausible solution to the problem is
to extract the $\mu$ dependence from the matrix element $\la
O(\mu)\ra$, and then combine it with the $\mu$-dependent Wilson
coefficients to form scale- and scheme-independent effective
coefficients $c_i^{\rm eff}$. The factorization approximation is
applied afterwards to the hadronic matrix element of the operator
$O$ at the tree level. However, it was pointed out recently in
\cite{Buras98} that $c_i^{\rm eff}$ suffer from gauge and infrared
ambiguities since an off-shell external quark momentum, which is
usually chosen to regulate the infrared divergence occurred in the
radiative corrections to the local 4-quark operators, will
introduce a gauge dependence.

A closely related problem is connected to the generalized
factorization approach in which the nonfactorized contribution to
the matrix element in $B\to PP,VP$ decays is lumped into the
effective number of colors $\nc$, called $1/\xi$ in \cite{BSW87}.
The deviation of $1/\nc$ from $1/N_c$ measures the nonfactorizable
effect. The unknown parameter $\nc$ is usually assumed to be
universal (i.e., channel independent) within the framework of
generalized factorization and it can be extracted from experiment.
However, as stressed by Buras and Silvestrini \cite{Buras98}, if
$c_i^{\rm eff}$ are gauge and infrared regulator dependent, then
the values of $\nc$ extracted from the data on two-body hadronic
decays are also gauge dependent and therefore they cannot have any
physical meaning. Recently, this controversy on gauge dependence
and infrared singularity connected with the effective Wilson
coefficients is resolved by Li and two of us (H.Y.C. and K.C.Y.)
\cite{CLY}: Gauge invariance of the decay amplitude is maintained
under radiative corrections by assuming on-shell external quarks.
The infrared pole emerged in a physical on-shell scheme signifies
the nonperturbative dynamics involved in a decay process and has
to be absorbed into a universal hadron wave function. As a
consequence, it is possible to construct the effective Wilson
coefficients which are not only renormalization scale- and
scheme-independent but also gauge invariant and infrared finite.

For penguin-dominated rare $B$ decays, there is another subtle
issue for the effective parameter $\nc$. As shown in \cite{CT98},
nonfactorizable effects in the matrix elements of $(V-A)(V+A)$
operators are {\it a priori} different from that of $(V-A)(V-A)$
operators, i.e. $\nc(LR)\neq \nc(LL)$. We will demonstrate in the
present work that the most recently measured $B^-\to\rho^0\pi^-$
decay together with the experimental information on the
tree-dominated modes $B^-\to\omega\pi^-$ clearly imply
$\nc(LL)<3$, while the CLEO measurement of $B^-\to\phi K^-$
indicates $\nc(LR)>3$. Contrary to the previous studies, we show
that the experimental data of $\rho^0\pi^\pm$ and $\phi K^\pm$
cannot be accommodated simultaneously by treating
$\nc(LL)=\nc(LR)$. This observation is very crucial for improving
the discrepancy between theory and experiment for $B\to\eta' K$
decays.

This paper is organized as follows. In Sec.~II we discuss the
gauge and infrared problems connected with the effective Wilson
coefficients and their solution. Input parameters necessary for
calculations such as quark mixing matrix elements, running quark
masses, decay constants, heavy-to-light form factors are
summarized in Sec. III. In Sec. IV we classify the factorized
decay amplitudes into six different classes. Results for branching
ratios and their implications are discussed in details in Sec. V
with special attention paid to $B\to \rho\pi,\omega\pi,\phi K,\phi
K^{*},\eta'K,K\pi$ modes; in particular, all possible sources of
theoretical uncertainties are summarized in Sec. V.G. The role of
final-state interactions played in charmless $B$ decays is
elaborated on in Sec. VI. For reader's convenience, we compare our
results with the literature in Sec. VII. Sec. VIII is for the
conclusion. Factorized amplitudes for all charmless $B\to
PP,VP,VV$ decays are tabulated in the Appendix.

\section{Framework}
The effective Hamiltonian is the standard starting point for
describing the nonleptonic weak decays of hadrons. The relevant
effective $\Delta B=1$ weak Hamiltonian for hadronic charmless $B$
decays is
\be
{\cal H}_{\rm eff}(\Delta B=1) &=& {G_F\over\sqrt{2}}\Bigg\{ V_{ub}V_{uq}^*
\Big[c_1(\mu)O_1^u(\mu)+c_2(\mu)O_2^u(\mu)\Big]+V_{cb}V_{cq}^*\Big[c_1(\mu)
O_1^c(\mu)+c_2(\mu)O_2^c(\mu)\Big]  \non \\
&& -V_{tb}V_{tq}^*\sum^{10}_{i=3}c_i(\mu)O_i(\mu)\Bigg\}+{\rm h.c.},
\en
where $q=d,s$, and
\be
&& O_1^u= (\bar ub)_\vma(\bar qu)_\vma, \qquad\qquad\qquad\qquad~~
O_2^u = (\bar u_\alpha b_\beta)_\vma(\bar q_\beta u_\alpha)_\vma, \non \\
&& O_1^c= (\bar cb)_\vma(\bar qc)_\vma, \qquad\qquad\qquad\qquad~~~
O_2^c = (\bar c_\alpha b_\beta)_\vma(\bar q_\beta c_\alpha)_\vma, \non \\
&& O_{3(5)}=(\bar qb)_\vma\sum_{q'}(\bar q'q')_{\vma(\vpa)}, \qquad  \qquad
O_{4(6)}=(\bar q_\alpha b_\beta)_\vma\sum_{q'}(\bar q'_\beta q'_\alpha)_{
\vma(\vpa)},   \\
&& O_{7(9)}={3\over 2}(\bar qb)_\vma\sum_{q'}e_{q'}(\bar q'q')_{\vpa(\vma)},
  \qquad O_{8(10)}={3\over 2}(\bar q_\alpha b_\beta)_\vma\sum_{q'}e_{q'}(\bar
q'_\beta q'_\alpha)_{\vpa(\vma)},   \non
\en
with $O_3$--$O_6$ being the QCD penguin operators,
$O_{7}$--$O_{10}$ the electroweak penguin operators and $(\bar
q_1q_2)_{_{V\pm A}}\equiv\bar q_1\gamma_\mu(1\pm \gamma_5)q_2$. In
order to ensure the renormalization-scale and -scheme independence
for the physical amplitude, the matrix elements of 4-quark
operators have to be evaluated in the same renormalization scheme
as that for Wilson coefficients and renormalized at the same scale
$\mu$.

   Although the hadronic matrix element $\la O(\mu)\ra$ can be directly
calculated in the lattice framework, it is conventionally evaluated under
the factorization hypothesis so that $\la O(\mu)\ra$ is factorized into
the product of two matrix elements of
single currents, governed by decay constants and form factors.
In spite of its tremendous simplicity, the naive factorization approach
encounters two major difficulties. One of them is that
the hadronic matrix element under factorization is
renormalization scale $\mu$ independent as the vector or axial-vector
current is partially conserved. Consequently, the amplitude $c_i(\mu)
\la O\ra_{\rm fact}$ is not truly physical as the scale dependence of
Wilson coefficients does not get compensation from the matrix elements.

A plausible solution to the aforementioned scale problem is to
extract the $\mu$ dependence from the matrix element $\la
O(\mu)\ra$, and combine it with the $\mu$-dependent Wilson
coefficient functions to form $\mu$-independent effective
coefficients. Schematically, we may write \be \label{ceff}
c(\mu)\la O(\mu)\ra=c(\mu)g(\mu)\la O\ra_{\rm tree}\equiv c^{\rm
eff}\la O\ra _{\rm tree}.
\en
The factorization approximation is applied afterwards to the
hadronic matrix element of the operator $O$ at the tree level.
Since the tree-level matrix element $\la O\ra_{\rm tree}$ is
renormalization scheme and scale independent, so are the effective
Wilson coefficients $c_i^{\rm eff}$. However, the problem is that
we do not know how to carry out first-principles calculations of
$\la O(\mu)\ra$ and hence $g(\mu)$. It is natural to ask the
question: Can $g(\mu)$ be evaluated at the quark level in the same
way as the Wilson coefficient $c(\mu)$ ? One of the salient
features of the operator product expansion (OPE) is that the
determination of the short-distance $c(\mu)$ is independent of the
choice of external states. Consequently, we can choose quarks as
external states in order to extract $c(\mu)$. For simplicity, we
consider a single multiplicatively renormalizable 4-quark operator
$O$ (say, $O_+$ or $O_-$) and assume massless quarks. The
QCD-corrected weak amplitude induced by $O$ in full theory is
\be
\label{full} A_{\rm full}=\left[1+{\alpha_s\over
4\pi}\left(-{\gamma\over 2}\,\ln{M_W^2 \over -p^2}+a \right)
\right]\la O\ra_q,
\en
where $\gamma$ is an anomalous dimension, $p$ is an off-shell momentum
of the external quark lines, which is introduced as an infrared cutoff,
and the non-logarithmic constant term $a$ in general depends on the
gauge chosen
for the gluon propagator. The subscript $q$ in (\ref{full}) emphasizes the
fact that the matrix element is evaluated between external quark states.
In effective theory, the renormalized $\la O(\mu)\ra_q$ is related to
$\la O\ra_q$ in full theory via
\be
\la O(\mu)\ra_q &=& \left[1+{\alpha_s\over
4\pi}\left(-{\gamma\over 2} \,\ln{\mu^2\over -p^2}+r \right)
\right]\la O\ra_q   \non \\ &\equiv & g'(\mu,-p^2,\lambda)\la
O\ra_q,
\en
where $g'$ indicates the perturbative corrections to the 4-quark
operator renormalized at the scale $\mu$. The constant term $r$ is
in general renormalization scheme and gauge dependent, and it has
the general expression \cite{Buras98}:
\be
r=r^{\rm NDR,HV}+\lambda r^\lambda,
\en
where NDR and HV stand for the naive dimension regularization and
't Hooft-Veltman renormalization schemes, respectively, and
$\lambda$ is a gauge parameter with $\lambda=0$ corresponding to
Landau gauge. Matching the effective theory with full theory,
$A_{\rm full}=A_{\rm eff}=c(\mu)\la O(\mu)\ra_q$, leads to
\be
c(\mu)=\,1+{\alpha_s\over 4\pi}\left(-{\gamma\over 2}\,\ln{M_W^2
\over \mu^2}+d \right),
\en
where $d=a-r$. Evidently, the Wilson coefficient is independent of the
infrared cutoff and it is gauge invariant as the gauge dependence is
compensated between $a$ and $r$. Of course, $c(\mu)$ is still
renormalization scheme and scale dependent.

   Since $A_{\rm eff}$ in full theory [Eq.~(2.4)] is $\mu$ and scheme
independent, it is obvious that
\be
c'^{\rm eff}=c(\mu)g'(\mu,-p^2,\lambda)
\en
is also independent of the choice of the renormalization scheme
and scale. Unfortunately, $c'^{\rm eff}$ is subject to the
ambiguities of the infrared cutoff and gauge dependence. As
stressed in \cite{Buras98}, the gauge and infrared dependence
always appears as long as the matrix elements of operators are
calculated between quark states. By contrast, the effective
coefficient $c^{\rm eff}=c(\mu)g(\mu)$ should not suffer from
these problems.

It was recently shown in \cite{CLY} that the above-mentioned
problems on gauge dependence and infrared singularity connected
with the effective Wilson coefficients can be resolved by
perturbative QCD (PQCD) factorization theorem. In this formalism,
partons, {\it i.e.}, external quarks, are assumed to be on shell,
and both ultraviolet and infrared divergences in radiative
corrections are isolated using the dimensional regularization.
Because external quarks are on shell, gauge invariance of the
decay amplitude is maintained under radiative corrections to all
orders. This statement is confirmed by an explicit one-loop
calculation in \cite{CLY}. The obtained ultraviolet poles are
subtracted in a renormalization scheme, while the infrared poles
are absorbed into universal nonperturbative bound-state wave
functions. The remaining finite piece is grouped into a hard decay
subamplitude. The decay rate is then factorized into the
convolution of the hard subamplitude with the bound-state wave
functions, both of which are well-defined and gauge invariant.
Explicitly, the effective Wilson coefficient has the expression
\begin{eqnarray}
c^{\rm eff}=c(\mu)g_1(\mu)g_2(\mu_f)\;, \label{nef}
\end{eqnarray}
where $g_1(\mu)g_2(\mu_f)$ is identified as the factor $g(\mu)$
defined in Eq. (\ref{ceff}). In above equation $g_1(\mu)$ is an
evolution factor from the scale $\mu$ to $m_b$, whose anomalous
dimension is the same as that of $c(\mu)$, and $g_2(\mu_f)$
describes the evolution from $m_b$ to $\mu_f$ ($\mu_f$ being a
factorization scale arising from the dimensional regularization of
infrared divergences), whose anomalous dimension differs from that
of $c(\mu)$ because of the inclusion of the dynamics associated
with spectator quarks. The infrared pole emerged in the physical
on-shell scheme signifies the nonperturbative dynamics involved in
a decay process and it has to be absorbed into the universal meson
wave functions.\footnote{For inclusive processes, the infrared
divergence due to radiative corrections is compensated by gluon
bremsstrahlung, leading to a well-defined and finite correction.
However, for exclusive hadronic decay processes the loop-induced
infrared divergence is not canceled by gluon bremsstrahlung in the
quark $\to$ three quarks decay process. In fact, the
bremsstrahlung contribution is irrelevant to the hadronic matrix
elements for exclusive decays. In the present framework of
perturbative QCD factorization theorem, the infrared pole is
absorbed by bound-state wave functions rather than canceled by the
bremsstrahlung process.} Hence, in the PQCD formalism the
effective Wilson coefficients are gauge invariant, infrared
finite, scheme and scale independent.

\begin{figure}[tb]
\psfig{figure=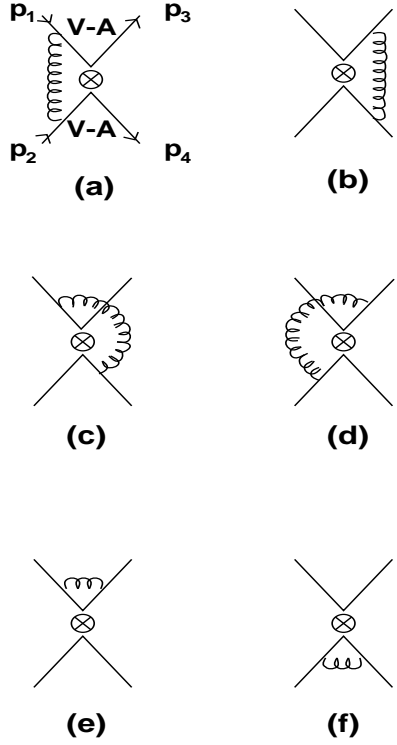,height=3.5in} \vspace{0.9cm}
    \caption{{\small
     Vertex corrections to the 4-quark operators $O_1-O_{10}$.}}
     \label{fig:4quark}
\end{figure}

In the above framework, $\la O\ra_{\rm tree}$ is related to the
meson wave function $\phi(\mu_f)$ (see \cite{CLY} for detail). For
our purposes of applying factorization, we will set $\mu_f=m_b$ to
compute $c^{\rm eff}$ and then evaluate the tree level hadronic
matrix element $\la O\ra_{\rm tree}$ using the factorization
approximation. It is straightforward to calculate $g_1(\mu)$ from
the vertex correction diagrams (see Fig. \ref{fig:4quark}) and
penguin-type diagrams for the 4-quark operators
$O_i~(i=1,\cdots,10)$. In general,
\be
\la O_i(\mu)\ra=\left[\,{\rm \openone}+{\alpha_s(\mu)\over
4\pi}\hat m_s(\mu)+{\alpha \over 4\pi}\hat m_e(\mu)\right]_{ij}\la
O_j\ra_{\rm tree},
\en
where the one-loop QCD and electroweak corrections to matrix
elements are parametrized by the matrices $\m_s$ and $\m_e$,
respectively. Hence,
\be
c_i^{\rm eff}=\left[\,{\rm \openone}+{\alpha_s(\mu)\over 4\pi}
\hat m_s^T(\mu)+{\alpha\over 4\pi}\hat
m_e^T(\mu)\right]_{ij}c_j(\mu),
\en
where the superscript $T$ denotes a transpose of the matrix.
Following the notation of \cite{Ali,AKL}, we
obtain\footnote{Unlike \cite{Ali,AKL}, we have included vertex
corrections to the electroweak coefficients $c_7-c_{10}$. It also
seems to us that a constant term ${2\over 3}$ is missed in
\cite{Ali,AKL} in the coefficient $\tilde G(m_i)$ in front of
$(c_4+c_6)$ in $C_p$ [see Eq. (\ref{ct})].}
\be
c_1^{\rm eff}\Big|_{\mu_f=m_b} &=& c_1(\mu)+{\alpha_s\over
4\pi}\left(\gamma^{(0)T}\ln{m_b\over \mu}+\hat
r^T\right)_{1i}c_i(\mu), \non \\ c_2^{\rm eff}\Big|_{\mu_f=m_b}
&=& c_2(\mu)+{\alpha_s\over 4\pi}\left(\gamma^{(0)T}\ln{m_b\over
\mu}+\hat r^T\right)_{2i}c_i(\mu), \non \\ c_3^{\rm
eff}\Big|_{\mu_f=m_b} &=& c_3(\mu)+{\alpha_s\over
4\pi}\left(\gamma^{(0)T}\ln{m_b\over \mu}+\hat
r^T\right)_{3i}c_i(\mu)-{\alpha_s\over 24\pi}(C_t+C_p+C_g), \non
\\ c_4^{\rm eff}\Big|_{\mu_f=m_b} &=& c_4(\mu)+{\alpha_s\over
4\pi}\left(\gamma^{(0)T}\ln{m_b\over \mu}+\hat
r^T\right)_{4i}c_i(\mu)+{\alpha_s\over 8\pi}(C_t+C_p+C_g), \non \\
c_5^{\rm eff}\Big|_{\mu_f=m_b} &=& c_5(\mu)+{\alpha_s\over
4\pi}\left(\gamma^{(0)T}\ln{m_b\over \mu}+\hat
r^T\right)_{5i}c_i(\mu)-{\alpha_s\over 24\pi}(C_t+C_p+C_g), \non
\\ c_6^{\rm eff}\Big|_{\mu_f=m_b} &=& c_6(\mu)+{\alpha_s\over
4\pi}\left(\gamma^{(0)T}\ln{m_b\over \mu}+\hat
r^T\right)_{6i}c_i(\mu)+{\alpha_s\over 8\pi}(C_t+C_p+C_g), \non \\
c_7^{\rm eff}\Big|_{\mu_f=m_b} &=& c_7(\mu)+{\alpha_s\over
4\pi}\left(\gamma^{(0)T}\ln{m_b\over \mu}+\hat
r^T\right)_{7i}c_i(\mu)+{\alpha\over 8\pi}C_e, \non \\ c_8^{\rm
eff}\Big|_{\mu_f=m_b} &=& c_8(\mu)+{\alpha_s\over
4\pi}\left(\gamma^{(0)T}\ln{m_b\over \mu}+\hat
r^T\right)_{8i}c_i(\mu), \non \\ c_9^{\rm eff}\Big|_{\mu_f=m_b}
&=& c_9(\mu)+{\alpha_s\over 4\pi}\left(\gamma^{(0)T}\ln{m_b\over
\mu}+\hat r^T\right)_{9i}c_i(\mu)+{\alpha\over 8\pi}C_e, \non \\
c_{10}^{\rm eff}\Big|_{\mu_f=m_b} &=& c_{10}(\mu)+{\alpha_s\over
4\pi}\left(\gamma^{(0)T}\ln{m_b\over \mu}+\hat
r^T\right)_{10i}c_i(\mu), \label{ceff1}
\en
where the matrices $\gamma^{(0)}$ as well as $\hat r$ arise from
the vertex corrections to the operators $O_1-O_{10}$ (see Fig.
\ref{fig:4quark}), $C_t$, $C_p$, $C_e$ and $C_g$ from the QCD
penguin-type diagrams of the operators $O_{1,2}$, the QCD
penguin-type diagrams of the operators $O_3-O_6$, the electroweak
penguin-type diagram of $O_{1,2}$, and tree-level diagram of the
dipole operator $O_g$, respectively:
\be
C_t &=& -\left({\lambda_u\over\lambda_t}\tilde
G(m_u)+{\lambda_c\over\lambda_t}\tilde G(m_c)\right)c_1, \non \\
C_p &=& [\tilde G(m_q)+\tilde G(m_b)]c_3+\sum_{i=u,d,s,c,b}\tilde
G(m_i)(c_4+c_6),  \non \\ C_g &=& -{2m_b\over \sqrt{\la k^2\ra}
}c^{\rm eff}_{g},  \non  \\ C_e &=& -{8\over
9}\left({\lambda_u\over\lambda_t}\tilde
G(m_u)+{\lambda_c\over\lambda_t}\tilde G(m_c)\right)(c_1+3c_2),
\non \\ \tilde G(m_q) &=& {2\over 3}\kappa-G(m_q,k,\mu),
\label{ct}
\en
with $\lambda_{q'}\equiv V_{q'b}V_{q'q}^*$, and $\kappa$ being a
parameter characterizing the $\gamma_5$ scheme dependence in
dimensional regularization, for example,
\be
\kappa=\cases{ 1 & NDR,  \cr 0 & HV.  \cr}
\en
The function $G(m,k,\mu)$ in Eq. (\ref{ct}) is given by \be
\label{G} G(m,k,\mu)=-4\int^1_0dx\,x(1-x)\ln\left(
{m^2-k^2x(1-x)\over \mu^2}\right),
\en
where $k^2$ is the momentum squared carried by the virtual gluon.
For $k^2>4m^2$, its analytic expression is given by
\be
&& {\rm Re}\,G=\,{2\over 3}\left(-\ln{m^2\over\mu^2}+{5\over
3}+4{m^2\over k^2}-(1+2{m^2\over k^2})\sqrt{1-4{m^2\over
k^2}}\,\ln\,{1+\sqrt{1-4{m^2 \over k^2}}\over 1-\sqrt{1-4{m^2\over
k^2}}}\right),   \non \\ && {\rm Im}\,G=\,{2\over
3}\pi\left(1+2{m^2\over k^2}\right)\sqrt{1-4{m^2 \over k^2}}.
\en
It should be remarked that although the penguin coefficients
$c_{3}-c_{10}$ are governed by the penguin diagrams with $t$ quark
exchange, the effective Wilson coefficients do incorporate the
perturbative effects of the penguin diagrams with internal $u$ and
$c$ quarks induced by the current-current operator $O_1$.

The matrix $\hat r$ in (\ref{ceff1}) gives momentum-independent
constant terms which depend on the treatment of $\gamma_5$ in
dimensional regularization. To compute the anomalous dimension
$\gamma^{(0)}$ and the matrix $\hat r$, we work in the on-shell
(massless) fermion scheme and assume zero momentum transfer
squared between color-singlet currents, i.e. $(p_1-p_3)^2=0$ for
$O_{\rm odd}$ operators and $(p_1-p_4)^2=0$ for $O_{\rm even}$
operators as required by the light final bound state, for which
the transferred energy squared is equal to the mass squared of the
bound state and hence is negligible. Then energy conservation
implies that $(p_1+p_2)^2=-(-p_2+p_3)^2$ for the case of massless
external fermions if $(p_1-p_3)^2=0$. However, the $b$ quark mass
cannot be ignored in the $b$ decay processes. Therefore, we should
have $(p_1+p_2)^2+(-p_2+p_3)^2=m_b^2$ for charmless $B$ decays.
Considering the possible spectator quark effects, the reasonable
kinematic range for $(p_1+p_2)^2$ and $(-p_2+p_3)^2$ lies in the
region between $m_b^2/2$ and $m_b^2$. Here we choose
$(p_1+p_2)^2\approx (-p_2+p_3)^2\approx m_b^2$ for $O_{\rm odd}$
operators and $(p_1-p_4)^2=0$, $(p_1+p_2)^2\approx
(-p_2+p_4)^2\approx m_b^2$ for $O_{\rm even}$ operators (see Fig.
\ref{fig:4quark} for notation). The results are insensitive to the
kinematics of the Mandelstam variables.

We obtain the logarithmic term $\ln(m_b/\mu)$ in Eq. (\ref{ceff1})
with the anomalous dimension \cite{Ali,Buras96}
\be
\label{gamma} \gamma^{(0)}=\left(\matrix{ -2 & 6 & 0 & 0 & 0 & 0 &
0 & 0 & 0 & 0 \cr 6 & -2 & 0 & 0 & 0 & 0 & 0 & 0 & 0 & 0 \cr 0 & 0
& -2 & 6 & 0 & 0 & 0 & 0 & 0 & 0 \cr  0 & 0 & 6 & -2  & 0 & 0 & 0
& 0 & 0 & 0 \cr 0 & 0 & 0 & 0 & 2 & -6 & 0 & 0 & 0 & 0 \cr 0 & 0 &
0 & 0 & 0 & -16 & 0 & 0 & 0 & 0 \cr 0 & 0 & 0 & 0 & 0 & 0 & 2 & -6
& 0 & 0 \cr  0 & 0 & 0 & 0 & 0 & 0 & 0 & -16 & 0 & 0 \cr 0 & 0 & 0
& 0 & 0 & 0 & 0 & 0 & -2 & 6 \cr  0 & 0 & 0 & 0 & 0 & 0 & 0 & 0 &
6 & -2 \cr}\right),
\en
and the matrix $\hat r$
\be
\label{rndr} \hat r_{\rm NDR}=\left(\matrix{ 3 & -9 & 0 & 0 & 0 &
0 & 0 & 0 & 0 & 0 \cr -9 & 3 & 0 & 0 & 0 & 0 & 0 & 0 & 0 & 0 \cr 0
& 0 & 3 & -9 & 0 & 0 & 0 & 0 & 0 & 0 \cr  0 & 0 & -9 & 3 & 0 & 0 &
0 & 0 & 0 & 0 \cr 0 & 0 & 0 & 0 & -1 & 3 & 0 & 0 & 0 & 0 \cr 0 & 0
& 0 & 0 & -3 & 1 & 0 & 0 & 0 & 0 \cr 0 & 0 & 0 & 0 & 0 & 0 & -1 &
3 & 0 & 0 \cr 0 & 0 & 0 & 0 & 0 & 0 & -3 & 1 & 0 & 0 \cr 0 & 0 & 0
& 0 & 0 & 0 & 0 & 0 & 3 & -9 \cr 0 & 0 & 0 & 0 & 0 & 0 & 0 & 0 &
-9 & 3 \cr}\right)
\en
in the NDR scheme, and
\be
\label{rhv} \hat r_{\rm HV}=\left(\matrix{ {7\over 3} & -7 & 0 & 0
& 0 & 0 & 0 & 0 & 0 & 0 \cr -7 & {7\over 3} & 0 & 0 & 0 & 0 & 0 &
0 & 0 & 0 \cr 0 & 0 & {7\over 3} & -7 & 0 & 0 & 0 & 0 & 0 & 0 \cr
0 & 0 & -7 & {7\over 3} & 0 & 0 & 0 & 0 & 0 & 0 \cr 0 & 0 & 0 & 0
& -3 & 9 & 0 & 0 & 0 & 0 \cr 0 & 0 & 0 & 0 & 1 & -{1\over 3} & 0 &
0 & 0 & 0 \cr 0 & 0 & 0 & 0 & 0 & 0 & -3 & 9 & 0 & 0 \cr 0 & 0 & 0
& 0 & 0 & 0 & 1 & -{1\over 3} & 0 & 0 \cr 0 & 0 & 0 & 0 & 0 & 0 &
0 & 0 & {7\over 3} & -7 \cr 0 & 0 & 0 & 0 & 0 & 0 & 0 & 0 & -7 &
{7\over 3} \cr}\right)
\en
in the HV scheme. It is interesting to note that the $(V-A)(V-A)$
operators $O_1-O_4$, $O_9$, $O_{10}$ have the same $\gamma^{(0)}$
and $\hat r$ matrix elements and likewise for $(V-A)(V+A)$
operators $O_5-O_8$.

From Eq. (\ref{ceff1}) it is clear that the effective Wilson
coefficients depend on CKM matrix matrix elements and the gluon's
momentum  $k^2$. Using the next-to-leading order (NLO) $\Delta
B=1$ Wilson coefficients obtained in the HV and NDR schemes at
$\mu={m}_b(m_b)$, $\Lambda^{(5)}_{\overline{\rm MS}}=225$ MeV and
$m_t=170$ GeV in Table 22 of \cite{Buras96}, we obtain the
numerical values of effective renormalization-scheme and -scale
independent, gauge invariant Wilson coefficients $c_i^{\rm eff}$
for $b\to s$, $b\to d$ and $\bar b\to\bar d$ transitions (see
Table \ref{tab:wcs}), where uses have been made of the Wolfenstein
parameters $\rho=0.175$ and $\eta=0.370$ (see Sec. III.A) and the
quark masses given in Eq. (\ref{qmass}). Note that effective
Wilson coefficients for $\bar b\to \bar s$ transition are the same
as that for $b\to s$ to the accuracy considered in Table
\ref{tab:wcs}. We see that $c_2^{\rm eff}=-0.365$ is quite
different from the NLO Wilson coefficients: $c_2^{\rm
NDR}(m_b)=-0.185$ and $c_2^{\rm HV}(m_b)=-0.228$ \cite{Buras96},
but close to the lowest order value $c_2^{\rm L.O.}(m_b)=-0.308$
\cite{Buras96}.

\vskip 0.4cm
\begin{table}[ht]
\caption{Numerical values of the effective Wilson coefficients
$c_i^{\rm eff}$ for $b\to s$, $b\to d$ and $\bar b\to\bar d$
transitions evaluated at $\mu_f=m_b$ and $k^2=m_b^2/2$, where use
of the Wolfenstein parameters $\rho=0.175$ and $\eta=0.370$ has
been made. \label{tab:wcs}}
\begin{center}
\begin{tabular}{ l c c c  }
 & $b\to s$,~$\bar b\to\bar s$ & $b\to d$ & $\bar b\to\bar d$  \\ \hline
$c_1^{\rm eff}$ & 1.168 & 1.168 & 1.168 \\
$c_2^{\rm eff}$ & $-0.365$ & $-0.365$ & $-0.365$ \\
$c_3^{\rm eff}$ & $0.0225+i0.0045$ & $0.0224+i0.0038$ & $0.0227+i0.0052$ \\
$c_4^{\rm eff}$ & $-0.0458-i0.0136$ & $-0.0454-i0.0115$ &
$-0.0464-i0.0155$ \\
$c_5^{\rm eff}$ & $0.0133+i0.0045$ &
$0.0131+i0.0038$ & $0.0135+i0.0052$ \\
$c_6^{\rm eff}$ &
$-0.0480-i0.0136$ & $-0.0475-i0.0115$ & $-0.0485-i0.0155$ \\
$c_7^{\rm eff}/\alpha$ & $-0.0303-i0.0369$ & $-0.0294-i0.0329$ &
$-0.0314-i0.0406$ \\
$c_8^{\rm eff}/\alpha$ & 0.055 & 0.055 &
0.055 \\
$c_9^{\rm eff}/\alpha$ & $-1.427-i0.0369$ &
$-1.426-i0.0329$ & $-1.428-i0.0406$ \\
$c_{10}^{\rm eff}/\alpha$ &
0.48 & 0.48 & 0.48 \\
\end{tabular}
\end{center}
\end{table}

Several remarks are in order. (i) There exist infrared double
poles, i.e., $1/\epsilon_{\rm IR}^2$, in some of the amplitudes in
Fig. \ref{fig:4quark}, but they are canceled out when summing over
all the amplitudes. (ii) Care must be taken when applying the
projection method to reduce the tensor products of Dirac matrices
to the form $\Gamma\otimes\Gamma$ with
$\Gamma=\gamma_\mu(1-\gamma_5)$. As shown in \cite{CLY}, sometimes
it is erroneous to apply the projection method without taking into
account the effect of evanescent operators. (iii) When quarks are
on their mass shell, it is straightforward to show that the gauge
dependent contributions to Figs. 1(a) and 1(b) are compensated by
that of Figs. 1(c) and 1(d), while the gauge dependent part of
Figs. 1(e) and 1(f) is canceled by that of the quark wave function
renormalization \cite{CLY}. (iv) For comparison with (\ref{rndr}),
the matrix $\hat r$ obtained in the NDR $\gamma_5$ scheme using
off-shell regularization and Landau gauge is given by\footnote{Our
expression for $\hat r^{\lambda=0}_{\rm NDR}$ is slightly
different from that given in \cite{Ali} for the matrix elements
$\hat r_{55}$ to $\hat r_{88}$.}
\be
\label{rndr1} \hat r_{\rm NDR}^{\lambda=0}=\left(\matrix{ {7\over
3} & -7 & 0 & 0 & 0 & 0 & 0 & 0 & 0 & 0 \cr -7 & {7\over 3} & 0 &
0 & 0 & 0 & 0 & 0 & 0 & 0 \cr 0 & 0 & {7\over 3} & -7 & 0 & 0 & 0
& 0 & 0 & 0 \cr  0 & 0 & -7 & {7\over 3} & 0 & 0 & 0 & 0 & 0 & 0
\cr 0 & 0 & 0 & 0 & -{2\over 3} & 2 & 0 & 0 & 0 & 0 \cr 0 & 0 & 0
& 0 & -2 & {34\over 3} & 0 & 0 & 0 & 0 \cr 0 & 0 & 0 & 0 & 0 & 0 &
-{2\over 3} & 2 & 0 & 0 \cr 0 & 0 & 0 & 0 & 0 & 0 & -2 & {34\over
3} & 0 & 0 \cr 0 & 0 & 0 & 0 & 0 & 0 & 0 & 0 & {7\over 3} & -7 \cr
0 & 0 & 0 & 0 & 0 & 0 & 0 & 0 & -7 & {7\over 3} \cr}\right).
\en
The numerical results of $c^{\rm eff}_i$ obtained using
(\ref{rndr1}) are similar to that listed in Table I except for
$c^{\rm eff}_{2,6,10}$: $c^{\rm eff}_2=-0.325$, $c^{\rm
eff}_6=-0.0560-i0.0136$ and $c^{\rm eff}_{10}=0.263\alpha$. We see
from Table I that, contrary to the commonly used value (but not
gauge invariant) Re\,$c_6^{\rm eff}\approx -0.060\sim -0.063$ in
the literature, the {\it gauge-invariant} effective penguin
coefficient Re\,$c_6^{\rm eff}\approx -0.048$ does not get much
enhancement, recalling that $c_6(m_b)=-0.041$ to NLO
\cite{Buras96}. (v) To check the scheme and scale independence of
effective Wilson coefficients, say $c^{\rm eff}_{1,2}$, it is
convenient to work in the diagonal basis in which the operators
$O_\pm={1\over 2}(O_1\pm O_2)$ do not mix under renormalization.
The Wilson coefficients in general have the expressions
\cite{Buras96}:
\be
c_\pm(\mu)=\left[1+{\alpha_s(\mu)\over 4\pi}J_\pm\right]
\left[{\alpha_s(m_W)\over
\alpha_s(\mu)}\right]^{\gamma^{(0)}_\pm/(2\beta_0)}
\left[1+{\alpha_s(m_W)\over 4\pi}(B_\pm-J_\pm)\right], \label{cw}
\en
where $c_\pm=c_1\pm c_2$, $\beta_0=11-{2\over 3}n_f$ with $n_f$
being the number of flavors between $m_W$ and $\mu$ scales,
$B_\pm$ specifies the initial condition of $c(m_W)$:
$c(m_W)=1+{\alpha_s(m_W)\over 4\pi}B_\pm$ and it is
$\gamma_5$-scheme dependent, and $J_\pm=\gamma^{(0)}_\pm
\beta_1/(2\beta_0^2)- \tilde\gamma^{(1)}_\pm /(2\beta_0)$ with
$\beta_1=102-38n_f/3$. The anomalous dimensions
$\tilde\gamma^{(1)}_\pm=\gamma^{(1)}_\pm-2\gamma_J$ are
$\gamma_5$-scheme dependent, where $\gamma^{(1)}_\pm$ are the
two-loop anomalous dimensions of $O_\pm$ and $\gamma_J$ is the
anomalous dimension of the weak current in full theory. (The
complete expression for $\gamma_\pm^{(1)}$ and $\gamma_J$ in
different schemes can be found in \cite{Buras96}, for example.) As
stressed in \cite{Buras98b}, $c(\mu)$ do not depend on the
external states; any external state can be used for their
extraction, the only requirement being that the infrared and mass
singularities are properly regularized. This means that the
short-distance Wilson coefficients calculated from Eq. (\ref{cw})
are independent of the fermion state, on-shell or off-shell. Since
$B_\pm-J_\pm$ is scheme independent \cite{Buras96,Buras98b}, the
scheme dependence of $c_\pm(\mu)$ is solely governed by $J_\pm$.
Using the fact that $\tilde \gamma^{(1)}_\pm$ are also free of the
external-state dependence, we have shown explicitly in \cite{CLY}
the renormalization scheme independence of $\hat r_\pm^T+J_\pm$
and hence $c^{\rm eff}_\pm$. It is also straightforward to show
that, to the leading logarithmic approximation, the scale
dependence of $c_\pm(\mu)$ arising from the
$\alpha_s(m_W)/\alpha_s(\mu)$ term in Eq. (\ref{cw}) is
compensated by the $\gamma^{(0)T}\ln(m_b/\mu)$ term in Eq.
(\ref{ceff1}).

\section{Input Parameters}
\subsection{Quark mixing matrix}
   It is convenient to parametrize the quark mixing matrix in terms of the
Wolfenstein parameters: $A,~\lambda,~\rho$ and $\eta$ \cite{Wolf},
\be
V=\left( \matrix{ 1-{1\over 2}\lambda^2 & \lambda & A\lambda^3(\rho-i\eta)
\cr    -\lambda & 1-{1\over 2}\lambda^2 & A\lambda^2 \cr
A\lambda^3(1-\rho-i\eta) & -A\lambda^2 & 1\cr }  \right)+{\cal O}(\lambda^4),
\en
where $\lambda=0.2205$ is equivalent to $\sin\theta_C$ with
$\theta_C$ being the Cabibbo angle. Note that this parametrization
is an approximation of the exact Chau-Keung parametrization
\cite{CK} of the quark mixing matrix. For the parameter $A$, we
fix it to $A=0.815$ corresponding to $|V_{cb}|=0.0396$. As for the
parameters $\rho$ and $\eta$, two different updated analyses
\cite{Parodi,Mele} have been performed using the combination of
the precise measurement of $\Delta M_d$, the mass difference in
the $B_d$ system, the updated limit on $\Delta M_s$, the mass
difference in the $B_s$ system, and the determination of
$|V_{ub}|$ from charmless semileptonic $B$ decays. The results
\be
\rho(1-{\lambda^2\over 2})=\,0.189\pm 0.074\,,   \qquad \eta(1-{
\lambda^2\over 2})=\,0.354\pm 0.045,
\en
and
\be
\rho=0.160^{+0.094}_{-0.070}, \qquad \quad \eta=0.381^{+0.061}_{-0.058}
\en
are cited in \cite{Parodi} and \cite{Mele}, respectively; they are
obtained by a simultaneous fit to all the available data. In
either fit, it is clear that $\sqrt{\rho^2+\eta^2}=0.41$ is
slightly larger than the previous analysis. For our purposes in
the present paper we will employ the values $\rho=0.175$ and
$\eta=0.370$; they correspond to the unitarity angles:
$\alpha=91^\circ$, $\beta=24^\circ$ and $\gamma=65^\circ$. We
shall see in Sec. V that some of hadronic rare $B$ decay data will
be much more easily accounted for if $\gamma>90^\circ$ or
$\rho<0$. Therefore, we shall use $\gamma=65^\circ$ as a
benchmarked value and then discuss the impact of a negative $\rho$
whenever necessary.

\subsection{Running quark masses}
    We shall see later that running quark masses appear in the matrix
elements of $(S-P)(S+P)$ penguin operators through the use of
equations of motion. The running quark mass should be applied at
the scale $\mu\sim m_b$ because the energy released in the
energetic two-body charmless decays of the $B$ meson is of order
$m_b$. Explicitly, we use \cite{Fusaoku}
\be
&& m_u(m_b)=3.2\,{\rm MeV},  \qquad m_d(m_b)=6.4\,{\rm MeV},
\qquad m_s(m_b)=90\,{\rm MeV},  \non \\ &&  m_c(m_b)=0.95\,{\rm
GeV},  \qquad m_b(m_b)=4.34\,{\rm GeV}, \label{qmass}
\en
in ensuing calculation, where $m_s(m_b)=90$ MeV corresponds to
$m_s=140$ MeV at $\mu=1$ GeV.
\subsection{Decay constants}
   For the decay constants we use $f_\pi=132$ MeV, $f_K=160$ MeV, $f_\rho=
216$ MeV, $f_{K^*}=221$ MeV, $f_\omega=195$ MeV and $f_\phi=237$
MeV. To determine the decay constants of the $\eta$ and $\eta'$
mesons, defined by $\la 0|\bar q\gamma_\mu\gamma_5
q|\etapp\ra=if^q_\etapp p_\mu$, we need to know the wave functions
of the physical $\eta'$ and $\eta$ states which are related to
that of the SU(3) singlet state $\eta_0$ and octet state $\eta_8$
by
\be
\eta'=\eta_8\sin\theta+\eta_0\cos\theta, \qquad \eta=\eta_8\cos\theta-\eta_0
\sin\theta.
\en
When the $\eta-\eta'$ mixing angle is $-19.5^\circ$, the
expressions of the $\eta'$ and $\eta$ wave functions become very
simple \cite{Chau1}:
\be
|\eta'\ra={1\over\sqrt{6}}|\bar uu+\bar dd+2\bar ss\ra, \qquad
|\eta\ra={1\over\sqrt{3}}|\bar uu+\bar dd-\bar ss\ra,
\en
recalling that
\be
|\eta_0\ra={1\over\sqrt{3}}|\bar uu+\bar dd+\bar ss\ra, \qquad
|\eta_8\ra={1\over\sqrt{6}}|\bar uu+\bar dd-2\bar ss\ra.
\en
At this specific mixing angle, $f_{\eta'}^u={1\over 2}f_{\eta'}^s$ in the
SU(3) limit. Introducing the decay constants $f_8$ and $f_0$ by
\be
\la 0|A_\mu^0|\eta_0\ra=if_0 p_\mu, \qquad \la 0|A_\mu^8|\eta_8
\ra=if_8 p_\mu,
\en
and noting that due to SU(3) breaking the matrix elements $\la 0|A_\mu^{0(8)}
|\eta_{8(0)}\ra$ do not vanish in general and they will induce a
two-angle mixing among the decay constants:
\be
f_{\eta'}^u={f_8\over\sqrt{6}}\sin\theta_8+{f_0\over\sqrt{3}}\cos\theta_0,
\qquad f_{\eta'}^s=-2{f_8\over\sqrt{6}}\sin\theta_8+{f_0\over\sqrt{3}}\cos
\theta_0.
\en
Likewise, for the $\eta$ meson
\be
f_{\eta}^u={f_8\over\sqrt{6}}\cos\theta_8-{f_0\over\sqrt{3}}\sin\theta_0,
\qquad f_{\eta}^s=-2{f_8\over\sqrt{6}}\cos\theta_8-{f_0\over\sqrt{3}}\sin
\theta_0.
\en
It must be accentuated that the two-mixing angle formalism
proposed in \cite{Leutwyler,Kroll2} applies to the decay constants
of the $\eta'$ and $\eta$ rather than to their wave functions.
Based on the ansatz that the decay constants in the quark flavor
basis follow the pattern of particle state mixing, relations
between $\theta_8,~\theta_0$ and $\theta$ are derived in
\cite{Kroll2}. It is found in \cite{Kroll2} that
phenomenologically
\be \label{theta}
\theta_8=-21.2^\circ, \qquad
\theta_0=-9.2^\circ, \qquad \theta=-15.4^\circ,
\en
and
\be \label{f80}
f_8/f_\pi=1.26,  \qquad f_0/f_\pi=1.17.
\en
Numerically, we obtain
\be   \label{fdecay}
f_{\eta}^u=78\,{\rm
MeV}, \quad f_\eta^s=-112\,{\rm MeV}, \quad f_{\eta'}^u= 63\,{\rm
MeV}, \quad f_{\eta'}^s=137\,{\rm MeV}.
\en

The decay constant $f_{\eta'}^c$, defined by $\la 0|\bar
c\gamma_\mu\gamma_5c|\eta'\ra=if_{\eta'}^c q_\mu$, has been
determined from theoretical calculations
\cite{Halperin,Ali2,Araki,Franz} and from the phenomenological
analysis of the data of $J/\psi\to\eta_c\gamma,\,J/\psi
\to\eta'\gamma$ and of the $\eta\gamma$ and $\eta'\gamma$
transition form factors
\cite{Ali,Kroll2,Petrov,Kroll,Cao,Ahmady1}; it lies in the range
--2.0 MeV $\leq f_{\eta'}^c\leq$ --18.4 MeV. In this paper we use
the values
\be
f_{\eta'}^c=-(6.3\pm 0.6)\,{\rm MeV},  \qquad f_\eta^c=-(2.4\pm
0.2) \,{\rm MeV},
\en
as obtained from a phenomenological analysis performed in
\cite{Kroll2}.

\subsection{Form factors}
As for form factors, we follow \cite{BSW85} to use the following
parametrization:
\be
\la 0|A_\mu|P(q)\ra &=& if_Pq_\mu, \qquad \la
0|V_\mu|V(p,\vp)\ra=f_Vm_V\vp_ \mu,   \non \\ \la
P'(p')|V_\mu|P(p)\ra &=& \left(p_\mu+p'_\mu-{m_P^2-m_{P'}^2\over
q^2}\,q_ \mu\right) F_1(q^2)+F_0(q^2)\,{m_P^2-m_{P'}^2\over
q^2}q_\mu,   \non \\ \la V(p',\vp)|V_\mu|P(p)\ra &=& {2\over
m_P+m_V}\,\epsilon_{\mu\nu\alpha \beta}\vp^{*\nu}p^\alpha p'^\beta
V(q^2),   \non \\ \la V(p',\vp)|A_\mu|P(p)\ra &=& i\Big[
(m_P+m_V)\vp^*_\mu A_1(q^2)-{\vp^*\cdot p\over
m_P+m_V}\,(p+p')_\mu A_2(q^2)    \non \\ && -2m_V\,{\vp^*\cdot
p\over q^2}\,q_\mu\big[A_3(q^2)-A_0(q^2)\big]\Big],
\en
where $q=p-p'$, $F_1(0)=F_0(0)$, $A_3(0)=A_0(0)$, and
\be
A_3(q^2)=\,{m_P+m_V\over 2m_V}\,A_1(q^2)-{m_P-m_V\over 2m_V}\,A_2(q^2).
\en
We consider two different form-factor models for heavy-to-light
form factors: the Bauer-Stech-Wirbel (BSW) model \cite{BSW85} and
the light-cone sum rule (LCSR) model \cite{Ball}. The relevant
form factors at zero momentum transfer are listed in Table II.

\vskip 0.4cm
\begin{table}[ht]
{\small Table II.~Form factors at zero momentum transfer for $B\to
P$ and $B\to V$ transitions evaluated in the BSW model
\cite{BSW85,BSW87}. The values given in the square brackets are
obtained in the light-cone sum rule (LCSR) analysis \cite{Ball}.
We have assumed SU(3) symmetry for the $B\to\omega$ form factors
in the LCSR approach. In realistic calculations we use Eq.
(\ref{Beta}) for $B\to\etapp$ form factors.  }
\begin{center}
\begin{tabular}{ l l c c c c }
Decay  & $F_1=F_0$ & $V$ & $A_1$ & $A_2$ & $A_3=A_0$ \\ \hline
$B\to\pi^\pm$ & 0.333~[0.305] & & & &  \\
$B\to K$ & 0.379~[0.341] & & & & \\
$B\to\eta$ & 0.168~[---] & & & & \\
$B\to\eta'$ & 0.114~[---] & & & & \\
$B\to\rho^\pm$ & & 0.329~[0.338] & 0.283~[0.261] & 0.283~[0.223] &
0.281~[0.372] \\
$B\to\omega$ & & 0.232~[0.239] & 0.199~[0.185] & 0.199~[0.158] &
0.198~[0.263] \\
$B\to K^*$ & & 0.369~[0.458] & 0.328~[0.337] & 0.331~[0.203] &
0.321~[0.470] \\
\end{tabular}
\end{center}
\end{table}

  It should be stressed that the form factors for $B\to\etapp$ transition
calculated by BSW \cite{BSW85} did not include the wave function
normalization and mixing angles. In the relativistic quark model
calculation of $B\to \etapp$ transition, BSW put in the $u\bar u$
constituent quark mass only. That is, the form factors considered
by BSW are actually $F_0^{B\eta_{u\bar u}}$ and
$F_0^{B\eta'_{u\bar u}}$. To compute the physical form factors,
one has to take into account the wave function normalizations of
the $\eta$ and $\eta'$:
\be
F_{0,1}^{B\eta}=\left({1\over\sqrt{6}}\cos\theta-{1\over\sqrt{3}}\sin\theta
\right)F_{0,1}^{B\eta_{u\bar u}},  \qquad &&
F_{0,1}^{B\eta'}=\left({1\over\sqrt{6}}\sin\theta+{1\over\sqrt{3}}\cos\theta
\right)F_{0,1}^{B\eta'_{u\bar u}}.
\en
Using $F_0^{B\eta_{u\bar u}}(0)=0.307$ and $F_0^{B\eta'_{u\bar
u}}(0)=0.254$ from BSW \cite{BSW87}, we find
$F_0^{B\eta}(0)=0.168$ and $F_0^{B\eta'}(0)=0.114$ as shown in
Table II. However, as we shall see in Sec.~V.D, the form factor
$F_0^{B\eta'}$ is preferred to be a bit larger in order to
accommodate the data of $B\to\eta' K$. Hence, we shall assume the
nonet symmetry relation
$\sqrt{3}F_0^{B\eta_0}(0)=\sqrt{6}F_0^{B\eta_8}(0)=F_0^{B\pi^\pm}(0)$
to obtain $F_0^{B\eta_0}$, $F_0^{B\eta_8}$ and then relate them to
the physical form factors via
\be
F_0^{B\eta}=\cos\theta F_0^{B\eta_8}-\sin\theta F_0^{B\eta_0}, \qquad
F_0^{B\eta'}=\sin\theta F_0^{B\eta_8}+\cos\theta F_0^{B\eta_0}.
\en
Numerically, we obtain
\be
\label{Beta} F_0^{B\eta}(0)=0.181, \qquad\quad
F_0^{B\eta'}(0)=0.148,
\en
for $F_0^{B\pi^\pm}(0)=0.33$.

For the $q^2$ dependence of form factors in the region where $q^2$ is not
too large, we shall use the pole dominance ansatz, namely,
\be
f(q^2)=\,{f(0)\over \left(1-{q^2/m^2_*}\right)^n},
\en
where $m_*$ is the pole mass given in \cite{BSW87}. A direct
calculation of $B\to P$ and $B\to V$ form factors at timelike
momentum transfer is available in the relativistic light-front
quark model \cite{CCH} with the results that the $q^2$ dependence
of the form factors $A_0,~A_2,~V,~F_1$ is a dipole behavior (i.e.
$n=2$), while $A_1,~F_0$ exhibit a monopole dependence ($n=1$).
Note that the original BSW model assumes a monopole behavior for
all the form factors. This is not consistent with heavy quark
symmetry for heavy-to-heavy transition. Therefore, in the present
paper we will employ the BSW model for the heavy-to-light form
factors at zero momentum transfer but take a different ansatz for
their $q^2$ dependence, namely a dipole dependence for
$F_1,A_0,A_2$ and $V$. In the light-cone sum rule analysis of
\cite{Ball}, the form-factor $q^2$ dependence is evaluated using
the parametrization
\be
f(q^2)=\,{f(0)\over 1-a(q^2/m_{B}^2)+b(q^2/m_{B}^2)^2},
\en
where the values of $a$ and $b$ are given in \cite{Ball}. The
hadronic charmless $B$ decays are in general insensitive to the
expressions of form-factor $q^2$ dependence because $q^2$ is
small. Nevertheless, we find that the decay rates of $B\to VV$
show a moderate dependence on the $q^2$ behavior of form factors.

\section{Factorized Amplitudes}
\subsection{Effective parameters and nonfactorizable effects}
It is known that the effective Wilson coefficients appear in the
factorizable decay amplitudes in the combinations $a_{2i}=
{c}_{2i}^{\rm eff}+{1\over N_c}{c}_{2i-1}^{\rm eff}$ and
$a_{2i-1}= {c}_{2i-1}^{\rm eff}+{1\over N_c}{c}^{\rm eff}_{2i}$
$(i=1,\cdots,5)$. Phenomenologically, the number of colors $N_c$
is often treated as a free parameter to model the nonfactorizable
contribution to hadronic matrix elements and its value can be
extracted from the data of two-body nonleptonic decays. As shown
in \cite{Cheng94,Kamal94,Soares}, nonfactorizable effects in the
decay amplitudes of $B\to PP,~VP$ can be absorbed into the
parameters $a_i^{\rm eff}$. This amounts to replacing $N_c$ in
$a^{\rm eff}_i$ by $(N_c^{\rm eff})_i$. Explicitly,
\be
a_{2i}^{\rm eff}={c}_{2i}^{\rm eff}+{1\over (N_c^{\rm
eff})_{2i}}{c}_{2i-1}^{ \rm eff}, \qquad \quad a_{2i-1}^{\rm eff}=
{c}_{2i-1}^{\rm eff}+{1\over (N_c^{\rm eff})_{2i-1}}{c}^{\rm
eff}_{2i}, \qquad (i=1,\cdots,5),
\en
where
\be
(1/N_c^{\rm eff})_i\equiv (1/N_c)+\chi_i\,,
\en
with $\chi_i$ being the nonfactorizable terms which receive main
contributions from color-octet current operators \cite{Neubert}.
In the absence of final-state interactions, we shall assume that
$\chi_i$ and hence $(\nc)_i$ are real. If $\chi_i$ are universal
(i.e. process independent) in charm or bottom decays, then we have
a generalized factorization scheme in which the decay amplitude is
expressed in terms of factorizable contributions multiplied by the
universal effective parameters $a_i^{\rm eff}$. For $B\to VV$
decays, this new factorization implies that nonfactorizable terms
contribute in equal weight to all partial wave amplitudes so that
$a_i^{\rm eff}$ {\it can} be defined. Phenomenological analyses of
the two-body decay data of $D$ and $B$ mesons indicate that while
the generalized factorization hypothesis in general works
reasonably well, the effective parameters $a_{1,2}^{\rm eff}$ do
show some variation from channel to channel, especially for the
weak decays of charmed mesons (see e.g. \cite{Cheng94}). A recent
updated analysis of $B\to D\pi$ data gives \cite{CY} \be
\label{a2} \nc(B\to D\pi)\sim (1.8-2.1),\qquad\quad \chi_2(B\to
D\pi)\sim (0.15-0.24).
\en

It is customary to assume in the literature that $(N_c^{\rm
eff})_1 \approx (N_c^{\rm eff})_2\cdots\approx (N_c^{\rm
eff})_{10}$ so that the subscript $i$ can be dropped; that is, the
nonfactorizable term is often postulated to behave in the same way
in penguin and tree decay amplitudes. A closer investigation shows
that this is not the case. We have argued in \cite{CT98} that
nonfactorizable effects in the matrix elements of $(V-A)(V+A)$
operators are {\it a priori} different from that of $(V-A)(V-A)$
operators. One primary reason is that the Fierz transformation of
the $(V-A)(V+A)$ operators $O_{5,6,7,8}$ is quite different from
that of $(V-A)(V-A)$ operators $O_{1,2,3,4}$ and $O_{9,10}$. As a
result, contrary to the common assertion, $\nc(LR)$ induced by the
$(V-A)(V+A)$ operators are theoretically different from $\nc(LL)$
generated by the $(V-A)(V-A)$ operators \cite{CT98}. Therefore, we
shall assume that
\be
&& N_c^{\rm eff}(LL)\equiv
\left(N_c^{\rm eff}\right)_1\approx\left(N_c^{\rm eff}\right)_2\approx
\left(N_c^{\rm eff}\right)_3\approx\left(N_c^{\rm eff}\right)_4\approx
\left(N_c^{\rm eff}\right)_9\approx
\left(N_c^{\rm eff}\right)_{10},   \non\\
&& N_c^{\rm eff}(LR)\equiv
\left(N_c^{\rm eff}\right)_5\approx\left(N_c^{\rm eff}\right)_6\approx
\left(N_c^{\rm eff}\right)_7\approx
\left(N_c^{\rm eff}\right)_8,
\en
and $N_c^{\rm eff}(LR)\neq N_c^{\rm eff}(LL)$ in general. In
principle, $N_c^{\rm eff}$ can vary from channel to channel, as in
the case of charm decay. However, in the energetic two-body $B$
decays, $\nc$ is expected to be process insensitive as supported
by the data \cite{Neubert}. From the data analysis in Sec. V, we
shall see that $\nc(LL)<3$ and $\nc(LR)>3$.

   The $\nc$-dependence of the effective parameters $a_i^{\rm eff}$
is shown in Table III for several representative values of $\nc$.
From the Table we see that (i) the dominant coefficients are
$a_1,\,a_2$ for current-current amplitudes, $a_4$ and $a_6$ for
QCD penguin-induced amplitudes, and $a_9$ for electroweak
penguin-induced amplitudes, and (ii) $a_1,a_4,a_6$ and $a_9$ are
$\nc$-stable, while the others depend strongly on $\nc$.
Therefore, for charmless $B$ decays whose decay amplitudes depend
dominantly on $\nc$-stable coefficients, their decay rates can be
reliably predicted within the factorization approach even in the
absence of information on nonfactorizable effects.

\begin{table}[ht]
{\small Table III. Numerical values for the effective coefficients
$a_i^{\rm eff}$ for $b\to s$ transition at $\nc=2,3,5,\infty$ (in
units of $10^{-4}$ for $a_3,\cdots,a_{10}$). For simplicity we
will drop the superscript ``eff'' henceforth. }
\begin{center}
\begin{tabular}{ c c c c c }
 & $N_c^{\rm eff}=2$ & $N_c^{\rm eff}=3$ & $N_c^{\rm eff}=5$  &
$N_c^{\rm eff}=\infty$ \\ \hline
$a_1$  &0.985 & 1.046  & 1.095 & 1.168  \\
$a_2$  &0.219 &  0.024 & --0.131 &  --0.365 \\
$a_3$ &$-4.15-22.8i$   & 72 & $133+18.1i$ &$225+45.3i$ \\
$a_4$ &$-345-113i$ & $-383-121i$  & $-413-127i$ & $-458-136i$ \\
$a_5$ &$-107-22.7i$ & $-27$  & $36.7+18.2i$ & $133+ 45.4i$ \\
$a_6$ &$-413-113i$ & $-435-121i$  & $-453-127i$ & $-480-136i$ \\
$a_7$ &$-0.22-2.73i$ & $-0.89-2.73i$  & $-1.43-2.73i$ & $-2.24-2.73i$
\\
$a_8$  &$2.93 -1.37i$ & $3.30-0.91i$  & $3.60-0.55i$ & 4   \\
$a_9$  &$-87.9- 2.71i$ & $-93.9-2.71i$  & $-98.6-2.71i$ &
$-105-2.71i$ \\
$a_{10}$ &$-17.3-1.36i$ & $0.32-0.90i$  &$14.4-0.54i$ & 36   \\
\end{tabular}
\end{center}
\end{table}

\subsection{Factorized amplitudes and their classification}
Applying the effective Hamiltonian (2.1), the factorizable decay
amplitudes of $B_u,~B_d\to PP,VP,VV$ obtained  within the
generalized factorization approach are tabulated in the Appendix.
Note that while our factorized amplitudes agree
with that presented in \cite{AKL}, we do include $W$-exchange,
$W$-annihilation and spacelike penguin matrix elements in the
expressions of decay amplitudes, though they are usually neglected
in practical calculations of decay rates. Nevertheless, whether or
not $W$-exchange and $W$-annihilation are negligible should be
tested and the negligence of spacelike penguins (i.e. the terms
$X^{(B,M_1M_2)}$ multiplied by penguin coefficients) is actually
quite questionable (see Sec. V.H. for discussion). Therefore, we
keep trace of annihilation terms and spacelike penguins in the Appendix.

All the penguin contributions to the decay amplitudes can be
derived from Table IV by studying the underlying $b$ quark weak
transitions \cite{CCT,Tseng}. To illustrate this, let
$X^{(BM_1,M_2)}$ denote the factorizable amplitude with the meson
$M_2$ being factored out:
\be
X^{(BM_1,M_2)}=\la M_2|(\bar q_2q_3)\vma|0\ra\la M_1|(\bar q_1b)_\vma|
\ov B\ra.
\en
In general, when $M_2$ is a charged state, only $a_{\rm even}$
penguin terms contribute. For example, from Table IV we obtain
\be
A(\ov B_d\to \pi^+ K^-)_{\rm peng} &\propto& \left[ a_4+a_{10}
+(a_6+a_8)R\right]X^{(\ov B_d \pi^+,K^-)},    \non \\
A(\ov B_d\to \pi^+K^{*-})_{\rm peng} &\propto& \left[ a_4+a_{10}
\right]X^{(\ov B_d \pi^{+},K^{*-})},    \non \\
A(\ov B_d\to \rho^+K^-)_{\rm peng} &\propto& \left[ a_4+a_{10}
-(a_6+a_8)R'\right]X^{(\ov B_d \rho^+,K^-)},
\en
with $R'\approx R\approx m_K^2/(m_bm_s)$. When $M_2$ is a flavor
neutral meson with $I_3=0$, namely, $M_2=\pi^0,\rho^0,\omega$ and
$\etapp$, $a_{\rm odd}$ penguin terms start to contribute. From
Table IV we see that the decay amplitudes of $\ov B\to M\pi^0,~\ov
B\to M\rho^0,~\ov B\to M\omega,~\ov B\to M\etapp$ ($\ov
B=B^-_u,\ov B_d$) contain the following respective factorizable
terms: \be \label{aodd} && {3\over 2}(-a_7+a_9)X_u^{(BM,\pi^0)},
\non \\ && {3\over 2}(a_7+a_9)X_u^{(BM,\rho^0)},   \non \\ &&
(2a_3+2a_5+{1\over 2}a_7+{1\over 2}a_9)X_u^{(BM,\omega)},   \non
\\ && (2a_3-2a_5-{1\over 2}a_7+{1\over 2}a_9)X_u^{(BM,\etapp)},
\en
where the subscript $u$ denotes the $u\bar u$ quark content of the
neutral meson:
\be
X^{(BM,\pi^0)}_u=\la \pi^0|(\bar uu)_\vma|0\ra\la M_1|(\bar q_1b)_\vma|\ov
B\ra.
\en
In deriving Eq.~(\ref{aodd}) we have used the fact that the $d\bar
d$ wave function in the $\pi^0,\rho^0$ ($\omega,\etapp$) has a
sign opposite to (the same as) that of the $u\bar u$ one. QCD
penguins contribute to all charmless $B_u$ and $B_d$ decays except
for $B_u^-\to\pi^-\pi^0,\rho^-\rho^0$ which only receive  $\Delta
I={3\over 2}$ contributions. Applying the rules of Table IV, it is
easily seen that
\be
A( B^-_u\to\pi^-\pi^0)_{\rm peng} &\propto & {3\over 2}\left[-a_7
+a_9+a_{10}+a_8{m^2_\pi\over m_d(m_b-m_d)}\right]X^{(B^-\pi^-,\pi^0)}.
\en

\vskip 0.4cm
\begin{table}[ht]
{\small Table IV. Penguin contributions to the factorizable $B\to
PP,~VP,VV$ decay amplitudes multiplied by
$-(G_F/\sqrt{2})V_{tb}V_{tq}^*$, where $q=d,s$. The notation $B\to
M_1,M_2$ means that the meson $M_2$ can be factored out under the
factorization approximation. In addition to the $a_{\rm even}$
terms, the rare $B$ decays also receive contributions from $a_{\rm
odd}$ penguin effects when $M_2$ is a flavor neutral meson. Except
for $\eta$ or $\eta'$ production, the coefficients $R$ and $R'$
are given by $R=2m_P^2/[(m_1+m_2)(m_b-m_3)]$ and
$R'=-2m_P^2/[(m_1+m_2)(m_b+m_3)]$, respectively. }
\begin{center}
\begin{tabular}{ c c c }
Decay  & $b\to qu\bar u,~b\to qc\bar c$ & $b\to qd\bar d,~b\to qs\bar s$
\\ \hline
$B\to P,P$ & $a_4+a_{10}+(a_6+a_8)R$ & $a_4-{1\over 2}a_{10}+(a_6-{1\over 2}
a_8)R$ \\
$B\to V,P$ & $a_4+a_{10}+(a_6+a_8)R'$ & $a_4-{1\over 2}a_{10}+(a_6-{1\over 2}
a_8)R'$ \\
$B\to P,V$ & $a_4+a_{10}$ & $a_4-{1\over 2}a_{10}$ \\
$B\to V,V$ & $a_4+a_{10}$ & $a_4-{1\over 2}a_{10}$ \\  \hline
$B\to P,P^0$ & $a_3-a_5-a_7+a_9$ & $a_3-a_5+{1\over 2}a_7-{1\over 2}a_9$ \\
$B\to V,P^0$ & $a_3-a_5-a_7+a_9$ & $a_3-a_5+{1\over 2}a_7-{1\over 2}a_9$ \\
$B\to P,V^0$ & $a_3+a_5+a_7+a_9$ & $a_3+a_5-{1\over 2}a_7-{1\over 2}a_9$ \\
$B\to V,V^0$ & $a_3+a_5+a_7+a_9$ & $a_3+a_5-{1\over 2}a_7-{1\over 2}a_9$ \\
\end{tabular}
\end{center}
\end{table}

   Just as the charm decays or $B$ decays into the charmed meson, the
tree-dominated amplitudes for hadronic charmless $B$ decays are customarily
classified into three classes \cite{BSW87}:
\begin{itemize}
\item Class-I for the decay modes dominated by the external $W$-emission
characterized by the parameter $a_1$. Examples are $\ov B_d\to \pi^+\pi^-,\,
\rho^+\pi^-,\,B^-_u\to K^{-(*)}K^{0(*)},\cdots$.
\item Class-II for the decay modes dominated by the color-suppressed
internal $W$-emission characterized by the parameter $a_2$. Examples are
$\ov B_d\to \pi^0\pi^0,\,\omega\pi^0,\cdots$.
\item Class-III decays involving both external and internal $W$ emissions.
Hence the class-III amplitude is of the form $a_1+ra_2$. Examples are
$B^-_u\to\pi^-\pi^0,\,\rho^-\pi^0,\,\omega\pi^-,\cdots$.
\end{itemize}
Likewise, penguin-dominated charmless $B$ decays can be classified
into three categories: \footnote{Our classification of factorized
penguin amplitudes is slightly different from that in \cite{AKL};
we introduce three new classes similar to the classification for
tree-dominated decays.}
\begin{itemize}
\item Class-IV for those decays whose amplitudes are governed by the QCD
penguin parameters $a_4$ and $a_6$ in the combination $a_4+Ra_6$,
where the coefficient $R$ arises from the $(S-P)(S+P)$ part of the
operator $O_6$. In general, $R=2m_{P_b}^2/[(m_1+m_2)(m_b-m_3)]$
for $B\to P_aP_b$ with the meson $P_b$ being factored out under
the factorization approximation,
$R=-2m_{P_b}^2/[(m_1+m_2)(m_b+m_3)]$ for $B\to V_aP_b$, and $R=0$
for $B\to P_aV_b$ and $B\to V_aV_b$. Note that $a_4$ is always
accompanied by $a_{10}$, and $a_6$ by $a_8$. In short, class-IV
modes are governed by $a_{\rm even}$ penguin terms. Examples are
$\ov B_d\to K^-\pi^+,\,K^-\rho^+, \,B^-_u\to \ov
K^0\pi^-,\,K^-K^0,\cdots$.
\item Class-V modes for those decays whose amplitudes are governed by the
effective coefficients $a_3,a_5,a_7$ and $a_9$ (i.e. $a_{\rm odd}$
penguin terms) in the combinations $a_3\pm a_5$ and/or $a_7\pm
a_9$ (see Table IV). Examples are $\ov B_d\to
\phi\pi^0,\,\phi\etapp,\,B^-_u\to\phi\pi^-,\phi\rho^-$.
\item Class-VI modes involving the interference of $a_{\rm even}$ and
$a_{\rm odd}$ terms, e.g. $\ov B_d\to \ov K^0\pi^0,\,\ov
K^0\phi,\, B_u^-\to K^-\pi^0,\,K^-\phi, \cdots$.
\end{itemize}

Sometimes the tree and penguin contributions are comparable. In
this case, the interference between penguin and spectator
amplitudes is at work. There are several such decay modes. For
example, $B^0\to\pi^0 \pi^0,~\etapp\etapp$ involve class-II and
-VI amplitudes, $B^-\to\rho^0K^-,\omega K^-$ consist of class-III
and -VI amplitudes, and $\ov B^0\to \rho^+ K^-$ receives
contributions from class-I and class-IV amplitudes (see Tables V
and VI).

Using the BSW model for form factors, we have computed the
relative magnitudes of tree, QCD and electroweak penguin
amplitudes for all charmless decay modes of $B_u$ and $B_d$ mesons
shown in Tables V-VII as a function of $\nc(LR)$ with two
different considerations for $\nc(LL)$: (a) $\nc(LL)$ being fixed
at the value of 2, and (b) $\nc(LL)=\nc(LR)$. Because of space
limitation, results for CP-conjugate modes are not listed in these
tables. For tree-dominated decays, we have normalized the tree
amplitude to unity. Likewise, the QCD penguin amplitude is
normalized to unity for penguin-dominated decays.

\section{Results for Branching ratios and Discussions}
With the factorized decay amplitudes tabulated in the Appendix and
the input parameters for decay constants, form factors,...,etc.,
shown in Sec.~III, it is ready to compute the decay rates given by
\be
\Gamma(B\to P_1P_2) &=& \frac{ p_c}{8\pi m_B^2}|A(B\to P_1P_2)|^2 \,, \non\\
\Gamma(B\to VP) &=& {p_c^3\over 8\pi m^2_V} |A(B\to VP)/(\vp\cdot
p_{_{B}})|^2,
\en
where
\be
p_c=\frac{\sqrt{[m_B^2-(m_1+m_2)^2][m_B^2-(m_1-m_2)^2]}}{2m_B}
\en
is the c.m. momentum of the decay particles.
For simplicity, we consider a single factorizable amplitude for $B\to VV$:
$A(B\to V_1V_2)=\alpha X^{(BV_1,V_2)}$. Then
\be
\Gamma(B\to V_1V_2)={p_c\over 8\pi m^2_{_{B}} }|\alpha(m_B+m_1)m_2 f_{V_2}
A_1^{BV_1}(m^2_2)|^2H,
\en
with
\be \label{H}
H = (a-bx)^2+2(1+c^2y^2),
\en
and
\be \label{abc}
&&  a={m_{B}^2-m_1^2-m_2^2\over 2m_1m_2}, \qquad b={2m^2_{B}p_c^2\over
m_1m_2(m_{B}+m_1)^2}, \qquad c={2m_{B}p_c\over (m_{B}+m_1)^2}, \non\\
&& x={A_2^{BV_1}(m_2^2)\over A_1^{BV_1}(m^2_2)}, \qquad \quad \qquad
y={V^{BV_1}(m_2^2)\over A_1^{BV_1}(m_2^2)},
\en
where $m_1$ ($m_2$) is the mass of the vector meson $V_1$ ($V_2$).

Branching ratios for all charmless nonleptonic two-body decays of
$B^-_u$ and $\ov B^0_d$ mesons are displayed in Tables VIII-X with
$\nc(LR)=2,3,5,\infty$ and two different considerations for
$\nc(LL)$. For the $B$ meson lifetimes, we use \cite{LEP}
\be
\tau(B^0_d)=(1.57\pm 0.03)\times 10^{-12}s, \qquad
\tau(B^-_u)=(1.67\pm 0.03)\times 10^{-12}s.
\en
Note that the branching ratios listed in Tables VIII-X are meant
to be averaged over CP-conjugate modes: \be \label{average} &&
{1\over 2}\left[ {\cal B}(B^-\to M_1M_2)+{\cal B}(B^+\to\ov M_1\ov
M_2) \right], \non \\ && {1\over 2}\left[ {\cal B}(B^0\to
M_1M_2)+{\cal B}(\ov B^0\to\ov M_1\ov M_2) \right].
\en

  To compute the decay rates we choose two representative form-factor
models: the BSW and LCSR models (see Sec. III.D). From Eq.~(A1) we
see that the decay rate of $B\to PP$ depends on the form factor
$F_0$, $B\to PV$ on $F_1$ and/or $A_0$. while $B\to VV$ on
$A_1,A_2$ and $V$. It is interesting to note that the branching
ratios of $B\to VV$ predicted by the LCSR are always larger than
that by the BSW model by a factor of $1.6\sim 2$ (see Table X).
This is because the $B\to VV$ rate is very sensitive to the
form-factor ratio $x=A_2/A_1$ at the appropriate $q^2$. This
form-factor ratio is almost equal to unity in the BSW model, but
it is less than unity in the LCSR (see Table II). Consider the
decay $\ov B^0\to K^{*-}\rho^+$ as an example. Its decay rate is
proportional to $A_1^{B\rho}(m^2_{K^*}) [(a-bx)^2+2(1+c^2y^2)]$,
where $a=19.3,~b=13.9,~c=0.72$, $x=A_2^{B\rho}(m^2
_{K^*})/A_1^{B\rho}(m^2_{K^*})$ and
$y=V^{B\rho}(m^2_{K^*})/A_1^{B\rho} (m^2_{K^*})$. We find $x=1.03$
and 0.87 in the BSW and LCSR models, respectively. It is easily
seen that the prediction of ${\cal B}(\ov B^0\to K^{*-}\rho^+)$ in
the LCSR is about 1.6 times as large as that in the BSW model (see
Table X).

\subsection{Spectator-dominated rare $B$ decays}
The class I-III charmless $B$ decays proceed at the tree level
through the $b$ quark decay $b\to u\bar ud$ and at the loop level
via the $b\to d$ penguin diagrams. Since
\be
V_{ub}V_{ud}^*=A\lambda^3(\rho-i\eta), \quad
V_{cb}V_{cd}^*=-A\lambda^3, \quad
V_{tb}V_{td}^*=A\lambda^3(1-\rho+i\eta), \label{btouud}
\en
in terms of the Wolfenstein parametrization [see Eq.~(3.1)], are
of the same order of magnitude, it is clear that the rare $B$
decays of this type are tree-dominated as the penguin
contributions are suppressed by the smallness of penguin
coefficients. As pointed out in \cite{AKL}, the decays
$B^0\to\pi^0\etapp$ are exceptional because their tree amplitudes
are proportional to
\be
a_2\left[ \la\etapp|(\bar uu)_\vma|0\ra\la\pi^0|(\bar db)_\vma|\ov B^0\ra+
\la\pi^0|(\bar uu)_\vma|0\ra\la\etapp|(\bar db)_\vma|\ov B^0\ra\right].
\en
The matrix element $\la\pi^0|(\bar db)_\vma|\ov B^0\ra$ has a sign
opposite to that of $\la\etapp|(\bar db)_\vma|\ov B^0\ra$ because
of the wave functions: $\pi^0=(\bar uu-\bar dd)/\sqrt{2}$ and
$\etapp\propto (\bar uu+\bar dd)$. The large destructive
interference of the tree amplitudes renders the penguin
contributions dominant (see Table V for the relative amplitudes).
This explains why $B^0\to\pi^0\eta'$ has the smallest branching
ratio, of order $10^{-7}$, in charmless $B\to PP$ decays.
Likewise, the branching ratios of $B^0\to\rho^0\etapp$ are also
very small. There is another exceptional one: $B^0\to\rho^0\omega$
whose tree amplitude is proportional to
\be
a_2\left[X_u^{(B_d\rho^0,\omega)}+X_u^{(B_d\omega,\rho^0)}\right].
\en
Again, a large destructive interference occurs because
$\rho^0=(\bar uu-\bar dd)/\sqrt{2}$ and $\omega=(\bar uu+\bar
dd)/\sqrt{2}$: the matrix element for $B_d\to\rho^0$ transition
has a sign opposite to that for $B_d\to\omega$. Consequently, this
decay is dominated by the penguin contribution and belongs to the
class-VI mode.

Experimentally, $B^-\to\rho^0\pi^-$ is the only tree-dominated
charmless $B$ decay that has been observed very recently. If
$\nc(LL)$ is treated as a free parameter, it is easily seen that
the decay rates of class-I modes increase with $\nc(LL)$ since
$a_1=c_1^{\rm eff}+c_2^{\rm eff}/\nc(LL)$ and $c_2^{\rm eff}$ is
negative. Because $a_2$ is positive at $\nc(LL)<3.2$ and it
becomes negative when $\nc(LL)>3.2$, the magnitude of $a_2$ has a
minimum at $\nc(LL)=3.2$. Therefore, the branching ratio of
class-II channels will decrease with $\nc(LL)$ until it reaches
the minimum at $1/\nc(LL)=0.31$ and then increases again. The
class-III decays involve interference between external and
internal $W$-emission amplitudes. It is obvious that the branching
ratios of class-III modes will decrease with $\nc(LL)$. On the
contrary, when $\nc(LL)$ is fixed, the branching ratios for most
of class-I to class-III modes are insensitive to $\nc(LR)$. This
means that penguin contributions are generally small.

Theoretically, some expectation on the effective parameters
$a_1^{\rm eff}$ and $a_2^{\rm eff}$ is as follows. We see from
Table III that $a_2$ is very sensitive to the nonfactorized
effects. Since the effective number of colors, $\nc(LL)$, inferred
from the Cabibbo-allowed decays $B\to (D,D^*)(\pi,\rho)$ is in the
vicinity of 2 (see Eq.~(\ref{a2}); for a recent work, see
\cite{CY}) and since the energy released in the energetic two-body
charmless $B$ decays is in general slightly larger than that in
$B\to D\pi$ decays, it is thus expected that \be \label{chi}
|\chi({\rm two-body~rare~B~decay})|\lsim |\chi(B\to D\pi)|,
\en
and hence $\nc(LL)\approx \nc(B\to D\pi)\sim 2$. This implies that
the values of $a_1$ and $a_2$ are anticipated to be $a_1\sim
0.986$ and $a_2\sim 0.22$\,.

\begin{figure}[tb]
\psfig{figure=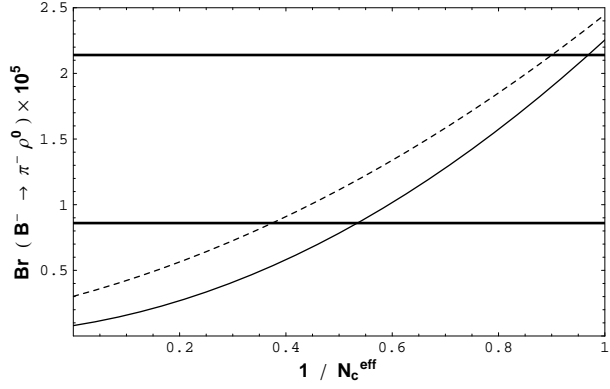,height=2.0in} \vspace{0.4cm}
    \caption{{\small The branching ratio of $B^-\to\rho^0\pi^-$ versus
    $1/\nc$. The solid (dotted) curve is calculated using the BSW (LCSR)
     model, while the solid thick lines are the CLEO measurements
     with one sigma errors.}}
     \label{fig:rho0pim}
\end{figure}

\begin{figure}[tb]
\psfig{figure=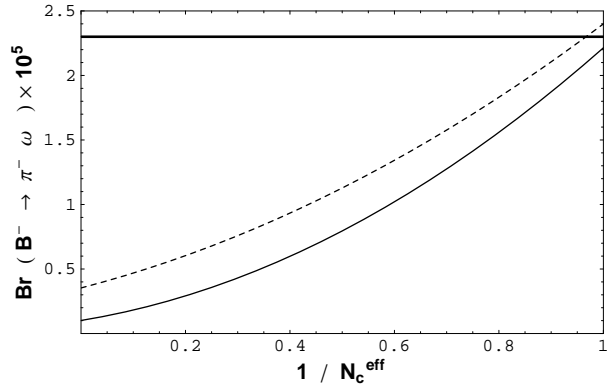,height=2.0in} \vspace{0.4cm}
    \caption{{\small The branching ratio of $B^-\to\pi^-\omega$ versus
    $1/\nc$. The solid (dotted) curve is calculated using the BSW (LCSR)
     model, while the solid thick line is the CLEO upper limit.}}
     \label{fig:omegapi}
\end{figure}

Very recently CLEO has made the first observation of a hadronic
$b\to u$ decay, namely $B^\pm\to\rho^0\pi^\pm$ \cite{Gao}. The
preliminary measurement yields:
\be
{\cal B}(B^\pm\to\rho^0\pi^\pm)=\,(1.5\pm 0.5\pm 0.4)\times
10^{-5}.
\en
From Fig. \ref{fig:rho0pim} or Table IX it is clear that this
class-III mode which receives external and internal $W$-emission
contributions is sensitive to $1/\nc$ if $\nc(LL)$ is treated as a
free parameter, namely, $\nc(LR)=\nc(LL)=\nc$; it has the lowest
value of order $1\times 10^{-6}$ and then grows with $1/\nc$. We
see from Fig.~\ref{fig:rho0pim} that $0.38\leq 1/\nc\leq 0.96$.
Since the tree diagrams make the dominant contributions, we then
have
\be
1.1\leq \nc(LL)\leq 2.6~~~{\rm from}~B^\pm\to\rho^0\pi^\pm.
\en
Therefore, $\nc(LL)$ is favored to be less than 3, as expected.

There is an additional experimental hint that favors the choice
$\nc(LL)\sim 2$ or a smaller $\nc$: the class-III decay
$B^\pm\to\pi^\pm\omega$. This mode is very similar to
$\rho^0\pi^\pm$ as its decay amplitude differs from that of
$\omega\pi^\pm$ only in the penguin terms proportional to
$X_u^{(B\pi^-,\omega)}$ (see Appendix E) which are not only small
but also subject to the quark-mixing angle suppression. Therefore,
the decay rates of $\omega\pi^\pm$ and $\rho^0\pi^\pm$ are very
similar. Although experimentally only the upper limit ${\cal
B}(B^\pm\to\pi^\pm\omega)<2.3\times 10^{-5}$ is quoted by CLEO
\cite{CLEOomega2}, the CLEO measurements ${\cal B}(B^\pm\to
K^\pm\omega)=(1.5^{+0.7}_{-0.6}\pm 0.2) \times 10^{-5}$ and ${\cal
B}(B^\pm\to h^\pm\omega)=(2.5^{+0.8}_{-0.7}\pm 0.3)\times 10^{-5}$
with $h=\pi,~K$ indicate that the central value of ${\cal
B}(B^\pm\to\pi^\pm\omega)$ is about $1\times 10^{-5}$. A fit of
the model calculations to this central value yields
$0.4<1/\nc(LL)<0.6$ (see Fig.~\ref{fig:omegapi}) or
$1.7<\nc(LL)<2.5$. The prediction for $\nc(LL)=2$ is ${\cal
B}(B^\pm\to\omega\pi^\pm)= 0.8\times 10^{-5}$ and $1.1\times
10^{-5}$ in the BSW model and the LCSR, respectively.

 In analogue to the decays $B\to D^{(*)}\pi(\rho)$, the ratio $a_2/a_1$
can be inferred from the interference effect of spectator
amplitudes in class-III charmless $B$ decays by measuring the
ratios of charged to neutral branching fractions: \be \label{Ri}
R_1 &\equiv & 2\, {{\cal B}(B^-\to\pi^-\pi^0)\over {\cal B}(\ov
B^0\to \pi^-\pi^+ )}, \qquad \qquad R_2 \equiv 2\, {{\cal
B}(B^-\to\rho^-\pi^0)\over {\cal B}(\ov B^0\to \rho^-\pi^+)}, \non
\\ R_3 &\equiv & 2\,{{\cal B}(B^-\to\pi^-\rho^0)\over {\cal B}(\ov
B^0\to \pi^-\rho^+)},  \qquad \qquad R_4 \equiv 2\, {{\cal
B}(B^-\to\rho^-\rho^0)\over {\cal B}(\ov B^0\to \rho^-\rho^+)}.
\en
Since penguin contributions to $R_i$ are small as we have checked
explicitly, to a good approximation we have
\be
R_1 &\cong& {\tau(B^-)\over\tau(B^0_d)}\left(1+{a_2\over
a_1}\right)^2, \non\\ R_2 &\cong&
{\tau(B^-)\over\tau(B^0_d)}\left(1+{f_\pi\over
f_\rho}\,{A_0^{B\rho} (m^2_\pi)\over
F_1^{B\pi}(m^2_\rho)}\,{a_2\over a_1}\right)^2,  \non\\ R_3
&\cong& {\tau(B^-)\over\tau(B^0_d)}\left(1+{f_\rho\over
f_\pi}\,{F_1^{B\pi} (m^2_\rho)\over
A_0^{B\rho}(m^2_\pi)}\,{a_2\over a_1}\right)^2,  \non \\ R_4
&\cong& {\tau(B^-)\over\tau(B^0_d)}\left(1+{a_2\over
a_1}\right)^2.
\en
Evidently, the ratios $R_i$ are greater (less) than unity when the
interference is constructive (destructive). From Table XI we see
that a measurement of $R_i$ (in particular $R_3$) will constitute
a very useful test on the effective number of colors $\nc(LL)$.

\vskip 0.4cm
\begin{table}[ht]
{\small Table XI. The predictions of the ratios $R_i$ at $\nc=2$
and $\nc=\infty$, respectively, in the BSW [LCSR] model.}
\begin{center}
\begin{tabular}{ l c c c c }
& $R_1$ & $R_2$ & $R_3$ & $R_4$ \\
\hline
$\nc=2$ & 1.52~[1.52] & 1.25~[1.34] & 2.27~[1.84] & 1.57~[1.57] \\
$\nc=\infty$ & 0.48~[0.48] & 0.86~[0.76] & 0.16~[0.35] & 0.50~[0.50] \\
\end{tabular}
\end{center}
\end{table}

\begin{figure}[tb]
%\vspace{0.15cm}
\psfig{figure=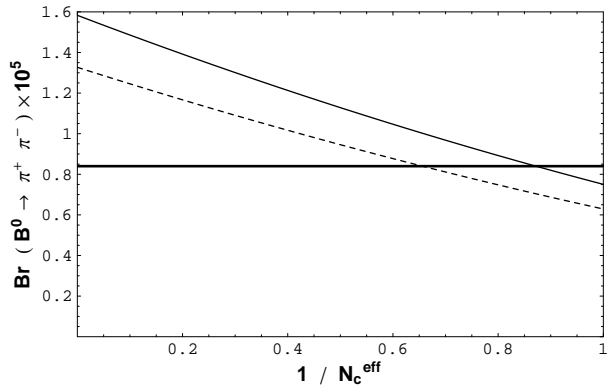,height=2.0in} \vspace{0.4cm}
    \caption{{\small The branching ratio of $B^0\to\pi^+\pi^-$ versus
    $1/\nc$ where use of $\nc(LL)=\nc(LR)=\nc$ has been made. The solid
    (dotted) curve is calculated using the BSW (LCSR)
     model, while the solid thick line is the CLEO upper limit.}}
     \label{fig:pippim}
\end{figure}

\begin{figure}[tb]
%\vspace{0.15cm}
\psfig{figure=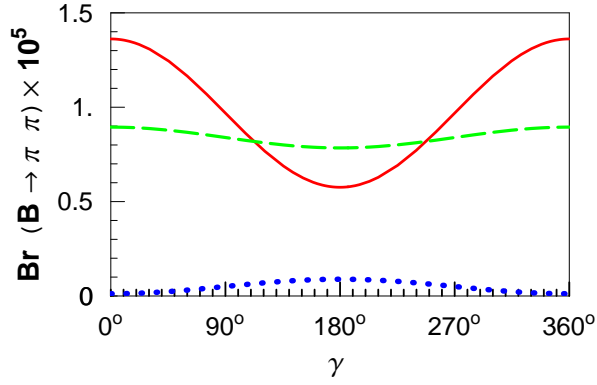,height=2.0in} \vspace{0.4cm}
    \caption{{\small Branching ratios of $B\to\pi \pi$ modes versus the
    unitarity angle $\gamma$, where the solid, dashed, and dotted
curves correspond to $\pi^+\pi^-$, $\pi^- \pi^0$, and $\pi^0
\pi^0$, respectively. Uses of $N_c^{\rm eff}(LL)=2$, $N_c^{\rm
eff}(LR)=5$ and the BSW model for form factors have been made.}}
     \label{fig:gamma}
\end{figure}

A very recent CLEO analysis of $B^0\to\pi^+\pi^-$ presents an
improved upper limit \cite{Roy}
\be
{\cal B}(B^0\to\pi^+\pi^-)<0.84\times 10^{-5}.
\en
It is evident from Fig.~\ref{fig:pippim} that $\nc(LL)$ is
preferred to be smaller and that the predicted branching ratio
seems to be too large compared to experiment. Indeed, most known
model predictions in the literature tend to predict a ${\cal
B}(B^0\to\pi^+\pi^-)$ much larger than the current limit. There
are several possibilities for explaining the data: (i) The CKM
matrix element $|V_{ub}|$ (or the value of $\sqrt{\rho^2+\eta^2}$)
and/or the form factor $F_0^{B\pi}(0)$ are smaller than the
conventional values. However, one has to bear in mind that the
product $V_{ub}F_0^{B\pi}$ is constrained by the measured
semileptonic $B\to\pi\ell\nu$ rate: A smaller $V_{ub}$ will be
correlated to a larger $B\to\pi$ form factor and vice versa. (ii)
Final-state interactions may play an essential role. We shall see
in Sec. VI that if the isospin phase shift difference is nonzero
and larger than $70^\circ$, the decay rate of $\pi^+\pi^-$ will be
significantly suppressed whereas the mode $\pi^0\pi^0$ is
substantially enhanced (see Fig. \ref{fig:delta}). (iii) The
unitarity angle $\gamma$ is larger than $90^\circ$, or the
Wolfenstein parameter $\rho$ is negative, an interesting
possibility pointed out recently in \cite{gamma}. It is clear from
Fig. \ref{fig:gamma} that the experimental limit of $\pi^+\pi^-$
can be accommodated by $\gamma>105^\circ$. From Eq. (\ref{btouud})
and Appendix B for the factorized amplitude of $B\to\pi^+\pi^-$,
it is easily seen that the interference between tree and penguin
amplitudes is suppressed when $\rho$ is negative. We also see from
Fig. \ref{fig:gamma} that the $\pi^-\pi^0$ mode is less sensitive
to $\gamma$ as it does not receive QCD penguin contributions and
electroweak penguins are small. The current limit on $\pi^-\pi^0$
is ${\cal B}(B^-\to\pi^-\pi^0)<1.6\times 10^{-5}$ \cite{Roy}.

Three remarks are in order before ending this section. First, it
is interesting to note that the tree-dominated class I-III modes
which have branching ratios of order $10^{-5}$ or larger must have
either one vector meson in the final state because of the larger
vector-meson decay constant $f_V>f_P$ or two final-state vector
mesons because of the larger spin phase space available due to the
existence of three different polarization states for the vector
meson. For example, it is expected that
\be
&& {\cal B}(\ov B^0\to \rho^-\rho^+)\sim {\cal B}(\ov B^0\to
\rho^-\pi^+)> {\cal B}(\ov B^0\to \pi^-\rho^+)\sim 1\times
10^{-5},  \non \\ && {\cal B}(B^-\to \rho^-\rho^0) \sim {\cal
B}(B^-\to \rho^-\pi^0)> {\cal B}(B^-\to \pi^-\rho^0)\sim 1\times
10^{-5},  \non \\ && {\cal B}(B^-\to \omega\rho^-)> {\cal
B}(B^-\to \omega\pi^-) \sim 1\times 10^{-5}.
\en
The $\rho^-\pi^+$ ($\rho^-\pi^0$) decay has a larger rate than the
$\rho^+ \pi^-$ ($\rho^0\pi^-$) mode mainly because of the
difference of the decay constants $f_\rho$ and $f_\pi$ so that
$f_\rho F_1^{B\pi}>f_\pi A_0^{B\rho}$. Second, it is well known
that the unitarity angle $\alpha$ can be determined from measuring
the CP asymmetry in $(B^0,\ov B^0)\to\pi^+\pi^-$ decays provided
that penguin contributions are negligible. But we see from Table V
that the QCD penguin contribution is important for
$B^0\to\pi^0\pi^0$ and moderate for $B^0\to\pi^+\pi^-$.
Nevertheless, if isospin is a good symmetry, an isospin analysis
of $B^0\to\pi^+\pi^-,\pi^0\pi^0$, $B^+\to \pi^+\pi^0$ and their
CP-conjugate modes can lead to the extraction of $2\alpha$ without
electroweak penguin pollution. However, isospin symmetry is broken
by electroweak penguins and also by the $u$ and $d$ quark mass
difference which will contaminate the model-independent
determination of $\sin 2\alpha$ \cite{Gardner}. Third, as
mentioned before, the branching ratio of the class-II modes is
very sensitive to the value of $\nc$: it has a minimum at
$\nc=3.2$\,. Our preferred prediction is made at the value
$\nc(LL)=2$ and hence the branching ratio is not very small.
Nevertheless, the decay rates of class-II channels are in general
significantly smaller than that of class-I and class-III ones. As
a result, $W$-exchange, $W$-annihilation and final-state
interactions, which have been neglected thus far, could be
important for class-II decays and they may even overwhelm the
usual factorized contributions.

\subsection{General features of QCD-penguin dominated $B$ decays}
For penguin-dominated class IV-VI decay modes, some general
observations are the following:

\vskip 0.4cm 1.~~ Class-IV modes involve the QCD penguin
parameters $a_4$ and $a_6$ in the combination $a_4+Ra_6$, where
$R>0$ for $B\to P_aP_b$, $R=0$ for $P_aV_b$ and $V_aV_b$ final
states, and $R<0$ for $B\to V_aP_b$, where $P_b$ or $V_b$ is
factorizable under the factorization assumption. Therefore, the
decay rates of class-IV decays are expected to follow the pattern:
\be \Gamma(B\to P_aP_b)>\Gamma(B\to P_aV_b)\sim\Gamma(B\to V_a
V_b)>\Gamma (B\to V_aP_b),
\en
as a consequence of various possibilities of interference between
the $a_4$ and $a_6$ penguin terms. From Tables VIII-X, we see that
\be \label{pattern} && {\cal B}(\ov B^0\to K^-\pi^+)
>{\cal B}(\ov B^0\to K^{*-}\pi^+)\sim {\cal B}(\ov B^0\to K^{*-}\rho^+)>
{\cal B}(\ov B^0\to K^-\rho^+),   \non \\
%&& \Gamma(\ov B^0\to \ov K^0\pi^0)
%>\Gamma(\ov B^0\to \ov K^{*0}\pi^0)\sim \Gamma(\ov B^0\to \ov K^{*0}\rho^0)>
%\Gamma(\ov B^0\to \ov K^0\rho^0),   \non \\
&& {\cal B}(B^-\to \ov K^0\pi^-) >{\cal B}(B^-\to \ov K^{*0}\pi^-)
\sim {\cal B}(B^-\to \ov K^{*0}\rho^-)>
{\cal B}(B^-\to \ov K^0\rho^-),   \non \\
&& {\cal B}(\ov B^0\to K^0\ov K^0)
>{\cal B}(\ov B^0\to K^{0}\ov K^{*0})\sim {\cal B}(\ov B^0\to
K^{*0}\ov K^{*0})>{\cal B}(\ov B^0\to K^{*0}\ov K^0), \non \\
&& {\cal B}(B^-\to K^-K^0) >{\cal B}(B^-\to K^- K^{*0})
\sim {\cal B}(B^-\to K^{*-} K^{*0})> {\cal B}(B^-\to K^{*-} K^0).
\en
Note that the above hierarchy is opposite to the pattern ${\cal
B}(B\to P_aV_b)>{\cal B}(B\to P_aP_b)$, as often seen in
tree-dominated decays. It implies that the spin phase-space
suppression of the penguin-dominated decay $B\to P_aP_b$ over
$B\to P_aV_b$ or $B\to V_aP_b$ is overcome by the constructive
interference between the penguin amplitudes in the former. Recall
that the coefficient $R$ is obtained by applying equations of
motion to the hadronic matrix elements of pseudoscalar densities
induced by penguin operators. Hence, a test of the hierarchy shown
in (\ref{pattern}) is important for understanding the penguin
matrix elements. \vskip 0.4cm 2.~~Contrary to tree-dominated
decays, the penguin-dominated charmless $B$ decays have the
largest branching ratios in the $PP$ mode. Theoretically, the
class-VI decay modes $B^-\to \eta' K^-,\, \ov B_d\to \eta'\ov K^0$
have branching ratios of order $4.5\times 10^{-5}$. These decay
modes receive two different sets of penguin terms proportional to
$a_4+Ra_6$ with $R>0$. The other penguin-dominated decay modes
which have branching ratios of order $10^{-5}$ are $\ov B^0\to
K^-\pi^+,~B^-\to K^-\pi^0, \ov K^0\pi^-$; all of them have been
observed by CLEO. \vskip 0.4cm 3.~~We will encounter hadronic
matrix elements of pseudoscalar densities when evaluating the
penguin amplitudes. Care must be taken to consider the
pseudoscalar matrix element for $\etapp\to$ vacuum transition: The
anomaly effects must be included in order to ensure a correct
chiral behavior for the pseudoscalar matrix element \cite{CT98}.
The results are \cite{Kagan,Ali} \be \label{anomaly}
\la\etapp|\bar s\gamma_5 s|0\ra &=& -i{m_{\etapp}^2\over
2m_s}\,\left(f_{ \etapp}^s-f^u_{\etapp}\right),  \non \\
\la\etapp|\bar u\gamma_5u|0\ra &=& \la\etapp|\bar
d\gamma_5d|0\ra=r_{\etapp} \,\la\etapp|\bar s\gamma_5s|0\ra,
\en
with \cite{CT98}
\be \label{reta}
r_{\eta'} &=&
{\sqrt{2f_0^2-f_8^2}\over\sqrt{2f_8^2-f_0^2}}\,{\cos\theta+
{1\over \sqrt{2}}\sin\theta\over \cos\theta-\sqrt{2}\sin\theta},
\non \\ r_{\eta} &=& -{1\over
2}\,{\sqrt{2f_0^2-f_8^2}\over\sqrt{2f_8^2-f_0^2}}\,
{\cos\theta-\sqrt{2}\sin\theta\over \cos\theta+{1\over\sqrt{2}}
\sin\theta}.
\en
\vskip 0.4cm 4. We shall see below that nonfactorized effects in
penguin-dominated decays are favored to be $\nc(LR)>3$, as implied
by the decay modes $B\to\phi K$ and $B\to\eta' K$, contrary to the
tree-dominated case where $\nc(LL)<3$. From Eqs.~(\ref{a2}) and
(\ref{chi}) it is anticipated that $\nc(LL)\approx \nc(B\to
D\pi)\sim 2$ and $\nc(LR)\sim 2-6$, depending on the sign of
$\chi$. Since $\nc(LR)> \nc(LL)$ implied by the data, therefore,
we conjecture that \cite{CCT}
\be
\nc(LR)\lsim 6.
\en

\subsection{$B\to \phi K,~\phi K^*$ decays}
\begin{figure}[tb]
%\vspace{0.5cm}
\psfig{figure=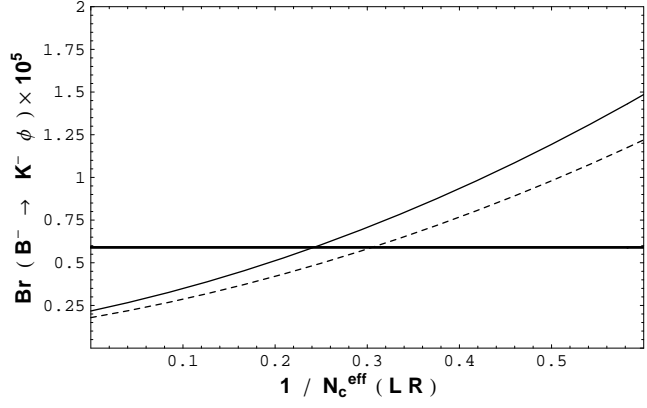,height=2.1in} \vspace{0.4cm}
    \caption{{\small The branching ratio of $B^-\to\phi K^-$ versus
    $1/\nc(LR)$ with $\nc(LL)$ being fixed at 2. The solid (dotted)
    curve is calculated using the BSW (LCSR)
     model, while the solid thick line is the CLEO upper limit.}}
     \label{fig:Kphi}
\end{figure}

\begin{figure}[tb]
\psfig{figure=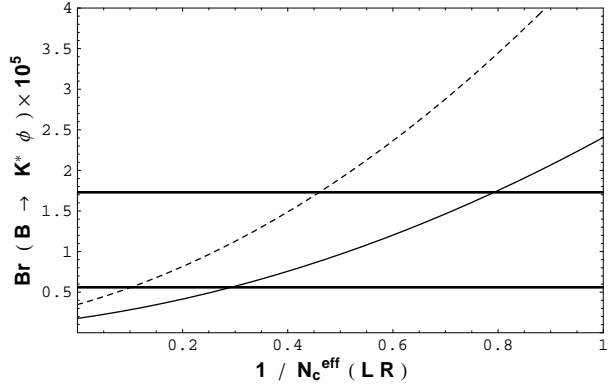,height=2.0in}
\vspace{0.4cm}
    \caption{{\small The branching ratio of $B\to\phi K^*$ vs $1/\nc(LR)$
with $\nc(LL)$ being fixed at the value of 2. The solid (dotted)
curve is calculated using the BSW (LCSR) model. The solid thick
lines are the CLEO measurements with one sigma errors.}}
\label{fig:VKphi}
\end{figure}

   The decay amplitudes of the class-VI penguin-dominated  modes
$B\to \phi K$ and $B\to \phi K^*$ are governed by the effective
coefficients $[a_3+a_4+a_5-{1\over 2}(a_7+a_9+a_{10})]$ (see
Appendixes C-G). As noted in passing, the QCD penguin coefficients
$a_3$ and $a_5$ are sensitive to $\nc(LL)$ and $\nc(LR)$,
respectively. We see from Figs.~\ref{fig:Kphi} and \ref{fig:VKphi}
that the decay rates of $B\to \phi K^{(*)}$ increase with
$1/\nc(LR)$ irrespective of the value of $\nc(LL)$. The new CLEO
upper limit \cite{Gao} \be \label{phiK} {\cal B}(B^\pm\to\phi
K^\pm)< 0.59\times 10^{-5}
\en
at 90\% C.L. implies that (see Fig. \ref{fig:Kphi}) \be
\label{ncLR} \nc(LR)\geq \cases{4.2 & BSW, \cr 3.2 & LCSR,  \cr}
\en
with $\nc(LL)$ being fixed at the value of 2. Note that this
constraint is subject to the corrections from spacelike penguin
and $W$-annihilation contributions. At any rate, it is safe to
conclude that $\nc(LR)>3>\nc(LL)$.

CLEO has seen a $3\sigma$ evidence for the decay $B\to\phi K^*$.
Its branching ratio, the average of $\phi K^{*-}$ and $\phi
K^{*0}$ modes, is reported to be \cite{CLEOomega2}
\be
\label{phiK*} {\cal B}(B\to\phi K^*)\equiv {1\over 2}\left[{\cal
B}(B^\pm\to\phi K^{*\pm}) +{\cal B}(B^0\to\phi K^{*0})\right]
=\left(1.1^{+0.6}_{-0.5}\pm 0.2\right)\times 10^{-5}.
\en
Using $\nc(LL)=2$ and the constraint (\ref{ncLR}), we find that
\be
{\cal B}(B\to\phi K^*)\leq \cases{ 0.4\times 10^{-5} & BSW, \cr
1.2\times 10^{-5} & LCSR,   \cr}
\en
and that the ratio $\Gamma(B\to\phi K^*)/\Gamma(B^\pm\to\phi
K^\pm)$ is 0.76 in the BSW model, while it is equal to 1.9 in the
LCSR. This is because $\Gamma(B\to \phi K^*)$ is very sensitive to
the form factor ratio $x=A_2^{BK^*}(m^2_\phi)/
A_1^{BK^*}(m^2_\phi)$, which is equal to 0.875 (1.03) in the LCSR
(BSW) model [see the discussion after Eq.~(\ref{average})]. In
particular, ${\cal B}(B\to\phi K^*)=0.74\times 10^{-5}$ is
predicted by the LCSR for $\nc(LL)=2$ and $\nc(LR)=5$, which is in
accordance with experiment. It is evident from Figs.
\ref{fig:Kphi} and \ref{fig:VKphi} that the data of $B\to\phi K$
and $B\to\phi K^*$ can be simultaneously accommodated in the LCSR
analysis. Therefore, the non-observation of $B\to\phi K$ does not
necessarily invalidate the factorization hypothesis; it could
imply that the form-factor ratio $A_2/A_1$ is less than unity. Of
course, it is also possible that the absence of $B\to\phi K$
events is a downward fluctuation of the experimental signal. At
any rate, in order to clarify this issue and to pin down the
effective number of colors $\nc(LR)$, measurements of $B\to\phi K$
and $B\to\phi K^*$ are urgently needed with sufficient accuracy.

\subsection{$B\to\eta' K^{(*)}$ and $\eta K^{(*)}$ decays}

\begin{figure}[tb]
\psfig{figure=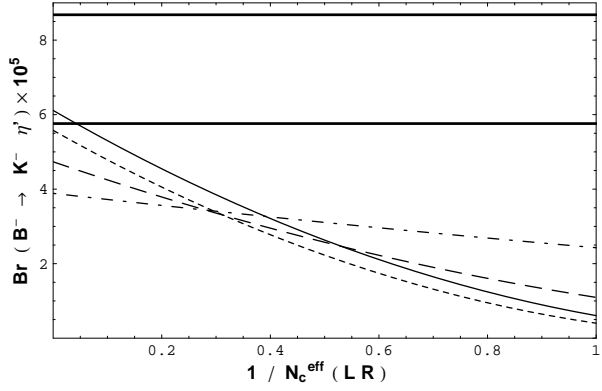,height=2.0in} \vspace{0.5cm}
    \caption{{\small The branching ratio of $B^\pm\to\eta' K^\pm$ as a
   function of $1/\nc(LR)$ with $\nc(LL)$ being fixed at the value of 2
   and $\eta=0.370$, $\rho=0.175$, $m_s(m_b)=90$ MeV. The
   calculation is done using the BSW model for form factors.
   The charm content of the $\eta'$ with $f_{\eta'}^c=-6.3\,{\rm MeV}$
   contributes to the solid curve but not to the dotted curve.
   The anomaly contribution to $\la\eta'|\bar s\gamma_5s|0\ra$
   is included. For comparison, predictions for
   $\nc(LL)=\nc(LR)$ as depicted by the dashed curve with $f_{\eta'}^c
   =0$ and dot-dashed curve with $f_{\eta'}^c=-6.3$ MeV are also shown.
    The solid thick lines are the preliminary updated
    CLEO measurements (\ref{Ketapp}) with one sigma errors.}}
   \label{fig:Ketap}
\end{figure}

The published CLEO results \cite{Behrens1} on the decay $B\to\eta'
K$
\be \label{Ketap}
{\cal B}(B^\pm\to\eta' K^\pm) &=&
\left(6.5^{+1.5}_{-1.4}\pm 0.9\right)\times 10^{-5}, \non \\
 {\cal B}(B^0\to\eta' K^0) &=& \left(4.7^{+2.7}_{-2.0}\pm 0.9
\right)\times 10^{-5},
\en
are several times larger than earlier theoretical predictions
\cite{Chau1,Kramer,Du} in the range of $(1-2)\times 10^{-5}$. It
was pointed out in past two years by several authors
\cite{Ali,Kagan,Deshpande1} that the decay rate of $B\to\eta' K$
will get enhanced because of the small running strange quark mass
at the scale $m_b$ and sizable $SU(3)$ breaking in the decay
constants $f_8$ and $f_0$.\footnote{To demonstrate how the decay
rate of $B^-\to\eta' K^-$ is enhanced, we first use the parameters
$F_0^{BK}(0)=0.38$,
$\sqrt{3}F_0^{B\eta_0}(0)=\sqrt{6}F_0^{B\eta_8}(0)=F_0^{B\pi}(0)$,
$m_s=140$ MeV, $f_0=f_8=f_\pi$,
$\theta_8=\theta_0=\theta=-20^\circ$, which in turn imply
$f_{\eta'}^u=53$ MeV, $f_{\eta'}^s=108$ MeV and $F^{B\eta'}_0(0)=
0.133$. With the above inputs, we obtain ${\cal B}(B^-\to\eta'
K^-)=(1.0-1.5)\times 10^{-5}$ at $0<1/\nc<0.5$ where
$\nc(LR)=\nc(LR)=\nc$. Then we consider some possible effects of
enhancement. First of all, the penguin amplitude of $B\to\eta' K$
proportional to $a_6$ and $a_8$ will get enhanced by a factor of
1.6 if $m_s=90$ MeV, the strange quark mass at $\mu=m_b$, instead
of $m_s=140$ MeV, the mass at 1 GeV, is employed. Second, SU(3)
breaking in the decay constants $f_8$ and $f_0$ [see
Eq.~(\ref{f80})] and the two-mixing angle formulation for the
decay constants $f_\eta$ and $f_{\eta'}$ [see Eq.~(\ref{theta})]
lead to $f_{\eta'}^u=63$ MeV and $f^s_{\eta'}=137$ MeV.
Consequently, the factorized terms $X_u^{(BK,\eta')}$ and
$X_s^{(BK,\eta')}$ (see Appendix C) are enhanced by a factor of
1.17 and 1.27, respectively. Third, for $\theta=-15.4^\circ$ [ see
Eq.~(\ref{theta})] we obtain $F_0^{B\eta'}(0)=0.148$. Thus,
$X^{(B\eta',K)}$ is increased by a factor of 1.11\,. As a result
of an accumulation of above several small enhancements, the
branching ratio eventually becomes ${\cal B}(B^-\to\eta'
K^-)=(2.7-4.7)\times 10^{-5}$.} Ironically, it was also realized
around a year ago that \cite{Kagan,Ali} the above-mentioned
enhancement is partially washed out by the anomaly effect in the
matrix element of pseudoscalar densities, an effect overlooked
before. Specifically, $\la\eta'|\bar s\gamma_5
s|0\ra=-i(m_{\eta'}^2/ 2m_s)\left(f_{\eta'}^s -f^u_{\eta'}\right)$
[see Eq.~(\ref{anomaly})], where the QCD anomaly effect is
manifested by the decay constant $f_{\eta'} ^u$. Since
$f_{\eta'}^u\sim {1\over 2}f_{\eta'}^s$ [cf. Eq.~(\ref{fdecay})],
it is obvious that the decay rate of $B\to\eta' K$ induced by the
$(S-P) (S+P)$ penguin interaction is suppressed by the anomaly
term in $\la\eta'|\bar s\gamma_5 s|0\ra$. As a consequence, the
net enhancement is not large. If we treat $\nc(LL)$ to be the same
as $\nc(LR)$, as assumed in previous studies, we would obtain
${\cal B}(B^\pm\to\eta' K^\pm)=(2.7-4.7)\times 10^{-5}$ at
$0<1/\nc<0.5$ for $m_s(m_b)=90$ MeV and $F_0^{BK}(0)=0.38$ (see
the dashed curve in Fig.~\ref{fig:Ketap}). It is easily seen that
the experimental branching ratios can be accommodated by a smaller
strange quark mass, say $m_s(m_b)=60$ MeV, and/or a large form
factor $F_0^{BK}$, for instance $F_0^{BK}(0)=0.60$. However, it is
very important to keep in mind that it is dangerous to adjust the
form factors and/or light quark masses in order to fit a few
particular modes; the comparison between theory and experiment
should be done using the same set of parameters for all channels
\cite{CT97}. Indeed, a too small $m_s(m_b)$ will lead to a too
large $B\to K\pi$, while a too large $F_0^{BK}(0)$ will break the
SU(3)-symmetry relation $F_0^{BK}=F_0^{B\pi}$ very badly as the
form-factor $F_0^{B\pi}(0)$ larger than 0.33 is disfavored by the
current limit on $B^0\to \pi^+\pi^-$ (see Sec.~V.B).

  What is the role played by the intrinsic charm content of the $\eta'$
to $B\to\eta' K$ ? It has been advocated that the new internal
$W$-emission contribution coming from the Cabibbo-allowed process
$b\to c\bar c s$ followed by a conversion of the $c\bar c$ pair
into the $\eta'$ via two gluon exchanges is potentially important
since its mixing angle $V_{cb}V_{cs}^*$ is as large as that of the
penguin amplitude and yet its Wilson coefficient $a_2$ is larger
than that of penguin operators. As noted in Sec.~III.C, the decay
constant $f_{\eta'}^c$ lies in the range --2.0 MeV $\leq
f_{\eta'}^c\leq$ --18.4 MeV. The sign of $f_{\eta'}^c$ is crucial
for the $\eta'$ charm content contribution. For a negative
$f_{\eta'} ^c$, its contribution to $B\to\eta' K$ is constructive
for $a_2>0$. Since $a_2$ depends strongly on $\nc(LL)$, we see
that the $c\bar c\to\eta'$ mechanism contributes constructively at
$1/\nc(LL)>0.31$ where $a_2>0$, whereas it contributes
destructively at $1/\nc(LL)<0.31$ where $a_2$ becomes negative. In
order to explain the abnormally large branching ratio of
$B\to\eta' K$, an enhancement from the $c\bar c\to\eta'$ mechanism
is certainly welcome in order to improve the discrepancy between
theory and experiment. This provides another strong support for
$\nc(LL)\approx 2$. If $\nc(LL)=\nc(LR)$ is adopted, then ${\cal
B}(B\to\eta' K)$ will be {\it suppressed} at $1/\nc\leq 0.31$ and
enhanced at $1/\nc>0.31$ (see the dot-dashed curve in
Fig.~\ref{fig:Ketap} for $f_{\eta'}^c=-6.3$ MeV). If the
preference for $\nc$ is $1/\nc\lsim 0.2$ (see e.g. \cite{AKL}),
then it is quite clear that the contribution from the $\eta'$
charm content will make the theoretical prediction even worse at
the small values of $1/\nc$ ! On the contrary, if $\nc(LL)\approx
2$, the $c\bar c$ admixture in the $\eta'$ will always lead to a
constructive interference irrespective of the value of $\nc(LR)$
(see the solid curve in Fig.~\ref{fig:Ketap}).

   At this point, we see that the branching ratio of $B\to K\eta'$
of order $(2.7-4.7)\times 10^{-5}$ at $0<1/\nc<0.5$ for
$\nc(LL)=\nc(LR)$ and it becomes $(3.5-3.8)\times 10^{-5}$ when
the $\eta'$ charm content contribution with $f_{\eta'}^c=-6.3$ MeV
is taken into account. However, the discrepancy between theory and
experiment is largely improved by treating $\nc(LL)$ and $\nc(LR)$
differently. Setting $\nc(LL)=2$, we find that (see
Fig.~\ref{fig:Ketap}) the decay rates of $B\to\eta' K$ are
considerably enhanced especially at small $1/\nc(LR)$.
Specifically, ${\cal B}(B^\pm\to\eta' K^\pm)$ at $1/\nc(LR)\leq
0.2$ is enhanced from $(3.6-3.8)\times 10^{-5}$ to $(4.6-6.1)
\times 10^{-5}$ due to three enhancements. First, the $\eta'$
charm content contribution $a_2X_c^{(BK,\eta')}$ now always
contributes in the right direction to the decay rate irrespective
of the value of $\nc(LR)$. Second, the interference in the
spectator amplitudes of $B^\pm\to\eta' K^\pm$ is constructive.
Third, the term proportional to
\be
2(a_3-a_5)X_u^{(BK,\eta')}+(a_3+a_4-a_5)X_s^{(BK,\eta')}
\en
is enhanced when $(\nc)_3=(\nc)_4=2$.

\begin{figure}[tb]
\vspace{0.2cm}
 \psfig{figure=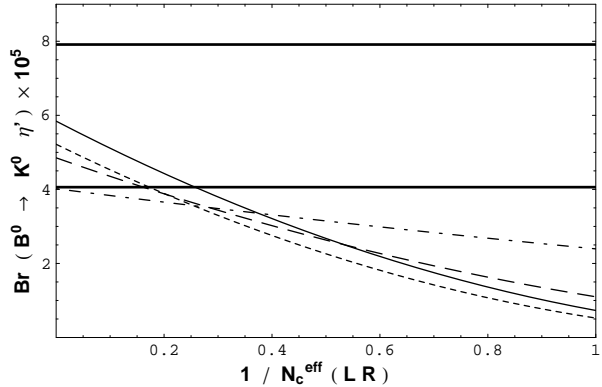,height=2.0in}
\vspace{0.4cm}
    \caption{{\small Same as Fig.~\ref{fig:Ketap} except for
    $B^0\to K^0\eta'$.}}
    \label{fig:K0etap}
\end{figure}

   A recent CLEO reanalysis of $B\to\eta' K$  using a
data sample 80\% larger than in previous studies yields the
preliminary results \cite{Behrens2,Roy}:
\be \label{Ketapp}
{\cal B}(B^\pm\to\eta' K^\pm) &=& \left(7.4^{+0.8}_{-1.3}\pm 1.0\right)
\times 10^{-5},  \non \\
{\cal B}(B^0_d\to\eta' K^0) &=&
\left(5.9^{+1.8}_{-1.6}\pm 0.9\right) \times 10^{-5},
\en
suggesting that the original measurements (\ref{Ketap}) were not
an upward statistical fluctuation.  It is evident from
Fig.~\ref{fig:K0etap} that the measurement of
$\overline{B}^0\to\eta'\overline{K}^0$ is well explained in the
present framework based on the Standard Model. Contrary to some
early claims, we see that it is not necessary to invoke some new
mechanisms, say the SU(3)-singlet contribution $S'$ \cite{Dighe},
to explain the data. The agreement with experiment provides
another strong support for $\nc(LL)\sim 2$ and for the relation
$\nc(LR)>\nc(LL)$.

\begin{figure}[tb]
%\vspace{0.15cm}
\psfig{figure=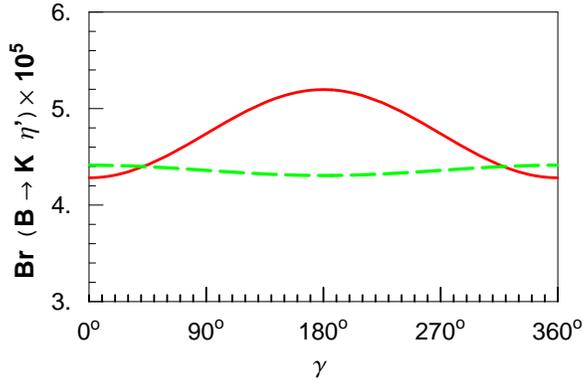,height=2.0in} \vspace{0.4cm}
    \caption{{\small Branching ratios of $B\to\eta' K$ modes versus the
    unitarity angle $\gamma$, where the solid and dashed
curves correspond to $\eta' K^-$ and $\eta' K^0$, respectively.
Uses of $N_c^{\rm eff}(LL)=2$, $N_c^{\rm eff}(LR)=5$ and the BSW
model for form factors have been made.}}
     \label{fig:RetapK}
\end{figure}

Thus far, the calculation is carried out using $m_s(m_b)=90$ MeV
and the prediction of ${\cal B}(B^-\to\eta' K^-)$ is on the lower
side of the experimental data. The discrepancy between theory and
experiment can be further improved by using a smaller strange
quark mass, say $m_s(m_b)=70$ MeV. However, as stressed before,
the calculation should be consistently carried out using the same
set of parameters for all channels \cite{CT97}. Indeed, a too
small $m_s(m_b)$ will lead to a too large $B\to K\pi$,

From the face values of the data, it appears that the branching
ratio of the charged mode $\eta' K^-$ is slightly larger than that
of the neutral mode $\eta' K^0$, though they are in agreement
within one sigma error. Note that the neutral mode does not
receive contributions from external $W$-emission and
$W$-annihilation diagrams. Since the external $W$-emission is
small due to small mixing angles, it is naively anticipated that
both decays should have very similar rates unless $W$-annihilation
plays some role. However, if the two branching values are
confirmed not to converge when experimental errors are improved
and refined in the future, a plausible explanation is ascribed to
a negative Wolfenstein's $\rho$ parameter. We see from Fig.
\ref{fig:RetapK} that the charged $\eta' K^-$ mode is
significantly enhanced at $\gamma>90^\circ$, whereas the neutral
$\eta' K^0$ mode remains steady.

Contrary to the abnormally large decay rate of $B\to\eta' K$, the
branching ratio of $B\to \eta K$ is very small because of the
destructive interference in penguin amplitudes due to the opposite
sign between the factorized terms $X^{(B\eta,K)}$ and
$a_6X_s^{(BK,\eta)}$; that is, the $(\bar uu+\bar dd)$ and $\bar
ss$ components interfere destructively for the $\eta$ but
constructively for the $\eta'$. From Table VIII we obtain
\be
{ {\cal B}(B\to\eta' K)\over {\cal B}(B\to\eta K)}=\cases{ 34 &
charged~$B$; \cr 58 & neutral~$B$. \cr}
\en
Since the sign of $a_6X_s^{(BK^*,\eta^{(')})}$ is flipped in
$B\to\eta^{(')} K^*$ decays, the interference effect becomes the
other way around: constructive in $B\to \eta K^*$ and destructive
in $B\to\eta' K^*$:
\be
 { {\cal B}(B\to\eta' K^*)\over {\cal B}(B\to\eta
K^*)}=\cases{ 0.13 & charged~$B$;  \cr 0.11 &neutral~$B$.  \cr}
\en
It has been argued in \cite{Halperin} that ${\cal B}(B\to\eta'
K^*)$ is about twice larger than that of $B\to\eta' K$, a
prediction not borne out by the current limit ${\cal
B}(B^0\to\eta' K^{*0})<2.0\times 10^{-5}$ \cite{Gao} and the
measurement of ${\cal B}(B^0\to\eta' K^0)$ (\ref{Ketapp}). Note
that it has been advocated that the two-gluon fusion mechanism may
account for the observed large decay rate of $B\to\eta' K$
\cite{Ahmady,Du98}. Using the same gluon-fusion mechanism, large
branching fractions of $B\to\eta' K^*$ of order $3\times 10^{-5}$
are found in \cite{Du99}, to be compared with $7\times 10^{-7}$ in
our calculations. Therefore, it is important to measure the
processes $B^-\to\eta' K^{*-}$ and $B^0\to\eta' K^{*0}$ to test
the two-gluon fusion mechanism.

\subsection{$B\to K\pi$ decays}
There are four $K\pi$ modes in $B_{u,d}$ decays: $\ov B^0\to
K^-\pi^+$, $B^-\to \ov K^0\pi^-$ , $B^-\to K^-\pi^0$, and $\ov
B^0\to \ov K^0 \pi^0$. Theoretically, the following pattern is
expected:
\be \Gamma(B^-\to \ov K^0\pi^-) \gsim \Gamma(\ov B^0\to
K^-\pi^+)> \Gamma(B^-\to K^-\pi^0)
> \Gamma(\ov B^0\to\ov K^0\pi^0).
\en
This pattern arises based on the following observations: (1) Since
the tree contributions are CKM suppressed, these decays are
penguin dominated. (2) Because of the $\pi^0$ wave function, it is
generally anticipated that the first two channels are larger than
the last two and ${\cal B}(B^-\to K^-\pi^0)/{\cal B}(B\to K
\pi^\pm)\approx 1/2$. (3) The small electroweak penguin effect
makes the first two processes almost the same. The slight
difference between $\ov K^0\pi^-$ and $K^-\pi^+$ comes from the
destructive interference between the tree and QCD penguin
amplitudes in the latter; such an interference is absent in the
former as it proceeds only through penguin diagrams. (4) Though it
can be neglected in the first two modes, the electroweak penguin
plays a role in the last two. With a moderate electroweak penguin
contribution, the constructive (destructive) interference between
electroweak and QCD penguins in $K^-\pi^0$ and $\ov K^0 \pi^0$
explains why the former is larger than the latter.

\begin{figure}[tb]
\psfig{figure=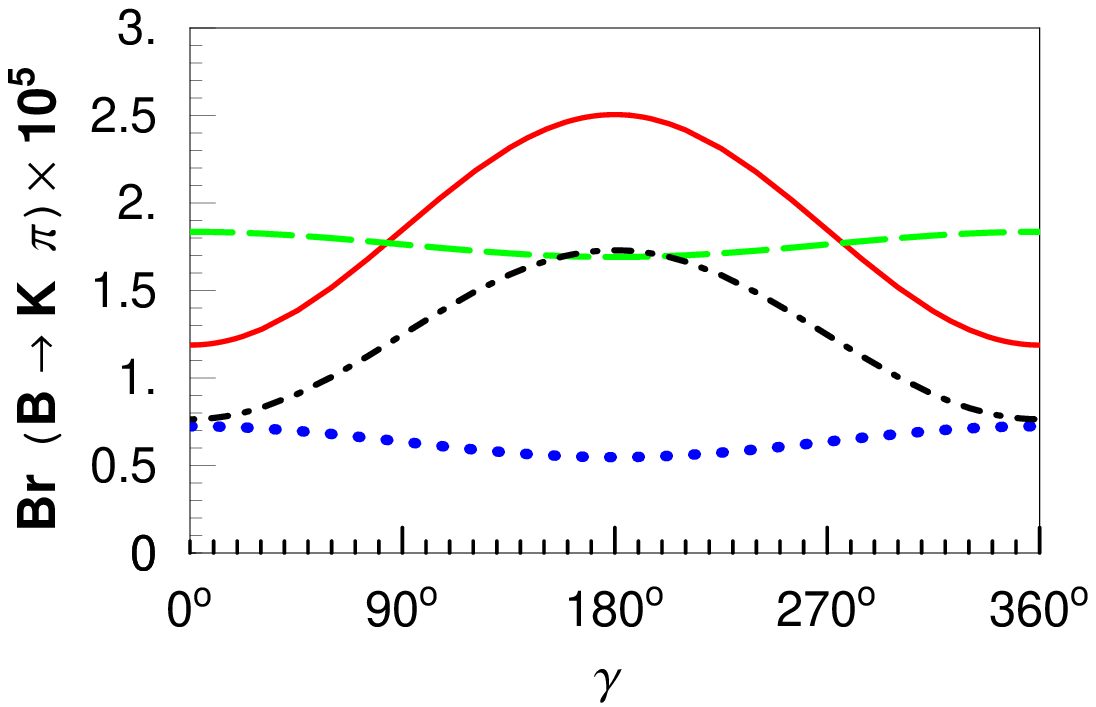,height=2.0in} \vspace{0.4cm}
    \caption{{\small Branching ratios of $K \pi$ modes versus
the  unitarity angle $\gamma$, where the solid, dashed, dotdashed
and dotted curves correspond to $B\to K^-\pi^+$, $\ov K^0\pi^-$,
$K^-\pi^0$ and $\ov K^0 \pi^0$, respectively. Uses of $N_c^{\rm
eff}(LL)=2$, $N_c^{\rm eff}(LR)=5$ and the BSW model for form
factors have been made.}} \label{fig:RpiK}
\end{figure}

Experimentally, a substantial difference in the first two decay
modes implied by the earlier data makes the Fleischer-Mannel bound
\cite{FM} on the unitarity angle $\gamma$ interesting. An
improvement of the data samples and a new decay mode observed by
CLEO \cite{CLEOpiK,Roy} indicate nearly equal branching ratios for
the three modes $K^-\pi^+$, $\ov K^0\pi^-$ and $K^-\pi^0$:
\be
&& {\cal B}(\ov B^0\to K^-\pi^+)=(1.4\pm 0.3\pm 0.2)\times
10^{-5}, \non \\ && {\cal B}(B^-\to \ov K^0\pi^-)=(1.4\pm 0.5\pm
0.2)\times 10^{-5}, \non\\ && {\cal B}(B^-\to K^-\pi^0)=(1.5\pm
0.4\pm 0.3)\times 10^{-5}.
\en
While the improvement on the first two decay modes is in
accordance with the theoretical expectation, the central value of
the new measured decay mode $B^-\to K^-\pi^0$ is larger than the
naive anticipation. Of course, one has to await the experimental
improvement to clarify this issue. If the present data persist, an
interesting interpretation based on the revived idea of a negative
$\rho$ is pointed out recently in \cite{He98}. To see the impact
of a negative $\rho$ or the dependence on the unitarity angle
$\gamma$, we plot in Fig. \ref{fig:RpiK} the branching ratios of
$K\pi$ modes versus $\gamma$. It is clear that (i) the
aforementioned pattern $\ov K^0\pi^- > K^-\pi^+ > K^-\pi^0$ is
modified to $K^-\pi^+>\ov K^0\pi^->K^-\pi^0$ when
$\gamma>90^\circ$, (ii) the decay rate of $K^-\pi^0$ is close to
that of $\ov K^0\pi^-$ when $\gamma$ approaches to $180^\circ$,
and (iii) the purely-penguin decay mode $\ov K^0\pi^-$ is
insensitive to the change of $\gamma$, as expected.

A rise of the $K^- \pi ^+$ and $K^- \pi^0$ decay rates from their
minima at $\gamma=0^\circ$ and $360^\circ$ (or $|\rho|=\rho_{\rm
max}=0.41$ and $\eta=0$) to the maxima at $\gamma=180^\circ$ (or
$\rho=0$ and $\eta=-0.41$) can be understood as follows:  The
interference between tree and penguin contributions in these two
decay processes is destructive for negative $\rho$ and becomes
largest at $\gamma=0^\circ$ and then decreases with increasing
$\gamma$. When the sign of $\rho$ is flipped, the interference
becomes constructive and has its maximal strength at
$\gamma=180^\circ$. It is obvious from the above discussion that a
negative $\rho$ alone is not adequate to explain the
nearly-equality of $ K\pi $ modes since an increase of $K^- \pi^0$
is always accompanied by a rise of $K^-\pi^+$. Therefore, final
state interactions are probably needed to explain the central
values of the data.

Finally we remark that it is anticipated that
$K^-\pi^+>K^{*-}\pi^+$ (likewise, $\ov K^0\pi^->\ov K^{*0}\pi^-$;
see Eq. (\ref{pattern})) owing to the absence of the $a_6$ penguin
term in the latter . The branching ratio of $K^{*-}\pi^+$ and $\ov
K^{*0}\pi^-$ is predicted to be of order $0.5\times 10^{-5}$ at
$\gamma=65^\circ$ (see Table IX) and $\sim 1.0\times 10^{-5}$ at
$\gamma=90^\circ$. As noted in passing, $K^-\rho^+$ and $\ov
K^0\rho^-$ have smaller branching ratios, typically of order
$1\times 10^{-6}$, as the $a_6$ penguin term contributes
destructively.

\subsection{$B^\pm\to \omega K^\pm$ and $B^\pm\to\rho^0 K^\pm$ decays}
The CLEO observation \cite{CLEOomega2} of a large branching ratio
for $B^\pm\to \omega K^\pm$
\be
{\cal B}(B^\pm\to\omega K^\pm)=\left(1.5^{+0.7}_{-0.6}\pm
0.2\right) \times 10^{-5},
\en
is rather difficult to explain at first sight. Its factorizable
amplitude is of the form (see Appendix E) \be  \label{omegaK}
A(B^-\to\omega K^-)  &=&
V_{ub}V_{us}^*\Big\{a_1X^{(B\omega,K)}+a_2X_u^{(BK,\omega)}\Big\}
-V_{tb}V_{ts}^*\Big\{ [a_4+a_{10}+R(a_6+a_8)]X^{(B\omega, K)} \non
\\ &+& [2a_3+2a_5+{1\over 2}(a_7+a_9)]X^{(BK,\omega)}+\cdots\Big\},
\en
with $R\cong -2m_K^2/(m_bm_s)$, where ellipses represent
contributions from $W$-annihilation and spacelike penguin
diagrams. It is instructive to compare this decay mode closely
with $B^-\to\rho^0K^-$:
\be   \label{rhoK} A(B^-\to\rho^0 K^-)&=&
V_{ub}V_{us}^*\Big\{a_1X^{(B\rho^0,K)}+a_2X_u^{(BK,\rho^0)}\Big\}
 \\ &-& V_{tb}V_{ts}^*\Big\{
[a_4+a_{10}+R'(a_6+a_8)]X^{(B\rho^0,K)}+{3\over
2}(a_7+a_9)X_u^{(BK,\rho^0)}+\cdots\Big\}, \non
\en
with $R'\cong -2m_\rho^2/(m_bm_s)$. Although the tree amplitude is
suppressed by the mixing angle,
$|V_{ub}V_{us}^*/V_{tb}V_{ts}^*|=\lambda^2$, the destructive
interference between $a_4$ and $a_6$ penguin terms renders the
penguin contribution small. Consequently, the relative weight of
tree and penguin contributions to $\omega K^-$ and $\rho^0K^-$
depends on the values of $\nc$ (see Table VI). At our favored
values $\nc(LL)=2$ and $\nc(LR)=5$, we see that the tree
contribution is important for both channels. It is also clear from
Table VI that the electroweak penguin contribution to $\rho^0K^-$
is as important as the tree diagram. The branching ratio of
$B^\pm\to\rho^0 K^\pm$ is estimated to be of order
$(0.5-0.9)\times 10^{-6}$ (see Table IX). This prediction is
relatively stable against $\nc$. While the current bound is ${\cal
B}(B^-\to\rho^0 K^-)<2.2\times 10^{-5}$ \cite{Gao}, the
preliminary measurement of $B^-\to\rho^0 K^-$ shows a large event
yield $14.8^{+8.8}_{-7.7}$ \cite{Gao}. If the branching ratio of
this decay is found to be, say, of order $0.5\times 10^{-5}$, then
it is a serious challenge to theorists.

\begin{figure}[tb]
\psfig{figure=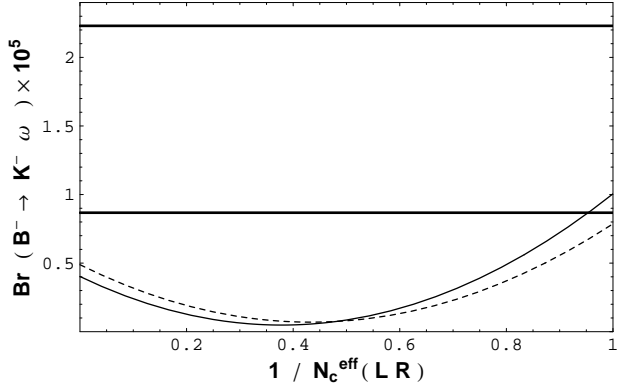,height=2.0in} \vspace{0.4cm}
    \caption{{\small The branching ratio of $B^-\to \omega K^-$ vs
    $1/\nc(LR)$
with $\nc(LL)$ being fixed at the value of 2. The solid (dotted)
curve is calculated using the BSW (LCSR) model. The solid thick
lines are the CLEO measurements with one sigma errors.}}
\label{fig:omegak}
\end{figure}

Since the $\omega K^-$ amplitude differs from that of $\rho^0 K^-$
only in the QCD penguin term proportional to $(a_3+a_5)$ and in
the electroweak penguin term governed by $a_9$, it is naively
anticipated that their branching ratios are similar if the
contributions from $a_3,a_5,a_9$ are negligible. The question is
then why is the observed rate of the $\omega K^-$ mode much larger
than the theoretical estimate of the $\rho^0 K^-$ mode ? By
comparing (\ref{omegaK}) with (\ref{rhoK}), it is natural to
contemplate that the penguin contribution proportional to
$(2a_3+2a_5)$ accounts for the large enhancement of $B^\pm\to
\omega K^\pm$. However, this is not the case: The coefficients
$a_3$ and $a_5$, whose magnitudes are smaller than $a_4$ and
$a_6$, are not large enough to accommodate the data unless
$\nc(LR) <1.2$ (see Fig.~\ref{fig:omegak}). It is evident that the
predicted branching ratio of $B^-\to\omega K^-$ is in general too
small if $\nc(LL)$ is fixed at the value of 2 and $1/\nc(LR)<0.5$.
If $\nc(LL)$ is assumed to be the same as $\nc(LR)$, then the
branching ratio can rise above $1\times 10^{-5}$ at the small
value of $1/\nc\cong 0$ \cite{AKL} since $a_3+a_5$ has its maximum
at $\nc=\infty$ (see Table III). However, it seems to us that
$\nc\to\infty$ for hadronic $B$ decays is very unlikely.

So far we have neglected three effects in the consideration of
$B^\pm\to\omega K^\pm,\rho^0 K^\pm$ decays: $W$-annihilation,
spacelike penguin diagrams and final-state interactions. The first
two mechanisms play the same role for both modes and they will
lead to the decay rate of $\omega K^-$ similar to $\rho^0 K^-$. If
the latter is observed to have a similar rate as the former, it is
plausible that $W$-annihilation and spacelike penguins could play
a prominent role to both modes. However, if ${\cal B}(B^-\to\rho^0
K^-)\ll {\cal B}(B^-\to\omega K^-)$ is observed experimentally,
then one possibility is that FSI may explain the disparity between
$\rho^0 K^-$ and $\omega K^-$ modes, as elaborated on in Sec. VI.
At any rate, it is crucial to measure the branching ratios of both
modes in order to understand their underlying mechanism.

\subsection{Electroweak penguins}

Electroweak penguin diagrams contribute to all charmless $B$
decays. The relative importance of electroweak penguin amplitudes
can be read directly from Tables V-VII. \footnote{The relative
importance of electroweak penguin effects in penguin-dominated $B$
decays is studied in \cite{AKL} by computing the ratio
\be
R_W=\,{{\cal B}(B\to h_1h_2)({\rm with}~a_7,\cdots,a_{10}=0)\over
{\cal B}(B\to h_1 h_2) }. \label{rw}
\en
However, because of variously possible interference of the
electroweak penguin amplitude with the tree and QCD penguin
contributions, $R_W$ is not the most suitable quantity for
measuring the relative importance of electroweak penguin effects;
see \cite{CCT} and an example in Sec. VII.} In order to study
their effects, we need to focus on those modes in which QCD
penguins do not contribute or their effects are small. It is known
that in the rare $B_s$ decays, the decay modes
\be
\label{Bs}
B_s\to\eta\pi,~\eta'\pi,~\eta\rho,~\eta'\rho,~\phi\pi,~\phi\rho
\en
do not receive any QCD penguin contributions \cite{Fleisher} (for
a detailed discussion, see \cite{CCT}) . Therefore, these six
decay modes are predominantly governed by the largest electroweak
penguin coefficient $a_9$. By contrast, there are only two
channels in charmless $B_u$ and $B_d$ decays that do not receive
QCD penguin contributions, namely $B^-\to\pi^-\pi^0$ and
$B^-\to\rho^- \rho^0$, and they are dominated by tree diagrams.
Nevertheless, there do exist several channels in which the QCD
penguin contribution is small. From the Appendix we see that the
amplitudes of the class-V decays
\be
B_d^0\to\phi\pi^0,\,\phi\eta,\,\phi\eta',\,\phi\rho^0,\,\phi\omega,\qquad
B^+\to\phi\pi^+,\,\phi\rho^+
\en
are proportional to $[a_3+a_5-{1\over 2}(a_7+a_9)]$. Since the
effective coefficients $a_3$ and $a_5$ are $\nc$ sensitive, the
decay rates depend very sensitively on $\nc$ and are governed by
electroweak penguins at $\nc(LL)\sim 2,~\nc(LR)\sim 5$ or
$\nc(LL)\sim\nc(LR)\sim 3$ where the QCD penguin contribution
characterized by $a_3+a_5$ is close to its minimum (see Table
III). Unfortunately, their branching ratios are very small (see
Tables VIII-X), of order $(1-6)\times 10^{-9}$. We also see that
the electroweak penguin in
\be
B^0_d\to K^0\rho^0,\qquad B^+\to K^+\rho^0
\en
is as important as the QCD penguin diagram because the latter is
proportional to $[a_4-2a_6m_K^2/(m_bm_s)]$ which involves a large
cancellation. The branching ratio of the above two modes is of
order $(0.5-1.0)\times 10^{-6}$.

\subsection{Theoretical uncertainties}
  The calculation of charmless hadronic $B$ decay rates suffers from many
theoretical uncertainties. Most of them have been discussed before
and it is useful to make a short summary below.

\begin{itemize}
\item
Heavy-to-light form factors and their $q^2$ dependence. We have
considered in the present paper two different form-factor models:
the BSW model and the LCSR approach. It turns out that ${\cal
B}(B\to VV)$ is very sensitive to the form-factor ratio $A_2/A_1$.
\item
Decay constants. Since the decay constants for light pseudoscalar and vector
mesons are well measured, the uncertainty due to this part is the least.
\item
Running quark masses at the scale $m_b$. The decay rates of
penguin-dominated charmless $B$ decays are generally sensitive to
the value of $m_s(m_b)$. The light quark masses arise in the decay
amplitude because equation of motion has been applied to the
matrix element of $(S-P)(S+P)$ interactions obtained from the
Fierz transformation of $(V-A)(V+A)$ penguin operators. Since the
current quark masses are not known precisely, this will result in
large uncertainties for branching ratios.\footnote{In order to
avoid the uncertainty originated from the light quark masses, an
attempt of evaluating the $(S-P)(S+P)$ matrix element using the
perturbative QCD method has been made in \cite{Du99}. It is found
that the results are comparatively smaller than that obtained
using equations of motion.} While the measured $B\to\eta' K$
favors a smaller strange quark mass, a too small value of
$m_s(m_b)$ will lead to a too large $B\to K\pi$.
\item
Quark mixing matrix elements parametrized in terms of the
parameters $\rho,\,\eta,\,A,\,\lambda$. The uncertainty due to the
values of $\rho$, $\eta$ and $A$ is reflected on the uncertainty
on the angles $\alpha,\beta, \gamma$ of the unitarity triangle.
\item
Nonfactorized contributions to hadronic matrix elements. The main
result of the present paper is to show that $\nc(LR)>3>\nc(LL)\sim
2$ implied by the bulk of the data.
\item
The magnitude of the gluon momentum transfer in the timelike
penguin diagram. We have employed $k^2=m_b^2/2$ for calculating
the effective Wilson coefficients, though in general $k^2$ lies in
the range $m^2_b/4 \lesssim k^2\lesssim m^2_b/2$
\cite{Deshpande90}. The common argument is that while CP violation
is sensitive to the value of $k^2$, this is not the case for the
decay rate.
\item
Final-state interactions (FSI). This is the part least known.
Nevertheless, some qualitative statement and discussion about FSI
still can be made, as shown in the next section.
\item
$W$-annihilation contribution. It is commonly believed that this
contribution is negligible due to helicity suppression. Moreover,
$W$-exchange is subject to both color and helicity suppression.
The helicity suppression is likely to work because of the large
energy released in rare $B$ decays.
\item
Spacelike penguin contribution. The spacelike penguin amplitude
gains a large enhancement by a factor of $m_B^2/(m_bm_{u,d})$ or
$m_B^2/(m_bm_s)$. Therefore, {\it a priori} there is no convincing
reason to ignore this effect that has been largely overlooked in
the literature. Unfortunately, we do not have a reliable method
for estimating the spacelike penguins.
\end{itemize}

\section{Final-state Interactions}

It is customarily argued that final-state interactions (FSI) are
expected to play only a minor role in rare hadronic $B$ decays due
to the large energy released in the decay process. Nevertheless,
phenomenologically their presence could be essential in some
cases: (i) Inelastic scattering may account for the observed large
branching ratio of $B^-\to\omega K^-$. (ii) Some channels, for
instance $B^0\to K^+ K^-$ receives direct contributions only from
$W$-exchange and penguin-annihilation diagrams, can be induced
from FSI. (iii) The tree-dominated neutral modes, e.g.,
$B^0\to\pi^0\pi^0,~\pi^0\rho^0,~\rho^0\rho^0$, may get large
enhancement from FSI.

  In general, the effects of FSI are important and dramatic for the weak
decay $B\to X$ if there exists a channel $B\to Y$ with a
sufficiently large decay rate, i.e. ${\cal B}(B\to Y)\gg{\cal
B}(B\to X)$ and if $X$ and $Y$ modes couple through FSI. A famous
example is the decay $D^0\to\ov K^0\pi^0$ which is naively
expected to be very suppressed but it gets a large enhancement
from the weak decay $D^0\to K^-\pi^+$ followed by the FSI:
$K^-\pi^+\to \ov K^0\pi^0$.

\vspace{0.4 cm} \noindent \underline{Inelastic scattering
contribution to $B^\pm\to\omega K^\pm$} \vspace{0.2cm}

\begin{figure}[tb]
\psfig{figure=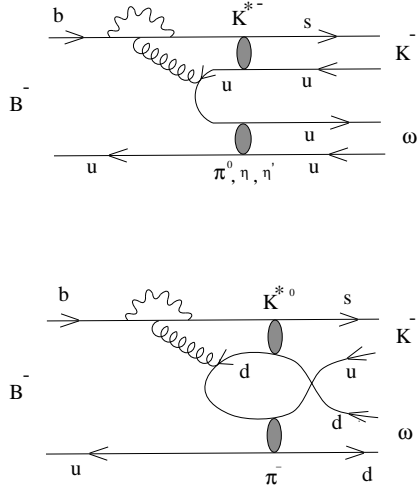,height=2.5in} \vspace{0.5cm}
\caption{Contributions to $B^-\to K^-\omega$ from final-state
interactions via the weak decays $B^-\to
K^{*-}\pi^0,~K^{*-}\etapp$ and $B^-\to K^{*0}\pi^-$ followed by
quark rescattering.} \label{fig:fsi}
\end{figure}

As shown in Sec. V.F, it is difficult to understand why the
observed branching ratio of $B^-\to\omega K^-$ is one order of
magnitude larger than the theoretical expectation $(1.2-1.8)\times
10^{-6}$ (see Table IX). There are three possible effects for
enhancement: $W$-annihilation, spacelike penguin diagrams and FSI.
If the pattern ${\cal B}(B^-\to\omega K^-)\gg {\cal
B}(B^-\to\rho^0 K^-)$ is observed experimentally, FSI may account
for the disparity between $\omega K^-$ and $\rho^0 K^-$ as the
first two mechanisms contribute equally to both modes. The weak
decays $B^-\to K^{*-}\pi^0,\,K^{*-}\etapp$ via the penguin process
$b\to su\bar u$ and
$B^-\to\{K^{*-}\pi^0,K^{*-}\etapp,K^{*0}\pi^-,K^-\rho^0,K^0\rho^-\}$
via $b\to sd\bar d$ followed by the quark rescattering reactions
$\{K^{*-}\pi^0,K^{*-}\etapp,K^{*0}\pi^-,K^-\rho^0,K^0\rho^-\}\to\omega
K^-$ contribute {\it constructively} to $B^-\to \omega K^-$ (see
Fig.~\ref{fig:fsi}), but {\it destructively} to $B^-\to\rho K^-$.
Since the branching ratios for $B^-\to K^{*-}\pi^0,~K^{*-}\etapp$
and $K^{*0}\pi^-$ are not small, of order $(0.3-0.7)\times
10^{-5}$, it is conceivable that a bulk of observed
$B^\pm\to\omega K^\pm$ arises from FSI via inelastic scattering.
However, it is not clear to us quantitatively if FSI are adequate
to enhance the branching ratio by one order of magnitude.

\vskip 5mm

\vspace{0.4 cm}
\noindent \underline{Inelastic scattering
contribution to $B^0\to K^+ K^-$} \vspace{0.2cm}

The decay $B^0\to K^+ K^-$ proceeds through the $W$-exchange and
penguin annihilation diagrams and its factorized amplitude given
in Appendix B is governed by the factorized term $\la K^+K^-|(\bar
qq)_\vma|0\ra\la 0|(\bar d b)_\vma|\ov B^0\ra$ with $q=u,s$. If
helicity suppression works, then this factorized term and hence
${\cal B}(B^0\to K^+K^-)$ is anticipated to be very suppressed.
Nevertheless, the final-state rescattering contribution to $B^0\to
K^+K^-$ from $\rho^+\rho^-,\pi^+\pi^-,\cdots$ intermediate states
could be sizable, in particular $B^0\to \rho^+\rho^-$ should have
a large branching ratio of order $(2-4)\times 10^{-5}$. Therefore,
this decay is expected to be dominated by the rescattering effect
\cite{Gronau}. A measurement of $B^0\to K^+K^-$ will provide
information on the inelastic FSI. The present limit is ${\cal
B}(B^0\to K^+ K^-)<2.3\times 10^{-6}$ \cite{Gao}. Another example
is the decay $B^0\to\phi\phi$ which proceeds via the spacelike
penguin diagram (see Appendix F). It receives indirect
contributions arising from the weak decays $B^0\to\etapp\etapp$
followed by the rescattering $\etapp\etapp\to\phi\phi$.

\vspace{0.4 cm} \noindent \underline{Elastic FSI on $B\to\pi\pi$}
\vspace{0.2cm}

\begin{figure}[tb]
%\vspace{0.15cm}
\psfig{figure=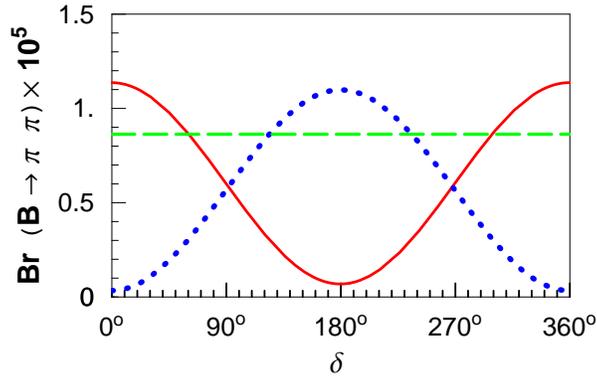,height=2.0in} \vspace{0.4cm}
    \caption{{\small Branching ratios of $B\to\pi \pi$ modes versus the
isospin phase shift difference $\delta$, where the solid, dashed,
and dotted curves correspond to $\pi^+\pi^-$, $\pi^- \pi^0$, and
$\pi^0 \pi^0$, respectively. Uses of $N_c^{\rm eff}(LL)=2$,
$N_c^{\rm eff}(LR)=5$ and the BSW model for form factors have been
made.}}
     \label{fig:delta}
\end{figure}

In order to understand the effect of FSI on $B\to\pi\pi$ decays,
we decompose the decay amplitudes into their isospin amplitudes
\be
{\cal M}(B^0\to\pi^0\pi^0) &=& \sqrt{1\over 3}
A_{0}e^{i\delta_0}-\sqrt{2\over 3}A_{2}e^{i\delta_2}, \non
\\ {\cal M}(B^0\to\pi^+\pi^-) &=&
\sqrt{2\over 3}A_{0}e^{i\delta_0}+\sqrt{1\over
3}A_{2}e^{i\delta_2}, \non \\ {\cal M}(B^-\to\pi^-\pi^0) &=&
\sqrt{3\over 2}A_{2}e^{i\delta_{2}}, \label{pipi}
\en
where $A_{0}$ and $A_{2}$ are isospin 0 and 2 amplitudes,
respectively, and $\delta_0$, $\delta_2$ are the corresponding
$S$-wave $\pi\pi$ scattering isospin phase shifts. Note that the
amplitudes (\ref{pipi}) for $\pi^+\pi^-$ and $\pi^-\pi^0$ are the
same as the usual invariant amplitudes, but ${\cal
A}(B^0\to\pi^0\pi^0)=\sqrt{2}{\cal M}(B^0\to\pi^0\pi^0)$. To
proceed we shall assume that inelasticity is absent or negligible
so that the isospin phase shifts are real and the magnitude of the
isospin amplitudes is not affected by elastic FSI. Theoretically,
$A_{0}$ and $A_{2}$ are of the same sign. As stressed in Sec. V.A,
model calculations tend to predict a branching ratio of
$B^0\to\pi^+\pi^-$ larger than the present limit. One possibility
is that the isospin phase difference $\delta=\delta_0-\delta_2$ is
nonzero. In Fig. \ref{fig:delta} we plot the branching ratios of
$\pi\pi$ modes versus $\delta$. It is evident that the suppression
of $\pi^+ \pi^-$ and enhancement of $\pi^0 \pi^0$ become most
severe when $\delta\gtrsim 70^\circ$ and furthermore the latter
becomes overwhelming at $\delta> 90^\circ$. Note that using the
Regge analysis, $\delta_{\pi\pi}$ is estimated to be $11^\circ$ in
\cite{Gerard}.

\section{Comparison with literature}
In this section we would like to compare our framework and results
with the excellent paper by Ali, Kramer and L\"u (AKL) \cite{AKL}
in which nonleptonic charmless $B$ decays are studied in a great
detail. Our expressions for the factorized decay amplitudes of all
two-body hadronic decays of $B_u$ and $B_d$ mesons are in
agreement with AKL except that we have also included $W$-exchange,
$W$-annihilation and spacelike penguin matrix elements in the
expressions of factorized decay amplitudes, though they are
usually neglected in the conventional calculation. Basically, our
framework differs from AKL in the choice of the input
parameters:\footnote{Using the same values of input parameters as
\cite{AKL}, we are able to reproduce all the branching ratios of
AKL except for the decays $B^0\to\rho^0\rho^0,~\rho^0\omega$. This
discrepancy is resolved after numerical corrections are made in
\cite{AKL} (private communication with C.D. L\"u).} (i) The
effective Wilson coefficients $c^{\rm eff}_i$ are obtained from
the $\mu$-dependent Wilson coefficient functions $c_i(\mu)$ at
$\mu=m_b$ in the present paper and at $\mu=m_b/2$ by AKL. Although
$c_i^{\rm eff}$ obtained by AKL and by us are scheme and scale
independent, our effective Wilson coefficients are gauge invariant
and free of the infrared singularity. (ii) As explained in detail
before, we treat $\nc(LL)$ and $\nc(LR)$ differently for
nonfactorized effects, while $\nc(LL)=\nc(LR)=\nc$ is assumed by
AKL with the preference $1/\nc\leq 0.2$ or $\nc\geq 5$. (iii) For
the Wolfenstein parameters $\rho$ and $\eta$, we use $\rho=0.175$
and $\eta=0.370$, corresponding to $(\rho^2+\eta^2)^{1/2}=0.41$,
while $\rho=0.12$, $\eta=0.34$ and $(\rho^2+\eta^2)^{1/2}=0.36$
are employed by AKL.  (iv) To evaluate the pseudoscalar matrix
element arising from the penguin interactions, we apply equations
of motion and use the light quark masses at $\mu=m_b$, while
$m_q(\mu=m_b/2)$ is employed by AKL. (v) We apply the usual
one-mixing angle formulation to the $\eta-\eta'$ mixing and
two-mixing angle formulation to the decay constants of the $\eta$
and $\eta'$, whereas AKL use the two-angle parametrization for
both $\eta-\eta'$ mixing and decay constants. (vi) The
pseudoscalar matrix elements of $\etapp$-vacuum transition,
characterized by the parameters $r_\eta$ and $r_{\eta'}$ in
(\ref{reta}), have different expressions in the present paper and
in AKL.

In spite of the differences in the aforementioned input
parameters, our work does agree with AKL in most cases. Some
noticeable differences are as follows:
\begin{enumerate}
\item While our expressions for factorized amplitudes agree
with AKL, we do have included $W$-annihilation and spacelike
penguin contributions. For example, the decay amplitudes of $
B^0\to K^{+(*)}K^{-(*)}$, which proceed only through
$W$-annihilation and spacelike penguin diagrams, are displayed in
our Tables.
\item Employing the same values of $\nc$ as AKL, our predictions of
branching ratios for tree-dominated decay modes are in general
larger than that of AKL by a factor of 1.3 due to the difference
in the use of $(\rho^2+\eta^2)$ or $|V_{ub}|^2$.
\item It was
advocated by AKL that the branching ratios of the decays
$B^-\to\phi K^-,~B^0\to\phi K^0$, $B^-\to\phi K^{*-},~B^0\to\phi
K^{*0}$ are almost equal in the factorization approach, whereas we
found that the decay rate of $B\to\phi K^*$ is very sensitive to
the form-factor ratio $x=A_2^{BK^*}(m^2_\phi)/
A_1^{BK^*}(m^2_\phi)$ and that the data of $B\to\phi K$ and
$B\to\phi K^*$ can be simultaneously accommodated in the
generalized factorization approach using the LCSR form factors
(see Figs. \ref{fig:Kphi} and \ref{fig:VKphi}).
\item We have argued that theoretically and
phenomenologically the effective number of colors for $(V-A)(V-A)$
and $(V-A)(V+A)$ four-quark operators should be treated
differently. The data of tree-dominated decays
$B^-\to\rho^0\pi^-,\omega\pi^-$ indicate $\nc(LL)<3$, while the
penguin-dominated modes $B^-\to\phi K^-,\eta' K^-$ clearly imply
$\nc(LR)>3$. If using $\nc(LL)=\nc(LR)=\nc$ as adopted by AKL, we
found that the data of $B^-\to\phi K^-$ and $B^-\to\rho^0\pi^-$
cannot be accommodated simultaneously.
\item Our prediction for $B\to\eta' K$ is significantly different
from that of AKL at the small value of $1/\nc$. As illustrated in
Fig.~\ref{fig:Ketap}, the branching ratio of $B^-\to\eta'K^-$
predicted by AKL for $\nc(LL)=\nc(LR)$, corresponding to the
dashed curve in Fig.~\ref{fig:Ketap}, is largely enhanced at small
$1/\nc(LR)$ provided that $\nc(LL)$ is fixed at the value of 2.
Therefore, without adjusting other input parameters, the
prediction of AKL will be significantly improved if $\nc(LL)$ and
$\nc(LR)$ are treated differently. Moreover, we have shown that it
is natural to have $\eta'K^\pm>\eta' K^0$ if $\gamma>90^\circ$.
\item We found that in the absence
of FSI, the branching ratio of $B^-\to\omega K^-$ is expected to
be of the same order as ${\cal B}(B^-\to\rho^0 K^-)\sim
(0.5-1.0)\times 10^{-6}$, whereas the branching ratio predicted by
AKL rises above $1\times 10^{-5}$ at the small values of $1/\nc$,
$1/\nc\approx 0$. We argue that if ${\cal B}(B^-\to\omega K^-)\gg
{\cal B}(B^-\to\rho^0 K^-)$ is observed experimentally, then
inelastic final-state rescattering may account for the disparity
between $\omega K^-$ and $\rho^0 K^-$.
\item  It is claimed by AKL that the decay $B^0\to\rho^0 K^0$
is completely dominated by the electroweak penguin transitions for
all values of $\nc$ and that a measurement of this mode will
enable one to determine the largest electroweak penguin
coefficient $a_9$. We found that the QCD penguin contribution to
$\rho^0 K^0$ is not small compared to the electroweak penguin. To
illustrate this point, we compute the ratio $R_W$ defined in Eq.
(\ref{rw}) and obtain $R_W(\rho^0K^0)=0.12$ averaged over
CP-conjugate modes for $\nc(LL)=\nc(LR)=2$, to be compared with
the value 0.08 predicted by AKL for the same values of $\nc$. It
thus appears that the $\rho^0K^0$ mode is almost completely
dominated by the electroweak penguin. However, at the amplitude
level, we found
\be
{\rm tree}:\,{\rm QCD~penguin}:\,{\rm electroweak~penguin}=
-0.18+0.54i:\,1:\,1.8-0.14i \label{amp1}
\en
for $\ov B^0\to\rho^0\ov K^0$ (see Table VI) and
\be
{\rm tree}:\,{\rm QCD~penguin}:\,{\rm electroweak~penguin}=
-0.32-0.47i:\,1:\,1.8-0.14i  \label{amp2}
\en
for $B^0\to\rho^0K^0$, where the QCD penguin amplitude has been
normalized to unity. It is evident that although Eqs. (\ref{amp1})
and (\ref{amp2}) lead to $R_W=0.12$, the electroweak penguin
contribution to the amplitude is largely contaminated by the QCD
penguin one. Therefore, we conclude that only the $B_s$ decay
modes listed in (\ref{Bs}) can provide a direct and unambiguous
determination of $a_9$.

\end{enumerate}
\section{Conclusions}
Using the next-to-leading order QCD-corrected effective
Hamiltonian and gauge-invariant, scheme- and scale-independent
effective Wilson coefficients, we have systematically studied
hadronic charmless two-body decays of $B_u$ and $B_d$ mesons
within the framework of generalized factorization. Nonfactorizable
effects are parametrized in terms of $\nc(LL)$ and $\nc(LR)$, the
effective numbers of colors arising from $(V-A)(V-A)$ and
$(V-A)(V+A)$ 4-quark operators, respectively. The branching ratios
are calculated as a function of $\nc(LR)$ with two different
considerations for $\nc(LL)$: (i) $\nc(LL)$ being fixed at the
value of 2, and (ii) $\nc(LL)=\nc(LR)$. Depending on the
sensitivity of the effective coefficients $a_i^{\rm eff}$ on
$\nc$, we have classified the tree and penguin transitions into
six different classes.  Our main results are the following:

\begin{itemize}

\item To avoid the gauge and infrared problems connected with
effective Wilson coefficients, we have worked in the on-shell
scheme to obtain gauge invariant, infrared finite $c_i^{\rm eff}$.
The infrared pole is consistently absorbed into universal
bound-state wave functions.

\item The relative magnitudes of tree, QCD penguin and electroweak
penguin amplitudes of all charmless $B$ decays are tabulated in
Tables V-VII for $\nc(LR)=2,3,5,\infty$ and $\nc(LL)=2$ as well as
$\nc(LL)=\nc(LR)$. The predicted branching ratios are summarized
in Tables VIII-X.
\item Hadronic charmless $B$ decays without strangeness in the final state
are dominated by the tree $b\to u\bar u d$ transition. The
exceptional modes are $B^0\to\pi^0\eta,\pi^0\eta',\rho^o\omega$
which proceed mainly through the penguin diagram. The first
measurement of the hadronic $b\to u$ decay $B^-\to\rho^0\pi^-$ by
CLEO indicates that $1.1\leq \nc(LL)\leq 2.6$. Therefore,
$\nc(LL)$ is preferred to be smaller than 3. Moreover,  the
current experimental information on $B^-\to\omega\pi^-$ and
$B^0\to\pi^+\pi^-$ also favors a small $\nc(LL)$. For example, the
former implies $1.7<\nc(LL)<2.5$. The fact that $\nc(LL)\approx 2$
is favored is also consistent with the nonfactorizable term
extracted from $B\to (D,D^*) (\pi,\rho)$ decays, $\nc(B\to
D\pi)\approx 2$. The measurement of the ratios $R_{1-4}$ of
charged to neutral branching fractions [see Eq.~(\ref{Ri})] is
useful for determining $\nc(LL)$.

\item The tree-dominated class I-III modes that
have branching ratios of order $10^{-5}$ or larger must have one
or two vector mesons in the final state. For example, it is
expected that ${\cal B}(\ov B^0\to \rho^-\rho^+)\sim {\cal B}(\ov
B^0\to \rho^-\pi^+)> {\cal B}(\ov B^0\to \pi^-\rho^+)\sim 1\times
10^{-5}$. By contrast, the decay rates of penguin-dominated
class-IV decays follow the pattern: $\Gamma(\ov B\to
P_aP_b)>\Gamma(\ov B\to P_aV_b)\sim\Gamma(\ov B\to V_a
V_b)>\Gamma(\ov B\to V_aP_b)$, where $P_b$ or $V_b$ is
factorizable under the factorization assumption, because of
various possibilities of interference between the penguin
amplitudes governed by the QCD penguin parameters $a_4$ and $a_6$,
Moreover, the penguin-dominated charmless $B$ decays have the
largest branching ratios in the $PP$ mode.

\item The present limit on $B^\pm\to\phi K^\pm$ implies that $\nc(LR)
\gtrsim 3.2$ and 4.2 in the BSW and LCSR models, respectively. The
data of $B\to \phi K$ and $B\to \phi K^*$ can be accommodated
simultaneously if the form-factor ratio $A_2^{BK^*}(m^2
_\phi)/A_1^{BK^*}(m^2_\phi)$ is less than unity. We found that the
ratio $\Gamma(B\to\phi K^*)/\Gamma(B^\pm\to\phi K^\pm)$ is 0.76 in
the BSW model, while it is equal to 1.9 in the LCSR analysis.

\item If $\nc(LL)$ is treated to be the same as $\nc(LR)$, we showed that
${\cal B}(B^-\to\eta' K^-)\sim (2.7-4.7)\times 10^{-5}$ at
$0<1/\nc<0.5$ and becomes even smaller at small $1/\nc$ when the
charm content contribution of the $\eta'$ is taken into account.
We have demonstrated that the discrepancy between theory and
experiment is significantly improved by setting $\nc(LL)\sim 2$.
In particular, the $\eta'$ charm content contribution is in the
right direction. Therefore, the data of $B\to K\eta'$ provide a
strong support for $\nc(LL)\sim 2$ and the relation
$\nc(LR)>\nc(LL)$. The mode $B\to\eta' K$ has the largest
branching ratio in the two-body charmless $B$ decays due mainly to
the constructive interference between the penguin contributions
arising from the $(\bar uu+\bar dd)$ and $\bar ss$ components of
the $\eta'$. By contrast, the destructive interference for the
$\eta$ production leads to a much smaller decay rate for $B\to\eta
K$. If the disparity between $\eta'K^\pm$ and $\eta'K^0$ is
confirmed in the future, it could be attributed to a negative
Wolfenstein's $\rho$ parameter.

\item The penguin-dominated class-V modes
$\ov B^0_d\to \phi\pi^0,\,\phi\eta,\,\phi\eta',\,\phi\rho^0,\,\phi\omega,~B
^+\to\phi\pi^+,\,\phi\rho^+$ depend very sensitively on $\nc$ and are
dominated by electroweak
penguins at $\nc(LL)\sim 2,~\nc(LR)\sim 5$ or $\nc(LL)\sim\nc(LR)\sim 3$.
The electroweak penguin effect in the decays $\ov B^0\to \ov K^0\rho^0,B^-\to
K^-\rho^0$ is as important as the QCD penguin contribution.

\item   Final-state interactions (FSI) are conventionally believed to
play only a minor role in hadronic charmless $B$ decays due to the
large energy released in the decay. We showed that in the absence
of FSI, the branching ratio of $B^+\to\omega K^+$ is expected to
be of the same order as ${\cal B}(B^+\to\rho^0 K^+)\sim
(0.5-1.0)\times 10^{-6}$, while experimentally it is of order
$1.5\times 10^{-5}$. We argued that $B^+\to\omega K^+$ may receive
a sizable final-state rescattering contribution from the
intermediate states
$K^{*-}\pi^0,K^{*-}\etapp,K^{*0}\pi^-,K^-\rho^0,K^0\rho^-$ which
interfere constructively, whereas the analogous rescattering
effect on $B^+\to\rho^0 K^+$ is very suppressed. However, if the
measured branching ratio $\rho^0 K^+$ is similar to that of
$\omega K^+$, then $W$-annihilation and spacelike penguins could
play a prominent role. Likewise, the decay mode $B^0\to K^+K^-$ is
expected to be dominated by inelastic rescattering from
$\rho^+\rho^-,\pi^+\pi^-$ intermediate states, and
$B^0\to\phi\phi$ is governed by the $\etapp\etapp$ intermediate
channels.

\item A negative Wolfenstein parameter $\rho$ or a unitarity angle
$\gamma$ larger than $90^\circ$ is helpful for explaining the
$\pi^+\pi^-$, $K\pi$ and $\eta' K$ data. All the known model
calculations predict a too large $\pi^+\pi^-$ rate compared to the
recently improved limit. We have shown that either
$\gamma>105^\circ$ or an isospin phase shift difference
$\delta_{\pi\pi}>70^\circ$ can account for the data of
$B\to\pi^+\pi^-$. Moreover, the disparity between the $\eta' K^-$
and $\eta' K^0$ modes can be accommodated by $\rho<0$. The
expected hierarchy pattern $\ov K^0\pi^->K^-\pi^+>K^-\pi^0$
predicted at $\gamma=65^\circ$ will be modified to $K^-\pi^+> \ov
K^0\pi^->K^-\pi^0$ at $\gamma>90^\circ$ and $K^-\pi^0$ becomes
close to $K^-\pi^+$ when $\gamma$ approaches to $180^\circ$.

\item Theoretical calculations suggest that the following decay
modes of $B^-_u$ and $\ov B^0_d$ have branching ratios are of
order $10^{-5}$ or in the range of a few times of $10^{-5}$:
$\eta'K^-,~\eta'K^0,~\rho^+\rho^-,~\rho^-\pi^+$,
$\rho^-\rho^0,~\rho^- \omega,~\rho^-\pi^0,~K^-\pi^+,~\bar
K^0\pi^-$,
$K^-\pi^0,~\rho^+\pi^-,~\rho^0\pi^-,~\omega\pi^-,~\rho^-\eta$.
Some of them have been observed and the rest will have a good
chance to be seen soon.

\end{itemize}

\vskip 1cm {\it Note added}\,:~~Recently CLEO has reported two new
measurements on the $\rho^\pm\pi^\mp$ and $K^{*\pm}\pi^\mp$ modes
of the neutral $B$ mesons \cite{Gao}: ${\cal B}(\ov
B^0\to\rho^+\pi^-+\rho^-\pi^+)=(3.5^{+1.1}_{-1.0}\pm 0.5)\times
10^{-5}$ and ${\cal B}(\ov B^0\to
K^{*-}\pi^+)=(2.2^{+0.8+0.4}_{-0.6-0.5})\times 10^{-5}$. We see
from Table IX that while the prediction ${\cal B}(\ov
B^0\to\rho^+\pi^-+\rho^-\pi^+)=3.7\times 10^{-5}$ is in good
agreement with experiment, the observation that $K^{*-}\pi^+\gsim
K^-\pi^+$ is opposite to the theoretical expectation (see the
discussion in Sec. V.E.).

\acknowledgements We are grateful to J. Smith for critically
reading the manuscript and for useful comments. Two of us (B.T.
and K.C.Y.) wish to thank the KEK theory group for their warm
hospitality. This work is supported in part by the National
Science Council of the Republic of China under Grants Nos.
NSC88-2112-M001-006 and NSC88-2112-M006-013.

%\newpage

%%%%%%%%%%%%%%%%%%%%%%%%%%%%%%%%%%%%%%%%%%%%%%%%%%%%%%%%

\newpage
\centerline{\bf APPENDIX}
\renewcommand{\thesection}{\Alph{section}}
\renewcommand{\theequation}{\thesection\arabic{equation}}
\setcounter{equation}{0} \setcounter{section}{1} \vskip 0.5 cm
\centerline{\bf A.}
\vskip 0.3cm

The factorized decay amplitudes of all charmless $B_{u,d}\to
PP,\,VP\,, VV$ decays are tabulated in the Appendix. The
factorized terms $X^{(BM_1,M_2)}$  have the expressions: \be
\label{term} X^{(B P_1,P_2)} &\equiv& \la P_2| (\bar{q}_2
q_3)_\vma|0\ra\la P_1|(\bar {q}_1b)_\vma|\ov B
\ra=if_{P_2}(m_{B}^2-m^2_{P_1}) F_0^{ B P_1}(m_{P_2}^2),  \non \\
X^{(BP,V)} &\equiv & \la V| (\bar{q}_2 q_3)_\vma|0\ra\la
P|(\bar{q}_1b)_\vma|\ov B \ra=2f_V\,m_V F_1^{ B
P}(m_{V}^2)(\vp^*\cdot p_{_{B}}),   \non \\ X^{( BV,P)} &\equiv &
\la P | (\bar{q}_2 q_3)_\vma|0\ra\la V|(\bar{q}_1b)_\vma|\ov B
\ra=2f_P\,m_V A_0^{ B V}(m_{P}^2)(\vp^*\cdot p_{_{B}}),  \non \\
X^{( BV_1,V_2)} &\equiv & \la V_2 | (\bar{q}_2 q_3)_\vma|0\ra\la
V_1|(\bar{q}_1b)_\vma|\ov B \ra =- if_{V_2}m_2\Bigg[
(\vp^*_1\cdot\vp^*_2) (m_{B}+m_{1})A_1^{ BV_1}(m_{2}^2)  \non \\
&-& (\vp^*_1\cdot p_{_{B}})(\vp^*_2 \cdot p_{_{B}}){2A_2^{
BV_1}(m_{2}^2)\over (m_{B}+m_{1}) } +
i\epsilon_{\mu\nu\alpha\beta}\vp^{*\mu}_2\vp^{*\nu}_1p^\alpha_{_{B}}
p^\beta_1\,{2V^{ BV_1}(m_{2}^2)\over (m_{B}+m_{1}) }\Bigg],
\en
where $\vp^*$ is the polarization vector of the vector meson $V$.
For a flavor-neutral $M_2$ with the quark content $(\bar
qq+\cdots)$, we will encounter the factorized term
\be
X^{(B M_1,M_2)}_q \equiv \la M_2| (\bar{q} q)_\vma|0\ra\la
M_1|(\bar{q}_1b)_\vma|\ov B\ra.
\en
For example,
\be
X^{(B^- K^-,\eta')}_s &=& \la \eta'|(\bar ss)_\vma|0\ra \la
{K}^-|(\bar sb)_ \vma| B^-\ra = if_{\eta'}^s(m_{B}^2-m^2_K)F_0^{B
K}(m_{\eta'}^2), \non\\ X^{(B^-\pi^-,\rho^0)}_u &=& \la \rho^0|
(\bar uu )_\vma|0\ra\la \pi^-|(\bar{d}b)_\vma| B^-
\ra=\sqrt{2}f_\rho\,m_\rho F_1^{ B \pi}(m_{\rho}^2)(\vp^*\cdot
p_{_{B}}).
\en
For $\ov B^0_d\to VV$ decays (see Appendix F), we have
distinguished spacelike penguin matrix elements arising from
$(V-A)(V+A)$ and $(V-A)(V-A)$ operators, e.g., \be
X_u^{(B^0,\rho^0\omega)} &=& \la 0|(\bar db)_\vma|\ov
B^0\ra\la\rho^0\omega|(\bar uu)_\vma|0\ra, \non \\ \ov
X_u^{(B^0,\rho^0\omega)} &=& \la 0|(\bar db)_\vma|\ov
B^0\ra\la\rho^0\omega|(\bar uu)_\vpa|0\ra.
\en
As stressed in Sec. IV.B, we have included $W$-exchange,
$W$-annihilation and spacelike penguin matrix elements in the
expressions of factorized decay amplitudes, though they are
usually neglected in practical calculations of branching ratios.

Note that the hadronic matrix elements of scalar and pseudoscalar
densities are conventionally evaluated by applying equations of
motion. However, we encounter in $\ov B_d\to PP,VV$ decays terms
like $\la \pi\pi|\bar dd|0\ra$ which cannot be directly related to
the matrix element $\la\pi\pi|\bar d\gamma_\mu d|0\ra$ via the use
of equation of motion
\be
-i\partial^\mu(\bar q_1\gamma_\mu q_2)=(m_1-m_2)\bar q_1q_2.
\en
Hence, the matrix element such as $\la \pi\pi|\bar dd|0\ra$ has to
be evaluated using a different technique. Unfortunately, chiral
perturbation theory, which has been employed to compute the same
matrix element occurred in $K\to\pi\pi$ decay, is no longer
applicable in energetic $B$ decays. Since $\la V|\bar
q_1q_2|0\ra=0$, $B\to VV$ decays do not receive factorizable contributions
from $a_6$ and $a_8$ penguin terms except for spacelike penguin
diagrams (see Appendixes F and G).

\def\etapp{{\eta^{(')}}}
\def\half{{1\over 2}}
\def\p{{(')}}

\newpage
All the amplitudes listed below should be multiplied by a
factor of $G_F/\sqrt{2}$.

\section{ $\ov B^0_{\lowercase{d}}\to PP$ decays}

\begin{eqnarray}
%%%%%%%%%%%%%%%%%%%%%%%%%%%%%%%%%%%%%%
A(\ov B^0_d \to  K^- \pi^+)= && \Bigg\{ V_{ub} V^{*}_{us} a_{1} -
V_{tb} V^{*}_{ts} \Big[a_{4} +a_{10}+ 2(a_6+a_8) {m^2_{K^-} \over
(m_{b} - m_{u}) (m_{u} + m_{s})}\Big]\Bigg\} X^{(\ov B^0_d
\pi^{+},K^{-})}\nonumber\\ && - V_{tb} V^{*}_{ts}\left[a_4-{1\over
2}a_{10}+(2a_{6}-a_{8}) {m^2_{\ov B^0_d}\over (m_{b} + m_{d})
(m_{s} - m_{d})}\right]
 X^{(\ov B^0_d, \pi^{+} K^{-})},
\end{eqnarray}
%%%%%%%%%%%%%%%%%%%%%%%%%%%%%%%%%%%%%%%
\begin{eqnarray}
A(\ov B^0_d \to  \ov K^0 \pi^0)= &&  \left[ V_{ub} V^{*}_{us}
a_{2}- V_{tb} V^{*}_{ts} (-{3\over 2}a_{7}+{3\over 2}
a_{9})\right] X_u^{(\ov B^0_d \ov K^0,\pi^0)}\nonumber\\ &&
-V_{tb} V^{*}_{ts} \Bigg\{\Big[ a_{4}-{1\over 2} a_{10} +2(a_6-{1
\over 2}a_8) {m^2_{\ov K^0} \over (m_{s}+m_{d})(m_{b} - m_{d})}
\Big] X^{(\ov B^0_d \pi^0 ,\ov K^0)}\nonumber  \\ &&
+\Big[a_4-{1\over 2}a_{10}+ (2a_{6}-a_{8}) {m^2_{\ov B^0_d} \over
(m_{b}+m_{d})(m_{s} - m_{d})} \Big] X^{(\ov B^0_d ,\pi^0 \ov
K^0)}\Bigg\},
\end{eqnarray}
%%%%%%%%%%%%%%%%%%%%%%%%%%%%%%%%%%%%%%%
\begin{eqnarray}
A(\ov B^0_d \to  \ov K^0 \eta^\p)= &&  V_{ub} V^{*}_{us} a_{2}
X_u^{(\ov B^0_d \ov K^0 ,\eta^\p)} + V_{cb} V^{*}_{cs} a_{2}
X^{(\ov B^0_d \ov K^0 ,\eta^\p)}_{c} \nonumber\\ && - V_{tb}
V^{*}_{ts} \Bigg\{ \left[ 2a_{3}-2a_{5} -{1\over 2}a_{7}
 +{1\over 2}a_{9}\right] X^{(\ov B^0_d \ov K^0 ,\eta^\p)}_{u}
\nonumber\\ && + \left[ a_{3} + a_{4} -a_{5} +(2a_{6}-a_{8})
{m^{2}_{\eta^\p}\over 2m_s (m_{b} -m_{s})}(1-{f^{u}_{\eta^\p}\over
f^{s}_{ \eta^\p}})\right. \nonumber\\ && \left. +{1\over 2}(
a_{7}-a_{9}-a_{10})\right] X^{(\ov B^0_d \ov K^0 ,\eta^\p)}_{s}+
\left[ a_{3} -a_{5}-a_{7}+a_{9}\right] X^{(\ov B^0_d \ov K^0
,\eta^\p)}_{c} \nonumber\\ &&+ \left[ a_4-{1\over
2}a_{10}+(2a_{6}-a_{8}) {m^{2}_{\ov K^0} \over (m_s + m_{d})
(m_{b} -m_{d})}\right] X^{(\ov B^0_d \eta^\p,\ov K^0)}\nonumber\\
&&+\left[a_{4} + a_{10}+2(a_6+a_8) {m_{\ov B^0_d }^{2} \over
{(m_s-m_d)(m_b +m_d)}}\right] X^{(\ov B^0_d ,\ov K^0 \eta^\p)}
\Bigg\},
\end{eqnarray}
%%%%%%%%%%%%%%%%%%%%%%%%%%%%%%%%%%%%%%%
\begin{eqnarray}
A(\ov B^0_d \to K^0 \ov{K}^0)= &&-V_{tb} V^{*}_{td} \Bigg\{ \left[
a_3+a_4+a_5-{1\over 2} (a_7+a_9+a_{10}) \right] X_d^{(\ov B^0_d
,K^0 \ov K^0)}\nonumber  \\ && + \left[ a_{3}+a_{5}- {1\over
2}a_{7} -{1\over 2} a_{9} \right] X^{(\ov B^0_d ,K^0 \ov
K^0)}_{s}\nonumber \\ && +\Big[ a_{4} -{1\over 2}a_{10}
 + (2a_6-a_{8}) {m^2_{\ov K^0} \over (m_{s}+m_{d})
(m_{b} - m_{s})}\Big] X^{(\ov B^0_d K^0,\ov K^0)}\nonumber \\
&&-(2a_6-a_8)\langle K^0\ov K^0| \bar d(1+\gamma_5)d|0\rangle
\langle 0|\bar d(1-\gamma_5)b|\bar B^0_d\rangle \Bigg\},
\end{eqnarray}
%%%%%%%%%%%%%%%%%%%%%%%%%%%%%%%%%%%%%%%
\begin{eqnarray}
A(\ov B^0_d \to K^+ K^-)= &&V_{ub} V^{*}_{ud} a_2 X_u^{(\ov B^0_d
,K^+ K^-)} -V_{tb} V^{*}_{td} \Bigg\{ \left[ a_3+a_5+a_7+a_9
\right] X_u^{(\ov B^0_d ,K^+ K^-)}\nonumber \\ && + \left[
a_{3}+a_{5}- {1\over 2}a_{7} -{1\over 2} a_{9} \right] X^{(\ov
B^0_d ,K^+ K^-)}_s \Bigg\},
\end{eqnarray}
%%%%%%%%%%%%%%%%%%%%%%%%%%%%%%%%%%%%%%%
\begin{eqnarray}
A(\ov B^0_d \to  \pi^+ \pi^-)= && V_{ub} V^{*}_{ud}\Big( a_{1}
X^{(\ov B^0_d \pi^+ ,\pi^-)} +a_{2}  X^{(\ov B^0_d ,\pi^+
\pi^-)}\Big)\nonumber\\ &&  -V_{tb} V^{*}_{td} \Bigg\{\left[ a_4
+a_{10}+2(a_6 +a_8 ) {m^{2}_{\pi^-} \over (m_d + m_u )(m_b - m_u
)}  \right] X^{(\ov B^0_d \pi^+ ,\pi^-)}\nonumber\\ &&  +
\left[2a_3 + a_4 + 2a_5+ {1\over 2} (a_7 +a_9-a_{10})\right]
X^{(\ov B^0_d ,\pi^+ \pi^-)}\nonumber \\ &&
-(2a_6-a_8)\la\pi^+\pi^-|\bar d(1+\gamma_5)d|0\ra\la 0|\bar
d(1-\gamma_5)b|\ov B^0_d\ra \Bigg\},
\end{eqnarray}
%%%%%%%%%%%%%%%%%%%%%%%%%%%%%%%%%%%%%%%
\begin{eqnarray}
A(\ov B^0_d \to  \pi^0 \pi^0) &=&  V_{ub} V^{*}_{ud}~2( a_2
X_u^{(\ov B^0_d \pi^0 ,\pi^0)} +a_{2}X_u^{(\ov B^0_d ,\pi^0
\pi^0)})\nonumber \\ &-&  V_{tb} V^{*}_{td} ~2\Bigg\{ \left[ -a_4
+ {3\over 2}(-a_7+a_9) +{1\over 2} a_{10}- (2a_6 - a_8 )
{m^{2}_{\pi^0} \over 2m_d (m_b - m_d )} \right] X_u^{(\ov B^0_d
\pi^0 ,\pi^0)}\nonumber \\ &+&  \left[2a_3 + a_4 + 2a_5+ {1\over
2}(a_7+ a_9 -a_{10})\right] X_u^{(\ov B^0_d ,\pi^0
\pi^0)}\nonumber \\ &-& (2a_6-a_8)\langle \pi^0 \pi^0| \bar
d(1+\gamma_5)d|0\rangle\langle 0|\bar d(1-\gamma_5)b| \bar
B^0_d\rangle  \Bigg\},
\end{eqnarray}
%%%%%%%%%%%%%%%%%%%%%%%%%%%%%%%%%%%%%%%
\begin{eqnarray}
A(\ov B^0_d \to  \pi^0 \eta^\p)= && V_{ub} V^{*}_{ud} a_2
(X_u^{(\ov B^0_d \pi^0 ,\eta^\p)} + X_u^{ (\ov B^0_d
\eta^\p,\pi^0)}) + V_{cb} V^{*}_{cd} a_2 X^{(\ov B^0_d
\pi^0,\eta^{(')})}_{c}\nonumber \\ && -V_{tb} V^{*}_{td} \Bigg\{
\Big[ 2a_3+ a_4 - 2a_5 -{1\over 2} (a_7 -a_9+a_{10}) \non\\ &&
+(2a_6 - a_8){m^{2}_{\eta^{(')}} \over 2m_s (m_b -m_d )}
({{f^{s}_{\eta^{(')}}\over f^{u}_{\eta^{(')}}}-1}) r_{\eta'} \Big]
X_u^{(\ov B^0_d \pi^0, \eta^{(')} )}\nonumber\\ && + \left[ a_3 -
a_5-a_7 + a_9 \right] X^{(\ov B^0_d \pi^0 ,
 \eta^{(')} )}_{c}
 + \left[ a_3 - a_5+ {1\over 2}a_7-{1\over 2}a_9 \right]
X^{(\ov B^0_d \pi^0 ,\eta^{(')} )}_{s}\nonumber\\ &&  + \left[ -
a_4-{3\over 2}(a_7 - a_9)+{1\over 2}a_{10} -(2a_6 -
a_8){m^{2}_{\pi^0} \over 2m_d (m_b -m_d )}
 \right] X_u^{(\ov B^0_d \eta^{(')},\pi^0 )}\nonumber\\
&& + \left[ a_3 - a_5-a_7 + a_9 \right] X^{(\ov B^0_d ,\pi^0
\eta^{(')} )}_{c} + \left[ a_3 - a_5+ {1\over 2}a_7-{1\over 2}a_9
\right] X^{(\ov B^0_d  ,\pi^0 \eta^{(')} )}_{s}\nonumber\\ && +
\left[ 2a_3 + a_4 + 2a_5 + {1\over 2} (a_7 +a_9- a_{10}) \right]
X^{(\ov B^0_d  ,\pi^0 \eta^{(')})} \nonumber\\
&&-(2a_6-a_8)\langle \etapp\pi^0| \bar
d(1+\gamma_5)d|0\rangle\langle 0|\bar d(1-\gamma_5)b| \bar
B^0_d\rangle\Bigg\},
\end{eqnarray}
%%%%%%%%%%%%%%%%%%%%%%%%%%%%%%%%%%%%%%%
\begin{eqnarray}
A(\ov B^0_d \to  \eta \eta^{'})= && V_{ub} V^{*}_{ud} a_2
\Big(X_u^{(\ov B^0_d \eta,\eta^{'})} +X_u^{(\ov B^0_d
\eta^{'},\eta)}+2X^{(\ov B^0_d ,\eta\eta^{'})}\Big) \non \\ && +
V_{cb} V^{*}_{cd} a_2 (X^{(\ov B^0_d \eta,\eta^{'})}_{c} +
X_c^{(\ov B^0_d \eta^{'},\eta)}+ 2X^{(\ov B^0_d
,\eta\eta^{'})}_{c}) \non \\ && -V_{tb} V^{*}_{td} \Bigg\{ \Big[
2a_3 + a_4 - 2a_5 -{1 \over 2} a_7 + {1\over 2}  a_9- {1\over 2}
a_{10}\nonumber  \\ && +(2a_6- a_8){m^{2}_{\eta^{'}} \over 2m_s
(m_b -m_d )} ( {f^{s}_{\eta^{'}} \over
f^{u}_{\eta^{'}}}-1)r_{\eta'} \Big] X_u^{(\ov B^0_d \eta, \eta^{'}
)}\nonumber\\ && + \Big[ 2a_3 + a_4 - 2a_5 -{1 \over 2} a_7 +
{1\over 2}  a_9- {1\over 2}  a_{10} \non \\ && +(2a_6-
a_8){m^{2}_{\eta} \over 2m_s (m_b -m_d )} ( {f^{s}_{\eta} \over
f^{u}_{\eta}}-1)r_{\eta} \Big] X_u^{(\ov B^0_d \eta', \eta
)}\nonumber\\ && + \left[ a_3 - a_5+{1\over 2}(a_7-a_9) \right]
(X^{(\ov B^0_d \eta,\eta' )}_{s} + X^{(\ov B^0_d \eta',\eta)}_{s})
\nonumber \\&& + \left[ a_3 - a_5-a_7 +a_9 \right] (X^{(\ov B^0_d
\eta,\eta^{'} )}_{c} + X^{(\ov B^0_d \eta^{'},
\eta)}_{c})\nonumber\\ && + \left[ a_3 + a_5- {1\over 2} (a_7+
a_9) \right] X^{(\ov B^0_d ,\eta \eta^{'} )}_{s} + \left[ a_3 +
a_5+a_7+a_9 \right] X^{(\ov B^0_d ,\eta \eta^{'} )}_{c}\nonumber
\\ && + \left[ 2a_3 + a_4 + 2a_5 + {1\over 2}  (a_7 +a_9-a_{10})
\right] X^{(\ov B^0_d ,\eta\eta' )}\nonumber \\ &&
-(2a_6-a_8)\langle \eta'\eta| \bar d(1+\gamma_5)d|0\rangle\langle
0|\bar d(1-\gamma_5)b| \bar B^0_d\rangle\Bigg\},
\end{eqnarray}
%%%%%%%%%%%%%%%%%%%%%%%%%%%%%%%%%%%%%%%
\begin{eqnarray}
A(\ov B^0_d \to  \eta \eta)= && V_{ub} V^{*}_{ud} 2a_2 (X_u^{(\ov
B^0_d \eta,\eta)} +X^{(\ov B^0_d ,\eta\eta)})+ V_{cb} V^{*}_{cd}
2a_2 (X^{(\ov B^0_d \eta, \eta)}_{c}+ X^{(\ov B^0_d
,\eta\eta)}_{c})\nonumber \\ && -V_{tb} V^{*}_{td} 2\Bigg\{ \left[
2a_3 + a_4 - 2a_5 -{1 \over 2} a_7 + {1\over 2}  a_9- {1\over 2}
a_{10} \right. \nonumber\\ && \left. +(2a_6- a_8){m^{2}_{\eta}
\over 2m_s (m_b -m_d )} ( {f^{s}_{\eta} \over
f^{u}_{\eta}}-1)r_{\eta} \right] X_u^{(\ov B^0_d \eta, \eta )}
\non \\ && + \left[ a_3 - a_5+{1\over 2}(a_7-a_9) \right] X^{(\ov
B^0_d \eta,\eta)}_{s}
 + \left[ a_3 - a_5-a_7 +a_9 \right] X^{(\ov B^0_d \eta,
\eta)}_{c}\nonumber \\ && + \left[ a_3 + a_5- {1\over 2} (a_7+
a_9) \right] X^{(\ov B^0_d ,\eta \eta)}_{s} + \left[ a_3 +
a_5+a_7+a_9 \right] X^{(\ov B^0_d ,\eta \eta )}_{c}\nonumber \\ &&
+ \left[ 2a_3 + a_4 + 2a_5 + {1\over 2}  (a_7 +a_9-a_{10}) \right]
X^{(\ov B^0_d  ,\eta\eta )} \nonumber \\ && -(2a_6-a_8)\langle
\eta \eta| \bar d(1+\gamma_5)d|0\rangle\langle 0|\bar
d(1-\gamma_5)b| \bar B^0_d\rangle\Bigg\},
\end{eqnarray}
%%%%%%%%%%%%%%%%%%%%%%%%%%%%%%%%%%%%%%%
\be
A(\ov B_d^0\to \eta'\eta')~{\rm is~obtained~from}~A(\ov
B_d^0\to\eta\eta)~{\rm  with}~\eta\to \eta'.
\en
%%%%%%%%%%%%%%%%%%%%%%%%%%%%%%%%%%

\section{ $B_{\lowercase{u}}^-\to PP$ decays}

\begin{eqnarray}
%%%%%%%%%%%%%%%%%%%%%%%%%%
A(B^-_u \to  \ov K^0 \pi^{-})= &&  - V_{tb} V^{*}_{ts}\Bigg\{\Big[
a_4+a_{10}+2(a_6+a_8) {m^{2}_{B^-_u}\over (m_s - m_u) (m_b + m_u
)} \Big] X^{(B^-_u ,\pi^- \ov K^0)} \nonumber\\ &&  + \Big[
a_4-{1\over 2} a_{10}+(2 a_6-a_8) {m^{2}_{\ov K^0}\over (m_s +
m_d) (m_b - m_d )}  \Big] X^{(B^-_u \pi^-,\ov K^0)}\Bigg\} \non\\
&& +V_{ub}V_{us}^*a_1X^{(B^-_u,\pi^-\ov K^0)},
\end{eqnarray}
%%%%%%%%%%%%%%%%%%%%%%%%%%%%%%%%%%%%%%%
\begin{eqnarray}
A(B^-_u \to  K^- \pi^0)= && V_{ub} V^{*}_{us}( a_1 X^{(B^-_u \pi^0
,K^-)} + a_1 X^{(B^-_u ,\pi^0 K^-)} + a_2 X_u^{(B^-_u K^- ,\pi^0)}
)\nonumber \\ &&  - V_{tb} V^{*}_{ts} \Bigg\{ \left[ a_4 +a_{10} +
2(a_6+a_8) {m^{2}_{K^-}\over (m_s + m_u) (m_b - m_u )} \right]
X^{(B^-_u \pi^0 ,K^-)} \nonumber\\ &&  +  {3\over 2}
\left[-a_7+a_9 \right] X_u^{(B^-_u K^-,\pi^0)}\non\\ && +\left[ a_4
+a_{10}+ 2( a_6+a_8) {m^{2}_{B^-_u}\over (m_s - m_u) (m_b + m_u )}
\right] X^{(B^-_u ,K^- \pi^0)} \Bigg\},
\end{eqnarray}
%%%%%%%%%%%%%%%%%%%%%%%%%%%%%%%%%%%%%%%
\begin{eqnarray}
A(B^-_u \to  K^- \eta^\p)= &&  V_{ub} V^{*}_{us} \Big[a_1
X^{(B^-_u \eta^\p,K^-)} + a_{2} X_u^{(B^-_u K^-,\eta^\p)} +a_1
X_u^{(B^-_u ,K^- \eta^\p)}+a_1 X_s^{(B^-_u ,K^-
\eta^\p)}\Big]\nonumber\\ && + V_{cb} V^{*}_{cs} a_2  X^{(B^-_u
K^- ,\eta^\p)}_{c}
 - V_{tb} V^*_{ts} \Bigg\{
\left[ 2a_3 - 2a_5 - {1\over 2}  a_7 + {1\over 2}  a_9 \right]
X_u^{(B^-_u K^- ,\eta^\p)}\nonumber\\ &&+\left[a_3-a_5 - a_7 +a_9
\right] X_c^{(B^-_u K^- ,\eta^\p)} + \Big[ a_3 +a_4- a_5  +{1\over
2}(a_7- a_9-a_{10})\non\\ && + (2 a_6 - a_8){m^{2}_{\eta^\p}\over
2m_s(m_b - m_s )} (1 - {f^{u}_{\eta^\p} \over f^{s}_{\eta^\p} })
 \Big] X_s^{(B^-_u K^- ,\eta^\p)}\nonumber \\
&&  + \left[ a_4 + a_{10}+ 2( a_6 +  a_8) {m^{2}_{K^-}\over (m_s +
m_u) (m_b - m_u )}  \right] X^{(B^-_u \eta^\p,K^-)}\nonumber \\ &&
+ \left[ a_4 +a_{10}+ 2( a_6 +  a_8) {m^{2}_{B^-_u}\over (m_s -
m_u) (m_b + m_u )}  \right] X^{(B^-_u ,K^- \eta^\p)} \Bigg\},
\end{eqnarray}
%%%%%%%%%%%%%%%%%%%%%%%%%%%%%%%%%%%%%%%
\begin{eqnarray}
A(B^-_u \to  K^- K^0) &=&
  \Bigg\{ V_{ub} V^{*}_{ud} a_1 - V_{tb} V^{*}_{td}
\Big[a_4 +a_{10}+ 2( a_6 +a_8) {m^{2}_{B^-_u}\over (m_d - m_u)
(m_b + m_u )}  \Big] \Bigg\} X^{(B^-_u ,K^- K^0)}\nonumber \\ &-&
V_{tb} V^{*}_{td}\left[ a_4 -{1\over 2} a_{10}+ (2 a_6 -a_8)
{m^{2}_{K^0}\over (m_d+m_s) (m_b-m_s )}   \right] X^{(B^-_u
K^-,K^0)},
\end{eqnarray}
%%%%%%%%%%%%%%%%%%%%%%%%%%%%%%%%%%%%%%%
\begin{eqnarray}
A(B^-_u \to  \pi^- \pi^0) &=&
 V_{ub} V^{*}_{ud} \Big[ a_1 X^{(B^-_u \pi^0 ,\pi^-)}
+ a_{2} X_u^{(B^-_u \pi^- ,\pi^0)} \Big]\nonumber \\ &-&   V_{tb}
V^{*}_{td} \Bigg\{ {3 \over 2}\left[-a_7+a_9 +a_{10} +2 a_8
{m^{2}_{\pi^0}\over (m_d + m_d)(m_b - m_d )} \right]  X^{(B^-_u
\pi^- ,\pi^0)}  \Bigg\},
\end{eqnarray}
%%%%%%%%%%%%%%%%%%%%%%%%%%%%%%%%%%%%%%%
\begin{eqnarray}
A(B^-_u \to  \pi^- \eta^\p)= &&  V_{ub} V^{*}_{ud} \left[
a_1(X_u^{(B^-_u \eta^\p,\pi^-)} +2X^{(B^-_u ,\pi^-\eta^\p)}) + a_2
X_u^{(B^-_u \pi^- ,\eta^\p)} \right]\nonumber \\ && + V_{cb}
V^{*}_{cd} a_2 X^{(B^-_u \pi^- ,\eta^\p)}_{c} -V_{tb} V^{*}_{td}
\Bigg\{ \Big[ 2a_3 + a_4 - 2a_5 +{1\over 2}(-a_7 + a_9-a_{10})
\nonumber \\ &&  +(2a_6 - a_8){m^{2}_{\eta^\p} \over 2m_s (m_b
-m_d )} ({f^s_{\eta^\p} \over  f^u_{\eta^\p}}-1)r_{\eta^\p} \Big]
X_u^{(B^-_u \pi^- , \eta^\p )}\non \\ &&  +\left[ a_3 -
a_5+{1\over 2}(a_7 -a_9) \right] X^{(B^-_u \pi^- ,\eta^\p )}_{s}
+\left[ a_3 - a_5- a_7 + a_9 \right] X^{(B^-_u \pi^- , \eta^\p
)}_{c}\nonumber \\ && + \left[ a_4 + a_{10} +2(a_6 + a_8)
{m^{2}_{\pi^-} \over (m_u + m_d) (m_b -m_u )} \right] X_u^{(B^-_u
\eta^\p,\pi^- )}\nonumber\\ && + \left[ a_4 + a_{10}-2(a_6 + a_8)
{m^{2}_{B^-_u} \over (m_b + m_u) (m_d -m_u )}
 \right] X^{(B^-_u ,\eta^\p\pi^- )}\Bigg\}.
\end{eqnarray}

\section{$\ov B_{\lowercase{d}}^0\to VP$ decays}

%%%%%%%%%%%%%%%%%%%%%%%%%%%%%%%%%%%%%%%%%%%%%%%%%%%%%%%%%%%%%
\begin{eqnarray}
A(\ov B^0_d \to \ov K^0 \rho^0)= &&  V_{ub} V^{*}_{us} a_2
X_u^{(\ov B^0_d \ov K^0 ,\rho^0)} - V_{tb} V^{*}_{ts} \Bigg\{
{3\over 2} (a_7+ a_9) X_u^{(\ov B^0_d \ov K^0,\rho^0)}\nonumber \\
&&  +\left[ a_4 - {1\over 2} a_{10}- (2a_6 - a_8) {m^{2}_{\ov K^0}
\over (m_s + m_d)(m_b + m_d)} \right] X^{(\ov B^0_d \rho^0 ,\ov
K^0)}\nonumber\\ && +\left[ a_4- {1\over 2} a_{10} - (2a_6 - a_8
){m^{2}_{\ov B^0_d } \over (m_s + m_d)(m_b + m_d)}  \right]
X^{(\ov B^0_d ,\ov K^0\rho^0)} \Bigg\},
\end{eqnarray}
%%%%%%%%%%%%%%%%%%%%%%%%%%%%%%%%%%%%%%%
\begin{eqnarray}
A(\ov B^0_d \to  K^- \rho^+)= &&  V_{ub} V^{*}_{us} a_1 X^{(\ov
B^0_d \rho^+ ,K^-)} \non \\ && -V_{tb} V^{*}_{ts}\Bigg\{ \Big[
a_4 - {1\over 2} a_{10} - (2a_6 - a_8 ) {m^{2}_{\ov B^0_d } \over
(m_s + m_d)(m_b + m_d)}  \Big] X^{(\ov B^0_d ,K^-
\rho^+)}\nonumber\\ &&  +\Big[ a_4 + a_{10}- 2(a_6 + a_8)
{m^{2}_{K^-} \over (m_s + m_u)(m_b + m_u)} \Big] X^{(\ov B^0_d
\rho^+ ,K^-)}\Bigg\},
\end{eqnarray}
%%%%%%%%%%%%%%%%%%%%%%%%%%%%%%%%%%%%%%%
\begin{eqnarray}
A(\ov B^0_d \to  K^{\ast -} \pi^+)= &&  V_{ub} V^{*}_{us} a_1
X^{(\ov B^0_d \pi^+ ,K^{\ast -})} -V_{tb} V^{*}_{ts}\Bigg\{ \left[
a_4 + a_{10}\right] X^{(\ov B^0_d \pi^+ ,K^{\ast -})} \nonumber\\
&&  +\left[ a_4 - {1\over 2} a_{10}- (2a_6 - a_8 ) {m^{2}_{\ov
B^0_d } \over (m_s + m_d)(m_b + m_d)}  \right] X^{(\ov B^0_d
,K^{\ast -} \pi^+)} \Bigg\},
\end{eqnarray}
%%%%%%%%%%%%%%%%%%%%%%%%%%%%%%%%%%%%%%%
\begin{eqnarray}
A(\ov B^0_d \to  \ov K^{\ast 0} \pi^0)= &&  V_{ub} V^{*}_{us} a_2
X^{(\ov B^0_d K^{\ast 0},\pi^0)} - V_{tb} V^{*}_{ts} \Bigg\{
\left[ - {3\over 2} a_7 + {3\over 2} a_9 \right] X^{(\ov B^0_d
K^{\ast 0},\pi^0)}\nonumber\\ &&  +\left[ a_4 - {1\over 2} a_{10}-
(2a_6 - a_8 ) {m^{2}_{\ov B^0_d } \over (m_s + m_d)(m_b + m_d)}
\right] X^{(\ov B^0_d ,K^{\ast 0}\pi^0)} \non \\ && +( a_4
-{1\over 2} a_{10})X^{(\ov B^0_d \pi^0 ,K^{\ast 0})}\Bigg\},
\end{eqnarray}
%%%%%%%%%%%%%%%%%%%%%%%%%%%%%%%%%%%%%%%
\begin{eqnarray}
A(\ov B^0_d \to  \ov K^0 \phi)= &&  - V_{tb} V^{*}_{ts} \Bigg\{
\left[  a_3 + a_4 + a_5 - {1\over 2} (a_7 + a_9 + a_{10} )\right]
X^{(\ov B^0_d \ov K^0,\phi)}\nonumber\\ &&  +\left[ a_4 - {1\over
2} a_{10}- (2a_6 - a_8 ) {m^{2}_{\ov B^0_d } \over (m_s + m_d)(m_b
+ m_d)}  \right] X^{(\ov B^0_d ,\ov K^0 \phi)} \Bigg\},
\end{eqnarray}
%%%%%%%%%%%%%%%%%%%%%%%%%%%%%%%%%%%%%%%
\begin{eqnarray}
A(\ov B^0_d \to  \ov K^{\ast 0} \etapp)= &&  V_{ub} V^{*}_{us} a_2
X^{(\ov B^0_d \ov K^{\ast 0},\etapp)} + V_{cb} V^{*}_{cs} a_2
X^{(\ov B^0_d \ov K^{\ast 0},\etapp)}_{c}\nonumber\\ &&  - V_{tb}
V^{*}_{ts} \Bigg\{ \left[ 2 (a_3 -  a_5) - {1\over 2} (a_7- a_9)
 \right] X^{(\ov B^0_d \ov K^{\ast 0},\etapp)}_{u}\nonumber\\
&&  +\left[ a_3 + a_4 -a_5 - {1\over 2}(-a_7 +  a_9 + a_{10})
\right.\nonumber\\ && - (2a_6 - a_8) {m^{2}_{\etapp} \over 2m_s
(m_b + m_s)}(1 - {f^{u}_{\etapp} \over f^{s}_{\etapp} }) \Big]
X^{(\ov B^0_d \ov K^{\ast 0},\etapp )}_{s}\nonumber \\ &&+\left[
a_3  - a_5 - a_7 + a_9 \right] X^{(\ov B^0_d \ov K^{\ast
0},\etapp)}_{c} +\left[ a_4 - {1\over 2} a_{10}\right] X^{(\ov
B^0_d \etapp ,\ov K^{\ast 0})}\nonumber\\ &&  +\left[ a_4 -
a_{10}- (2a_6 - a_8 ) {m^{2}_{\ov B^0_d } \over (m_s + m_d)(m_b +
m_d)}  \right] X^{(\ov B^0_d ,\ov K^{\ast 0}\etapp)} \Bigg\},
\end{eqnarray}
%%%%%%%%%%%%%%%%%%
\begin{eqnarray}
A( \ov B^0_d \to  \ov K^0 \omega)= &&  V_{ub} V^{*}_{us} a_2
X^{(\ov B^0_d \ov K^0 ,\omega)}_{u} - V_{tb} V^{*}_{ts} \Bigg\{
\left[ 2 a_3 + 2 a_5 + {1\over 2} (a_7+  a_9) \right] X^{(\ov
B^0_d \ov K^0 ,\omega)}_{u}\nonumber\\ &&  +\left[ a_4 - {1\over
2} a_{10}- (2a_6 - a_8) {m^{2}_{K} \over (m_s + m_d)(m_b + m_d)}
\right] X^{(\ov B^0_d \omega ,\ov K^0)} \nonumber\\ &&  +\left[
a_4 - {1\over 2} a_{10}- (2a_6 - a_8 ) {m^{2}_{\ov B^0_d } \over
(m_s + m_d)(m_b + m_d)}  \right] X^{(\ov B^0_d ,\ov K^0 \omega)}
\Bigg\},
\end{eqnarray}
%%%%%%%%%%%%%%%%%%%%%%%%%%%%%%%%%%%%%%%
\begin{eqnarray}
A(\ov B^0_d \to  \rho^- \pi^+)= &&  V_{ub} V^{*}_{ud} \Bigg\{ a_1
X^{(\ov B^0_d \pi^+,\rho^-)} + a_2 X^{(\ov B^0_d ,\rho^-
\pi^+)}\Bigg\}-V_{tb} V^{*}_{td}\Bigg\{ ( a_4 +
a_{10})X^{ (\ov B^0_d \pi^+ ,\rho^-)} \nonumber \\ &&
+\Big[2a_3 + a_4 - 2a_5 +{1\over 2}(- a_7 +a_9 - a_{10}) \non\\ &&
- (2a_6 - a_8 ){m^{2}_{\ov B^0_d } \over 2m_d(m_b + m_d)}
 \Big]X^{(\ov B^0_d ,\rho^- \pi^+)}\Bigg\},
\end{eqnarray}
%%%%%%%%%%%%%%%%%%%%%%%%%%%%%%%%%%%%%%%
\begin{eqnarray}
A(\ov B^0_d \to  \rho^+ \pi^-)= &&  V_{ub} V^{*}_{ud} \Bigg\{ a_1
X^{(\ov B^0_d \rho^+,\pi^-)} + a_2 X^{(\ov B^0_d ,\rho^+
\pi^-)}\Bigg\} -V_{tb} V^{*}_{td}\Bigg\{ \Big[ 2a_3 + a_4 - 2a_5
\nonumber \\ && -{1\over 2} (a_7- a_9 + a_{10}) - (2a_6 - a_8
){m^{2}_{\ov B^0_d } \over 2m_d(m_b + m_d)} \Big] X^{(\ov B^0_d
,\rho^+ \pi^-)}\nonumber\\ &&+\left[ a_4 + a_{10}- 2(a_6 + a_8)
{m^{2}_{\pi^-}\over (m_u + m_d)(m_b + m_u)} \right] X^{(\ov B^0_d
\rho^+ ,\pi^-)}\Bigg\},
\end{eqnarray}
%%%%%%%%%%%%%%%%%%%%%%%%%%%%%%%%%%%%%%%
\begin{eqnarray}
A(\ov B^0_d \to K^{*0} \ov K^0)= &&  - V_{tb} V^{*}_{td} \Bigg\{
\left[ a_4  - {1\over 2} a_{10} \right] X^{(\ov B^0_d \ov K^0,
\bar{K}^{\ast 0})} +\Big[ a_3 + a_4 + a_5- {1\over 2} (a_7 +  a_9
+ a_{10}) \non \\ && -(2a_6 -a_8) {m^{2}_{\ov B^0_d } \over {2m_d
(m_b + m_d)}} \Big] X^{(\ov B^0_d  ,\ov K^0 \bar{K}^{\ast 0})}
\Bigg\},
\end{eqnarray}
%%%%%%%%%%%%%%%%%%%%%%%%%%%%%%%%%%%%%%%
\begin{eqnarray}
A( \ov B^0_d \to \ov K^{*0} K^{0})= &&  - V_{tb} V^{*}_{td}
\Bigg\{ \Big[ a_4  - {1\over 2} a_{10}-(2a_6 - a_8){m^{2}_{\ov
K^0} \over (m_s + m_d ) (m_b + m_s)}  \Big] X^{(\ov B^0_d \ov
K^{\ast 0},K^0)}\nonumber \\ &&+\Big[ a_3 + a_4 + a_5- {1\over 2}
(a_7+  a_9 +  a_{10})\non\\ && -(2a_6 -a_8) {m^{2}_{\ov B^0_d
}\over {2m_d (m_b + m_d)}}\Big] X^{(\ov B^0_d  ,\ov K^{\ast 0}
K^0)}\Bigg\},
\end{eqnarray}
%%%%%%%%%%%%%%%%%%%%%%%%%%%%%%%%%%%%%%%
\begin{eqnarray}
A(\ov B^0_d \to  K^{*-} K^{+})= &&  - V_{tb} V^{*}_{td} \Bigg\{
\Big[ a_3 - a_5 - a_7 + a_9 + 2(a_6 + a_8){m^{2}_{\ov B^0_d }
\over 2m_u (m_b + m_d)}  \Big] X^{(\ov B^0_d  ,K^+ K^{*-})}
\non\\ && +\Big[a_3-a_5  - {1\over 2} (a_7+  a_9 ) + (2a_6 -
a_8){m^{2}_{\ov B^0_d } \over 2m_s (m_b + m_d)} \Big] X^{(\ov
B^0_d ,K^+ K^{*-})}_{s} \Bigg\} \non\\ && +V_{ub} V^{*}_{ud} a_2
X^{(\ov B^0_d  ,K^+ K^{*-})},
\end{eqnarray}
%%%%%%%%%%%%%%%%%%%%%%%%%%%%%%%%%%%%%%%
\begin{eqnarray}
A(\ov B^0_d \to  K^{-} K^{*+})= &&  - V_{tb} V^{*}_{td} \Bigg\{
\Big[ a_3 - a_5 - a_7 + a_9 + 2(a_6 + a_8){m^{2}_{\ov B^0_d }
\over 2m_u (m_b + m_d)}  \Big] X^{(\ov B^0_d  ,K^- K^{*+})}
\non\\ && +\Big[a_3-a_5 - {1\over 2} (a_7+  a_9 ) + (2a_6 -
a_8){m^{2}_{\ov B^0_d } \over 2m_s (m_b + m_d)} \Big] X^{(\ov
B^0_d ,K^-K^{*+})}_{s} \Bigg\} \non \\ &&+V_{ub} V^{*}_{ud} a_2
X^{(\ov B^0_d ,K^- K^{*+})},
\end{eqnarray}
%%%%%%%%%%%%%%%%%%%%%%%%%%%%%%%%%%%%%%%
\begin{eqnarray}
A(\ov B^0_d \to  \phi  \eta^{(')})= &&  -V_{tb} V^{*}_{td}\Bigg\{
\left[ a_3 + a_5 -{1\over 2} (a_7 + a_9) \right] X^{(\ov B^0_d
\eta^{(')},\phi)} \non\\ && +\left[ a_3 - a_5 +{1\over 2} (a_7 -
a_9) \right] X^{(\ov B^0_d  ,\phi\eta^{(')})}_{s} \Bigg\},
\end{eqnarray}
%%%%%%%%%%%%%%%%%%%%%%%%%%%%%%%%%%%%%%%
\begin{eqnarray}
A(\ov B^0_d \to  \phi  \pi^0)= &&  V_{tb} V^{*}_{td}\Bigg\{
 a_3 + a_5-{1\over 2} (a_7 + a_9) \bigg\}
X^{(\ov B^0_d \pi^0,\phi)},
\end{eqnarray}
%%%%%%%%%%%%%%%%%%%%%%%%%%%%%%%%%%%%%%%
\begin{eqnarray}
A(\ov B^0_d \to  \rho^0 \pi^0)= &&  V_{ub} V^{*}_{ud} a_2 \Bigg\{
X^{(\ov B^0_d ,\rho^0 \pi^0)}
 +  X^{(\ov B^0_d \rho^0,\pi^0)}_{u}
+ X^{(\ov B^0_d \pi^0,\rho^0)}_{u} \Bigg\}\nonumber \\ &&  -V_{tb}
V^{*}_{td}\Bigg\{ \Big[ 2a_3 + a_4 - 2a_5 -{1\over 2} (a_7 - a_9+
a_{10}) \non \\ && - (2a_6 - a_8 ){m^{2}_{\ov B^0_d } \over
2m_d(m_b + m_d)} \Big]X^{(\ov B^0_d ,\rho^0 \pi^0)}\nonumber\\ &&
+\left[- a_4 + (2a_6 - a_8 ) {m^{2}_{\pi^0} \over 2m_d(m_b + m_d)}
-{3\over 2} a_7 +{3\over 2} a_9 + {1\over 2} a_{10} \right]
X_u^{(\ov B^0_d \rho^0,\pi^0)}\nonumber\\ &&  +\left[- a_4
+{3\over 2} a_7 +{3\over 2} a_9 + {1\over 2} a_{10} \right]
X_u^{(\ov B^0_d \pi^0,\rho^0)}\Bigg\},
\end{eqnarray}
%%%%%%%%%%%%%%%%%%%%%%%%%%%%%%%%%%%%%%%
\begin{eqnarray}
A(\ov B^0_d \to  \rho^0 \eta^\p)=
&&  V_{ub} V^{*}_{ud} a_2
\Bigg\{ X^{(\ov B^0_d ,\rho^0\eta^\p)} +  X^{(\ov B^0_d \rho^0
,\eta^\p)}_{u} + X^{(\ov B^0_d \eta^\p,\rho^0)}_{u} \Bigg\} +
V_{cb} V^{*}_{cd} a_2 X^{(\ov B^0_d
\rho^0,\eta^\p)}_{c}\nonumber\\ &&  -V_{tb} V^{*}_{td}\Bigg\{
\Big[ 2a_3 + a_4 - 2a_5 -{1\over 2} (a_7 - a_9 + a_{10}) \non \\
&&- (2a_6 -a_8 ) {m^{2}_{\ov B^0_d } \over 2m_d(m_b + m_d)}
\Big]X^{(\ov B^0_d ,\rho^0 \eta^\p)}\nonumber\\ &&  +\Big[ 2a_3 +
a_4 - 2a_5 - (2a_6 - a_8 ){m^{2}_{\eta^\p} \over 2m_s(m_b + m_d)}
({{f^{s}_{\eta^\p}}\over f^{u}_{\eta^\p}} - 1) r_{\eta^\p}
\\ &&  -{1\over 2} (a_7 - a_9 + a_{10}) \Big] X^{(\ov
B^0_d \rho^0,\eta^\p)}_{u} +\left[ -a_4+{1\over 2} a_{10} +{3\over
2}( a_7 + a_9 ) \right] X^{(\ov B^0_d
\eta^\p,\rho^0)}_{u}\nonumber\\ &&  +\left[ a_3 - a_5 +{1\over 2}
a_7 -{1\over 2} a_9 \right] X^{(\ov B^0_d
\rho^0,\eta^\p)}_{s}+\left[ a_3 - a_5 - a_7 + a_9 \right] X^{(\ov
B^0_d \rho^0,\eta^\p)}_{c}\Bigg\}, \non
\end{eqnarray}
%%%%%%%%%%%%%%%%%%%%%%%%%%%%%%%%%%%%%%%
\begin{eqnarray}
A(\ov B^0_d \to  \omega \pi^0)= &&  V_{ub} V^{*}_{ud} a_2 \Bigg\{
X^{(\ov B^0_d ,\omega\pi^0)} +  X^{(\ov B^0_d \omega,\pi^0)}_{u} +
X^{(\ov B^0_d \pi^0,\omega)}_{u} \Bigg\}\nonumber\\ &&  -V_{tb}
V^{*}_{td}\Bigg\{ \Big[ 2a_3 + a_4 - 2a_5 -{1\over 2} (a_7 - a_9 +
a_{10}) \non\\ && - (2a_6 - a_8 ){m^{2}_{\ov B^0_d } \over
2m_d(m_b + m_d)} \Big] X^{(\ov B^0_d ,\omega\pi^0)}\nonumber\\ &&
+\left[ a_4-{3\over 2} a_7 +{3\over 2} a_9 - {1\over 2} a_{10} -
(2a_6 - a_8 ){m^{2}_{\pi^0} \over 2m_d(m_b + m_d)} \right]X^{(\ov
B^0_d \omega,\pi^0)}_{d}\nonumber\\ &&  +\left[ 2a_3 + a_4 + 2a_5
+{1\over 2} (a_7 + a_9 -  a_{10}) \right] X^{(\ov B^0_d
\pi^0,\omega)}_{u} \Bigg\},
\end{eqnarray}
%%%%%%%%%%%%%%%%%%%%%%%%%%%%%%%%%%%%%%%
\begin{eqnarray}
A(\ov B^0_d \to  \omega \eta^{(')}) &=&
V_{ub} V^{*}_{ud} a_2
\Bigg\{ X^{(\ov B^0_d ,\omega\eta^{(')})} +  X^{(\ov B^0_d
\omega,\eta^{(')})}_{u} + X^{(\ov B^0_d \eta^{(')},\omega)}_{u}
\Bigg\} + V_{cb} V^{*}_{cd} a_2 X^{(\ov B^0_d
\omega,\eta^{(')})}_{c}\nonumber\\
&-&  V_{tb} V^{*}_{td}\Bigg\{2\Big[ 2a_3 + a_4 - 2a_5 -
(2a_6 - a_8 ) {m^{2}_{\ov B^0_d } \over 2m_d(m_b + m_d)}  \non \\
&+& {1\over 2}(- a_7 + a_9 -  a_{10}
)\Big] X^{(\ov B^0_d ,\omega\eta^{(')})}_{u} +\Big[
2a_3 + a_4 - 2a_5 +{1\over 2}(- a_7 + a_9 - a_{10})\nonumber \\
&-& (2a_6 - a_8 ) {m^{2}_{\eta^{(')}} \over 2m_d(m_b + m_d)}
({f^{s}_{\eta^{(')}}\over f^{u}_{\eta^\p}}-1 ) r_{\eta^{(')}}
\Big] X^{(\ov B^0_d \omega,\eta^{(')})}_{u}\nonumber \\
&+& \left[ a_3 - a_5 +{1\over 2}( a_7 - a_9) \right] X^{(\ov B^0_d
\omega,\eta^{(')})}_{s} +\left[ a_3 - a_5- a_7 + a_9 \right]
X^{(\ov B^0_d \omega,\eta^{(')})}_{c}\nonumber\\
&+&  \left[ 2a_3
+ a_4 + 2a_5 +{1\over 2} (a_7+ a_9 - a_{10}) \right] X^{(\ov B^0_d
\eta^{(')},\omega)}_{u} \Bigg\}.
\end{eqnarray}
%%%%%%%%%%%%%%%%%%%%%%%%%%%%%%%%%%%%%%%

\section{$B_{\lowercase{u}}^-\to VP$ decays}

%%%%%%%%%%%%%%%%%%%%%%%%%%%%%%%%%%%%%%%
\begin{eqnarray}
A(B^-_u \to  K^- \rho^0)= &&  V_{ub} V^{*}_{us} \Bigg\{ a_1
X^{(B^-_u \rho^0,K^-)} +a_1 X^{(B^-_u ,\rho^0 K^-)} + a_2
X^{(B^-_u K^-,\rho^0)}_{u} \Bigg\}\nonumber\\ &&  -V_{tb}
V^{*}_{ts}\Bigg\{{3\over 2} \left[  a_7+ a_9 \right] X^{(B^-_u
K^-,\rho^0)}_{u}\nonumber\\ &&  + \left[ a_4+ a_{10} - 2(a_6 +
a_8) {m^{2}_{K^-} \over (m_s + m_u)(m_b + m_u)} \right] X^{(B^-_u
\rho^0 ,K^-)}\nonumber\\ &&  +\left[ a_4 + a_{10} - 2(a_6 + a_8 )
{m^{2}_{B^-_u } \over (m_s + m_u)(m_b + m_u)} \right]X^{(B^-_u
,\rho^0 K^-)} \Bigg\},
\end{eqnarray}
%%%%%%%%%%%%%%%%%%%%%%%%%%%%%%%%%%%%%%%
\begin{eqnarray}
A(B^-_u \to  K^{\ast -} \pi^0)= &&  V_{ub} V^{*}_{us} \Bigg\{ a_1
X^{(B^-_u \pi^0,K^{\ast -})} +a_1 X^{(B^-_u ,\pi^0 K^{\ast -})} +
a_2 X^{(B^-_u K^{\ast -},\pi^0)}\Bigg\}\nonumber\\ &&  -V_{tb}
V^{*}_{ts}\Bigg\{ \left[ -{3\over 2} a_7+{3\over 2} a_9 \right]
X^{(B^-_u K^{\ast -},\pi^0)}_{u} + ( a_4 + a_{10})
X^{(B^-_u \pi^0, K^{\ast -})}\nonumber\\ && + \left[ a_4 + a_{10}-
2(a_6 + a_8) {m^{2}_{B^-_u } \over (m_s + m_u)(m_b + m_u)} \right]
X^{(B^-_u ,\pi^0 K^{\ast -})}\Bigg\},
\end{eqnarray}
%%%%%%%%%%%%%%%%%%%%%%%%%%%%%%%%%%%%%%%
\begin{eqnarray}
A(B^-_u \to  \ov K^0 \rho^-)= &&  V_{ub} V^{*}_{us} a_1 X^{(B^-_u
,K^0 \rho^-)}\nonumber  \\ && -V_{tb} V^{*}_{ts}\Bigg\{ \left[ a_4
-{1\over 2} a_{10}- (2a_6 - a_8 ) {m^{2}_{K^0} \over (m_s +
m_d)(m_b + m_d)}
 \right] X^{(B^-_u \rho^-, K^0)}\nonumber\\
&&  +\left[ a_4 + a_{10}- 2(a_6 + a_8) {m^{2}_{B^-_u } \over (m_s
+ m_u)(m_b + m_u)} \right] X^{(B^-_u ,\rho^- K^0)}\Bigg\},
\end{eqnarray}
%%%%%%%%%%%%%%%%%%%%%%%%%%%%%%%%%%%%%%%
\begin{eqnarray}
A(B^-_u \to  \ov K^{\ast 0} \pi^-)= &&  V_{ub} V^{*}_{us} a_1
X^{(B^-_u ,K^{\ast 0}\pi^-)} -V_{tb} V^{*}_{ts}\Bigg\{\left[ a_4
-{1\over 2} a_{10} \right] X^{(B^-_u \pi^-, K^{\ast
0})}\nonumber\\ &&  +\left[ a_4 + a_{10}- 2(a_6 + a_8)
{m^{2}_{B^-_u } \over (m_s + m_u)(m_b + m_u)} \right] X^{(B^-_u
,\pi^- K^{\ast 0})}\Bigg\},
\end{eqnarray}
%%%%%%%%%%%%%%%%%%%%%%%%%%%%%%%%%%%%%%%
\begin{eqnarray}
A(B^-_u \to  K^- \omega)= &&  V_{ub} V^{*}_{us} \Bigg\{ a_1
X^{(B^-_u \omega,K^-)} +a_1 X^{(B^-_u ,\omega K^-)} + a_2
X_u^{(B^-_u K^-,\omega)}\Bigg\}\nonumber\\ &&  -V_{tb}
V^{*}_{ts}\Bigg\{ \left[ 2a_3 + 2a_5 +{1\over 2} a_7 +{1\over 2}
a_9 \right] X_u^{(B^-_u K^-,\omega)}\nonumber\\ &&  +\left[ a_4 +
a_{10}- 2(a_6 + a_8 ) {m^{2}_{K^-} \over (m_s + m_u)(m_b + m_u)}
 \right] X^{(B^-_u \omega, K^-)}\nonumber\\
&&  +\left[ a_4 + a_{10}- 2(a_6 + a_8) {m^{2}_{B^-_u } \over (m_s
+ m_u)(m_b + m_u)} \right] X^{(B^-_u ,\omega K^-)} \Bigg\},
\end{eqnarray}
%%%%%%%%%%%%%%%%%%%%%%%%%%%%%%%%%%%%%%%
\begin{eqnarray}
A(B^-_u \to  K^- \phi)= &&  V_{ub} V^{*}_{us} a_1 X^{(B^-_u ,\phi
K^-)}\nonumber\\ &&-V_{tb} V^{*}_{ts}\Bigg\{ \left[ a_4 + a_{10}-
2(a_6 + a_8) {m^{2}_{B^-_u } \over (m_s + m_u)(m_b + m_u)} \right]
X^{(B^-_u ,\phi K^-)}\nonumber\\ &&  +\left[ a_3 + a_4 + a_5
-{1\over 2}( a_7+ a_9 + a_{10}) \right] X^{(B^-_u
K^-,\phi)}\Bigg\},
\end{eqnarray}
%%%%%%%%%%%%%%%%%%%%%%%%%%%%%%%%%%%%%%%
\begin{eqnarray}
A(B^-_u \to  K^{*-} \eta^\p) &=&
 V_{ub} V^{*}_{us} \Big[a_1 X^{(B^-_u \eta^\p,K^{\ast -})}+
a_{2} X_u^{(B^-_u K^{\ast -},\eta^\p)}  \non\\ &+& a_1 X_u^{(B^-_u
,K^{\ast -} \eta^\p)} +a_1 X_s^{(B^-_u ,K^{\ast -} \eta^\p)}\Big]
+ V_{cb} V^{*}_{cs} a_2  X^{(B^-_u K^{\ast -},\eta^\p)}_{c}\non\\
&-&  V_{tb} V^*_{ts} \Bigg\{\left[ 2a_3 - 2a_5 - {1\over 2}  a_7 +
{1\over 2}  a_9 \right] X_u^{(B^-_u K^{\ast -},\eta^\p)}\nonumber
\\ &+& (a_3-a_5 - a_7 +a_9 ) X_c^{(B^-_u K^{\ast -},
\eta^\p)} + \Big[ a_3 +a_4- a_5  +{1\over 2}(a_7-
a_9-a_{10})\nonumber \\ &-&  (2 a_6 - a_8){m^{2}_{\eta^\p}\over
2m_s(m_b + m_s )} (1 - {f^{u}_{\eta^\p} \over f^{s}_{\eta^\p} })
 \Big] X_s^{(B^-_u K^{\ast -},\eta^\p)}
+ ( a_4 + a_{10}) X^{(B^-_u \eta^\p,K^{\ast -})}\nonumber \\ &+&
\left[ a_4 +a_{10}- 2( a_6 +  a_8) {m^{2}_{B^-_u } \over (m_s +
m_u) (m_b + m_u )}  \right] X^{(B^-_u ,K^{\ast -}\eta^\p)}
\Bigg\},
\end{eqnarray}
%%%%%%%%%%%%%%%%%%%%%%%%%%%%%%%%%%%%%%%
\begin{eqnarray}
A(B^-_u \to  \rho^- \pi^0)= &&  V_{ub} V^{*}_{ud} \Bigg\{ a_1
X^{(B^-_u \pi^0,\rho^-)} +a_1 X^{(B^-_u ,\pi^0 \rho^-)} + a_2
X^{(B^-_u \rho^-,\pi^0)}_{u} \Bigg\}\nonumber\\ &&  -V_{tb}
V^{*}_{td}\Bigg\{ \left[ a_4 +{3\over 2} a_7 -{3\over 2} a_9 -
{1\over 2} a_{10} - (2 a_6 - a_8 ) {m^{2}_{\pi^0} \over 2m_d (m_b
+ m_d )} \right] X^{(B^-_u \rho^-,\pi^0)}_{d} \nonumber\\ &&
+\left[ a_4 + a_{10} \right] X^{(B^-_u \pi^0, \rho^-)}\nonumber\\
&&+ \left[ a_4 + a_{10}- 2(a_6 + a_8) {m^{2}_{B^-_u } \over (m_u +
m_d)(m_b + m_u)} \right] X^{(B^-_u ,\pi^0 \rho^-)}\Bigg\},
\end{eqnarray}
%%%%%%%%%%%%%%%%%%%%%%%%%%%%%%%%%%%%%%%
\begin{eqnarray}
A(B^-_u \to  \rho^0 \pi^-)= &&  V_{ub} V^{*}_{ud} \Bigg\{ a_1
X^{(B^-_u \rho^0,\pi^-)} +a_1 X^{(B^-_u ,\pi^- \rho^0)} + a_2
X^{(B^-_u \pi^-,\rho^0)}_{u} \Bigg\}\nonumber\\ &&  -V_{tb}
V^{*}_{td}\Bigg\{ \left[- a_4 +{3\over 2} a_7 +{3\over 2} a_9 +
{1\over 2} a_{10} \right] X^{(B^-_u \pi^- ,\rho^0)}_{u}\nonumber\\
&&  + \left[ a_4+ a_{10} - 2(a_6 + a_8 ) {m^{2}_{\pi^-} \over (m_d
+ m_u)(m_b + m_u)}
 \right] X^{(B^-_u \rho^0,\pi^-)}\nonumber \\
&&  + \left[ a_4+ a_{10} - 2(a_6 + a_8) {m^{2}_{B^-_u } \over (m_u
+ m_d)(m_b + m_u)} \right] X^{(B^-_u ,\pi^- \rho^0)}\Bigg\},
\end{eqnarray}
%%%%%%%%%%%%%%%%%%%%%%%%%%%%%%%%%%%%%%%
\begin{eqnarray}
A(B^-_u \to  \pi^- \omega)= &&  V_{ub} V^{*}_{ud} \Bigg\{ a_1
X^{(B^-_u \omega,\pi^-)} +a_1 X^{(B^-_u ,\pi^- \omega)} + a_2
X^{(B^-_u \pi^-,\omega)}\Bigg\}\nonumber\\ &&  -V_{tb}
V^{*}_{td}\Bigg\{ \left[ 2a_3 + a_4 +  2a_5 +{1\over 2} (a_7+ a_9
- a_{10})\right]X_u^{(B_u^-\pi^-, \omega)}\nonumber  \\ && +\left[
a_4 + a_{10} - 2(a_6 +a_8 ) {m^{2}_{\pi^-} \over (m_d + m_u)(m_b +
m_u)} \right] X_u^{(B^-_u \omega,\pi^-)}\nonumber\\ && + \left[
a_4 + a_{10}- 2(a_6 + a_8) {m^{2}_{B^-_u } \over (m_u + m_d)(m_b +
m_u)} \right] X^{(B^-_u ,\pi^- \omega)}\Bigg\},
\end{eqnarray}
%%%%%%%%%%%%%%%%%%%%%%%%%%%%%%%%%%%%%%%
\begin{eqnarray}
A(B^-_u \to  \rho^- \eta^\p)= &&  V_{ub} V^{*}_{ud} \left[
a_1(X_u^{(B^-_u \eta^\p,\rho^-)} +X^{(B^-_u ,\rho^-\eta^\p)}) +
a_2 X_u^{(B^-_u \rho^-,\eta^\p)} \right]\nonumber  \\ && + V_{cb}
V^{*}_{cd} a_2 X^{(B^-_u \rho^-,\eta^\p)}_{c}
 -V_{tb} V^{*}_{td} \Bigg\{
\Big[ 2a_3 +a_4- 2a_5 +{1\over 2}(-a_7 + a_9- a_{10}) \nonumber \\
&&  -(2a_6 - a_8){m^{2}_{\eta^\p} \over 2m_s (m_b + m_d )}
({f^s_{\eta^\p} \over  f^u_{\eta^\p}}-1)r_{\eta^\p} \Big]
X_u^{(B^-_u \rho^-, \eta^\p )}\nonumber\\ &&  +\left[ 2a_3 +a_4 -
2a_5-{1\over 2}(a_7 -a_9) \right] X^{(B^-_u
\rho^-,\eta^\p)}_{s}\non\\ && +\left[ a_3 - a_5- a_7 + a_9 \right]
X^{(B^-_u \rho^-,\eta^\p )}_{c}
 + ( a_4 + a_{10} ) X_u^{(B^-_u \eta^\p,\rho^- )}\nonumber\\
&& + \left[ a_4 + a_{10}-2(a_6 + a_8) {m^{2}_{B^-_u } \over (m_b +
m_u) (m_d +m_u )}
 \right] X^{(B^-_u ,\eta^\p\rho^- )}\Bigg\},
\end{eqnarray}
%%%%%%%%%%%%%%%%%%%%%%%%%%%%%%%%%%%%%%%
\begin{eqnarray}
A(B_u^-\to\pi^-\phi)= && -V_{tb}V^*_{td}\left\{ a_3+a_5-{1\over
2}(a_7+a_9)\right\} X^{(B^-\pi^-,\phi)},
\end{eqnarray}
%%%%%%%%%%%%%%%%%%%%%%%%%%%%%%%%%%%%%%%
\begin{eqnarray}
A(B^-_u \to  K^{\ast -} K^0)= &&  V_{ub} V^{*}_{ud} a_1 X^{(B^-_u
,K^{\ast -} K^0)} -V_{tb} V^{*}_{td}\Bigg\{ \Big[ a_4-{1\over 2}
a_{10} \non \\ && - (2a_6 - a_8 ) {m^{2}_{K^0} \over (m_s +
m_d)(m_b + m_s)}
 \Big] X^{(B^-_u K^{\ast -},K^0)}\nonumber\\
&&  + \left[ a_4 + a_{10}- 2(a_6 + a_8) {m^{2}_{B^-_u } \over (m_u
+ m_d)(m_b + m_u)} \right]X^{(B^-_u ,K^{\ast -} K^0)}\Bigg\},
\end{eqnarray}
%%%%%%%%%%%%%%%%%%%%%%%%%%%%%%%%%%%%%%%
\begin{eqnarray}
A(B^-_u \to  K^- K^{\ast 0})= &&  V_{ub} V^{*}_{ud} a_1 X^{(B^-_u
,K^- K^{\ast 0})} -V_{tb} V^{*}_{td}\Bigg\{ \left[ a_4 -{1\over 2}
a_{10} \right] X^{(B^-_u K^{-},K^{\ast 0})}\nonumber\\ &&  +
\left[ a_4  + a_{10}- 2(a_6 + a_8) {m^{2}_{B^-_u } \over (m_u +
m_d)(m_b + m_u)}\right] X^{(B^-_u ,K^- K^{* 0})}\Bigg\}.
\end{eqnarray}
%%%%%%%%%%%%%%%%%%%%%%%%%%%%%%%%%%%%%%%

\section{$\ov B_{\lowercase{d}}^0\to VV$ decays}
\vskip 0.3 cm
%%%%%%%%%%%%%%%%%%%%%%%%%%%%%%%%%%%%%%%
\begin{eqnarray}
A(\ov B^0_d \to K^{*-} \rho^+)= &&  V_{ub} V^{*}_{us} a_{1}
X^{(\bar B^{0}_{d}\rho^+,K^{*-}) }\non\\ && - V_{tb} V^{*}_{ts}
\Bigg\{(a_{4}+a_{10}) X^{(\bar B^{0}_{d}\rho^+,K^{*-}) }
+(a_4-{1\over 2}a_{10}) X^{(\bar B^0_d, K^{*-} \rho^+ )}
\nonumber\\ &&+(-2a_6+a_8)\langle  K^{*-}\rho^+ | \bar
s(1+\gamma_5)d|0\rangle \langle 0|\bar d(1-\gamma_5)b|\bar
B^0_d\rangle \Bigg\},
\end{eqnarray}
%%%%%%%%%%%%%%%%%%%%%%%%%%%%%%%%%%%%%%%
\begin{eqnarray}
A(\ov B^0_d \to \ov K^{*0} \rho^0)= &&  V_{ub} V^{*}_{us} a_{2}
X_u^{(\bar B^0_d \bar K^{*0},\rho^0)} - V_{tb} V^{*}_{ts} \Bigg\{
 {3\over 2}(a_7 + a_9) X_u^{(\bar B^0_d \bar K^{*0},\rho^0)}\non\\
&& +(a_4-{1\over2}a_{10})X^{(\bar B^0_d \rho^0,\bar K^{*0})}
+(a_4-{1\over 2}a_{10}) X^{(\bar B^0_d,  \bar K^{*0} \rho^0
)}\nonumber\\ &&+(-2a_6+a_8)\langle  \bar K^{*0}\rho^0 | \bar
s(1+\gamma_5)d|0\rangle \langle 0|\bar d(1-\gamma_5)b|\bar
B^0_d\rangle \Bigg\},
\end{eqnarray}
%%%%%%%%%%%%%%%%%%%%%%%%%%%%%%%%%%%%%%%
\begin{eqnarray}
A(\ov B^0_d \to \ov K^{*0}\omega)= &&  V_{ub} V^{*}_{us} a_{2}
X_u^{(\bar B^0_d \bar K^{*0},\omega)}
 - V_{tb} V^{*}_{ts} \Bigg\{
 (2a_3+2a_5+{1\over 2}a_7 + {1\over 2}a_9) X_u^{(\bar B^0_d \bar
K^{*0},\omega)}\nonumber\\ &&+(a_4-{1\over2}a_{10})X^{(\bar B^0_d
\omega,\bar K^{*0})} +(a_4-a_{10}) X^{(\bar B^0_d,  \bar K^{*0}
\omega )}\nonumber\\ &&+(-2a_6+a_8)\langle  \bar K^{*0} \omega|
\bar s(1+\gamma_5)d|0\rangle \langle 0|\bar d(1-\gamma_5)b|\bar
B^0_d\rangle \Bigg\},
\end{eqnarray}
%%%%%%%%%%%%%%%%%%%%%%%%%%%%%%%%%%%%%%%
\begin{eqnarray}
A(\ov B^0_d \to \ov K^{*0} \phi)= && - V_{tb} V^{*}_{ts}\Bigg\{
\big[ a_3+a_4+a_5-{1\over2}(a_7+a_9+a_{10}) \big] X_s^{(\bar B^0_d
\bar K^{*0},\phi)} \non\\ && +(a_4-{1\over 2}a_{10}) X^{(\bar
B^0_d, \bar K^{*0}\phi)}\nonumber\\ &&+(-2a_6+a_8) \langle \bar
K^{*0}\phi|\bar s(1+\gamma_5)d|0\rangle \langle 0|\bar
d(1-\gamma_5)b|\bar B^0_d\rangle \Bigg\},
\end{eqnarray}
%%%%%%%%%%%%%%%%%%%%%%%%%%%%%%%%%%%%%%%
\begin{eqnarray}
A(\ov B^0_d \to K^{*0} \ov K^{*0})= &&-V_{tb} V^{*}_{td} \Bigg\{
\left[ a_3+a_4+a_5-{1\over 2} (a_7+a_9+a_{10}) \right] X_d^{(\ov
B^0_d ,K^{*0} \bar K^{*0})}\nonumber \\ && + \left[ a_{3}+a_{5}-
{1\over 2}a_{7} -{1\over 2} a_{9} \right] X^{(\ov B^0_d ,K^{*0}
\bar K^{*0})}_{s} +(a_{4} -{1\over 2}a_{10}) X^{(\ov B^0_d
K^{*0},\bar K^{*0})}\nonumber \\ &&-(2a_6-a_8)\langle K^{*0}\bar
K^{*0}| \bar d(1+\gamma_5)d|0\rangle \langle 0|\bar
d(1-\gamma_5)b|\bar B^0_d\rangle \Bigg\},
\end{eqnarray}
%%%%%%%%%%%%%%%%%%%%%%%%%%%%%%%%%%%%%%%
\begin{eqnarray}
A(\ov B^0_d \to K^{*+} K^{*-})= &&V_{ub} V^{*}_{ud} a_2 X_u^{(\bar
B^0_d, K^{*+} K^{*-})}
 - V_{tb} V^{*}_{td}\Bigg\{ (a_3 + a_5 + a_7 + a_9)
X_u^{(\bar B^0_d, K^{*+} K^{*-})}\nonumber \\ && +(a_3 + a_5 -
{1\over 2}a_7 - {1\over 2}a_9) X^{(\bar B^0_d, K^{*+} K^{*-})}_{s}
\Bigg\},
\end{eqnarray}
%%%%%%%%%%%%%%%%%%%%%%%%%%%%%%%%%%%%%%%
\begin{eqnarray}
A(\ov B^0_d \to \rho^+ \rho^-)= &&V_{ub} V^{*}_{ud} \Big[a_{1}
X^{(\bar B^0_d \rho^+,\rho^-)} +a_2 X_u^{(\bar B^0_d, \rho^+
\rho^-)}\Big] - V_{tb} V^{*}_{td} \Bigg\{ (a_4+a_{10})X^{(\bar
B^0_d \rho^+,\rho^-)} \nonumber\\ &&+[2a_3+a_4+{1\over
2}(a_9-a_{10})] X_u^{(\bar B^0_d, \rho^+ \rho^-)} +(2a_5+{1\over
2}a_7)\ov X_u^{(\bar B^0_d, \rho^+ \rho^-)}\nonumber\\
&&+(-2a_6+a_8)\langle \rho^+\rho^-| \bar d(1+\gamma_5)d|0\rangle
\langle 0|\bar d(1-\gamma_5)b|\bar B^0\rangle \Bigg\},
\end{eqnarray}
%%%%%%%%%%%%%%%%%%%%%%%%%%%%%%%%%%%%%%%%%%%%%%%%%%%%%%%%%%%%%%%%%
\begin{eqnarray}
A(\ov B^0_d \to \rho^0\rho^0)= &&V_{ub} V^{*}_{ud} a_{2} \,2\Big[
X_u^{(\bar B^0_d \rho^0,\rho^0)} +X_u^{(\bar B^0_d,
\rho^0\rho^0)}\Big]\nonumber\\
 && -V_{tb} V^{*}_{td}\, 2\Bigg\{
[-a_4+{1\over 2}(3a_7+3a_9+a_{10})]X_u^{(\bar B^0_d
\rho^0,\rho^0)}\nonumber \\ && +\left[2a_3+a_4+{1\over 2}(a_9-
a_{10})\right]X_u^{(\bar B^0_d, \rho^0\rho^0)}+(2a_5+{1\over
2}a_7)\ov X_u^{(\bar B^0_d, \rho^0\rho^0)}\nonumber\\ &&
+(-2a_6+a_8)\langle \rho^0\rho^0| \bar d(1+\gamma_5)d|0\rangle
\langle 0|\bar d(1-\gamma_5)b|\bar B^0\rangle \Bigg\},
\end{eqnarray}
%%%%%%%%%%%%%%%%%%%%%%%%%%%%%%%%%%%%%%%
\begin{eqnarray}
A(\ov B^0_d \to \rho^0\omega)= &&V_{ub} V^{*}_{ud} a_2
\Big[X_u^{(\bar B^0_d \rho^0,\omega)} +X_u^{(\bar B^0_d
\omega,\rho^0)} +X_u^{(\bar B^0_d, \omega \rho^0)}\Big]\nonumber\\
&&-V_{tb} V^{*}_{td} \Bigg\{ \left[2a_3+a_4+2a_5+{1\over 2}a_7
+{1\over 2}a_9 -{1\over 2}a_{10}\right]X_u^{(\bar B^0_d
\rho^0,\omega)}\nonumber\\ &&+\left[-a_4+{3\over 2}a_7+{3\over
2}a_9 +{1\over 2}a_{10}\right] X_u^{(\bar B^0_d \omega,\rho^0)}
+(-a_4+{3\over 2}a_9 +{1\over 2}a_{10}) X_u^{(\bar B^0_d,
\omega\rho^0)}\nonumber\\ && +{3\over 2}a_7\ov X_u^{(\bar B^0_d,
\omega\rho^0)} +(-2a_6+a_8)\langle \omega\rho^0| \bar
d(1+\gamma_5)d|0\rangle \langle 0|\bar d(1-\gamma_5)b|\bar
B^0\rangle \Bigg\},
\end{eqnarray}
%%%%%%%%%%%%%%%%%%%%%%%%%%%%%%%%%%%%%%%
\begin{eqnarray}
A(\ov B^0_d \to \omega \omega)= &&V_{ub} V^{*}_{ud}2 a_{2} (
X_u^{(\bar B^0_d \omega,\omega)} + X_u^{(\bar B^0_d,
\omega\omega)})\nonumber\\ &&- V_{tb} V^{*}_{td} \Bigg\{
(4a_3+2a_4+4a_5+a_7+a_9-a_{10})X_u^{(\bar B^0_d
\omega,\omega)}\nonumber\\ &&+\Big[2a_3+a_4+{1\over
2}(a_9-a_{10})\Big] X_u^{(\bar B^0_d,\omega\omega)} +(2a_5+{1\over
2}a_7)\ov X_u^{(\bar B^0_d,\omega\omega)}\nonumber \\ &&
+(-2a_6+a_8)\langle \omega\omega| \bar d(1+\gamma_5)d|0\rangle
\langle 0|\bar d(1-\gamma_5)b|\bar B^0\rangle \Bigg\},
\end{eqnarray}
%%%%%%%%%%%%%%%%%%%%%%%%%%%%%%%%%%%%%%%
\begin{eqnarray}
A(\ov B^0_d \to \rho^0\phi)= && - V_{tb} V^{*}_{td}
\left[a_3+a_5-{1\over 2}a_7-{1\over 2}a_9\right] X_s^{(\bar B^0_d
\rho^0,\phi)},
\end{eqnarray}
%%%%%%%%%%%%%%%%%%%%%%%%%%%%%%%%%%%%%%%
\begin{eqnarray}
A(\ov B^0_d \to \omega\phi)= && - V_{tb} V^{*}_{td}
\left[a_3+a_5-{1\over 2}a_7-{1\over 2}a_9\right]X_s^{(\bar B^0_d
\omega,\phi)},
\end{eqnarray}
%%%%%%%%%%%%%%%%%%%%%%%%%%%%%%%%%%%%%%%
\begin{eqnarray}
A(\ov B^0_d \to \phi\phi)= && - V_{tb} V^{*}_{td}
\left[a_3+a_5-{1\over 2}a_7-{1\over 2}a_9\right]X_s^{(\bar B^0_d,
\phi\phi)}.
\end{eqnarray}
%%%%%%%%%%%%%%%%%%%%%%%%%%%%%%%%%%%%%%%

\section{$B^-_{\lowercase{u}} \to VV$ decays}

%%%%%%%%%%%%%%%%%%%%%%%%%%%%%%%%%%%%%%%
\begin{eqnarray}
A(B_u^- \to K^{*-} \rho^0)= &&  V_{ub} V^{*}_{us} \Big[a_{1}
X^{(B^{-}\rho^0,K^{*-})} +a_{2} X_u^{( B^- K^{*-},\rho^0)}+a_1
X^{(B^{-}, \rho^0 K^{*-})} \Big]\nonumber\\ && - V_{tb} V^{*}_{ts}
\Bigg\{ (a_{4}+a_{10}) X^{( B^{-}\rho^0,K^{*-})} +{3\over 2}(a_7 +
a_9)  X_u^{(B^- K^{*-},\rho^0)}\nonumber\\
&&+(a_4+a_{10})X^{(B^{-}, \rho^0 K^{*-})}\nonumber\\
&&-2(a_6+a_8)\langle K^{*-}\rho^0| \bar s(1+\gamma_5)u|0\rangle
\langle 0|\bar u(1-\gamma_5)b|B^-\rangle \Bigg\},
\end{eqnarray}
%%%%%%%%%%%%%%%%%%%%%%%%%%%%%%%%%%%%%%%
\begin{eqnarray}
A(B_u^- \to \ov K^{*0}\rho^-)= &&  - V_{tb}
V^{*}_{ts}\Bigg\{(a_4-{1\over2}a_{10})X^{( B^- \rho^-,\bar
K^{*0})}+(a_4+a_{10})X^{( B^-, \rho^-\bar K^{*0})}\nonumber\\ &&
-2(a_6+a_8)\langle \rho^-\bar K^{*0}| \bar s(1+\gamma_5)u|0\rangle
\langle 0|\bar u(1-\gamma_5)b|B^-\rangle \Bigg\} \non\\
&& +V_{ub}V_{us}^*a_1X^{(B^-,\rho^-\bar K^{*0})},
\end{eqnarray}
%%%%%%%%%%%%%%%%%%%%%%%%%%%%%%%%%%%%%%%
\begin{eqnarray}
A( B_u^- \to  K^{*-} \omega)= &&  V_{ub}V_{us}^*\Big(
a_1X^{(B^-\omega,K^{*-})} +a_2 X_u^{(B^-
K^{*-},\omega)}+a_{1}X^{(B^-, \omega K^{*-})}\Big)\nonumber\\ && -
V_{tb} V^{*}_{ts} \Bigg\{ (a_4+a_{10})X^{(B^- \omega,K^{*-})} +
\left[2a_3+2a_5+{1\over 2}a_7 + {1\over 2}a_9\right] X_u^{(B^-
K^{*-},\omega)}\nonumber\\ &&+(a_4+a_{10})X^{(B^{-},  \omega
K^{*-})}\nonumber\\ &&-2(a_6+a_8)\langle K^{*-} \omega| \bar
s(1+\gamma_5)u|0\rangle \langle 0|\bar u(1-\gamma_5)b|B^-\rangle
\Bigg\},
\end{eqnarray}
%%%%%%%%%%%%%%%%%%%%%%%%%%%%%%%%%%%%%%%
\begin{eqnarray}
A(B_u^- \to  K^{*-} \phi)= && V_{ub} V^{*}_{us} a_1 X^{(B^-_u
,\phi K^{*-})} - V_{tb} V^{*}_{ts}\Bigg\{ \Big[ a_3+a_4+a_5 \non\\
&& -{1\over2}(a_7+a_9+a_{10})\Big] X_s^{( B^-  K^{*-},\phi)}
+(a_4+a_{10}) X^{( B^-,  K^{*-}\phi)}\nonumber\\ &&-2(a_6+a_8)
\langle  K^{*-}\phi|\bar s(1+\gamma_5)u|0\rangle \langle 0|\bar
u(1-\gamma_5)b| B^-\rangle \Bigg\},
\end{eqnarray}
%%%%%%%%%%%%%%%%%%%%%%%%%%%%%%%%%%%%%%%
\begin{eqnarray}
A( B_u^- \to  K^{*-} K^{*0}) = &&
V_{ub}V_{ud}^*a_1X^{(B^-,K^{*-}K^{*0})}\nonumber\\ &&- V_{tb}
V^{*}_{td} \Bigg\{ (a_{4}-{1\over 2}a_{10})X^{(B^- K^{*-},
K^{*0})} +(a_4+a_{10})X^{( B^-, K^{*-} K^{*0})}\nonumber\\
&&-2(a_6+a_8)\langle K^{*-} K^{*0}| \bar d(1+\gamma_5)u|0\rangle
\langle 0|\bar u(1-\gamma_5)b| B^-\rangle \Bigg\},
\end{eqnarray}
%%%%%%%%%%%%%%%%%%%%%%%%%%%%%%%%%%%%%%%
\begin{eqnarray}
A(B_u^-\to\rho^-\phi)= && -\sqrt{2}A(\ov B_d^0\to \rho^0\phi),
\end{eqnarray}
%%%%%%%%%%%%%%%%%%%%%%%%%%%%%%%%%%%%%%%
\begin{eqnarray}
A(B_u^- \to \rho^- \rho^0)= &&V_{ub} V^{*}_{ud} \Bigl[ a_1X^{(B^-
\rho^0,\rho^-)}
 +a_2  X_u^{(B^-\rho^-,\rho^0)}\Bigl]\nonumber\\
&&- V_{tb} V^{*}_{td} \Bigg\{ {3\over 2}(a_7+a_9+a_{10})X^{(B^-
\rho^0,\rho^-)} +(a_4+a_{10})X_u^{(B^-,\rho^0\rho^-)}\nonumber\\
&&-2(a_6+a_8)\langle \rho^0\rho^-| \bar d(1+\gamma_5)u|0
\rangle\langle 0|\bar u(1-\gamma_5)b|B^-\rangle \Bigg\},
\end{eqnarray}
%%%%%%%%%%%%%%%%%%%%%%%%%%%%%%%%%%%%%%%
\begin{eqnarray}
A(B_u^- \to \rho^- \omega) &=& V_{ub} V^{*}_{ud}\Big[a_1 X^{(B^-
\omega,\rho^-)}+a_2 (X_u^{(B^- \rho^-, \omega)}+X_u^{B^-,
\rho^-\omega)}) \Big] \non \\ &-&
V_{tb}V^{*}_{td}\Bigg\{(a_4+a_{10})X^{(B^- \omega,\rho^-)}\non\\
&+& \left[2a_3+a_4+2a_5+{1\over 2}a_7 + {1\over 2}a_9-{1\over
2}a_{10}\right]X_u^{(B^- \rho^-,\omega)}  +
(a_4+a_{10})X_u^{(B^-, \rho^- \omega )}\non \\
&-& 2(a_6+a_8)\langle \rho^-
\omega| \bar d(1+\gamma_5)u|0\rangle \langle 0|\bar
u(1-\gamma_5)b|B^-\rangle \Bigg\}.
\end{eqnarray}
%%%%%%%%%%%%%%%%%%%%%%%%%%%%%%%%%%%%%%%

\newpage
{\squeezetable
\begin{table}[ht]
{\small Table V. Relative magnitudes of tree, QCD penguin and
electroweak penguin amplitudes for charmless $B_{u,d}\to PP$
decays shown in first, second and third entries, respectively.
Predictions are made for $k^2=m_b^2/2$, $\eta=0.370,~\rho=0.175$,
and $\nc(LR)=2,3,5,\infty$ with $\nc(LL)$ being fixed to be 2 in
the first case and treated to be the same as $\nc(LR)$ in the
second case. The BSW model is used for heavy-to-light form
factors. Results for CP-conjugate modes are not listed here. For
tree-dominated decays, the tree amplitude is normalized to unity.
Likewise, the QCD penguin amplitude is normalized to unity for
penguin-dominated decays. Our preferred predictions are those for
$\nc(LL)=2$ and $\nc(LR)=5$.}
\begin{center}
\begin{tabular}{l l c c c c c c c c }
 &  & \multicolumn{4}{c}{$\nc(LL)=2$}
 &   \multicolumn{4}{c}{$\nc(LL)=\nc(LR)$}  \\ \cline{3-6} \cline{7-10}
\raisebox{2.0ex}[0cm][0cm]{Decay} &
\raisebox{2.0ex}[0cm][0cm]{Class} & 2 & 3 & 5 & $\infty$ & 2 & 3 &
5 & $\infty$  \\ \hline
 $\ov B^0_d \to \pi^+ \pi^-$&        I
 & 1 & 1 & 1 & 1 & 1 & 1 & 1 & 1 \\
 && -0.04+0.17i & -0.04+0.17i &-0.05+0.18i & -0.05+0.18i &-0.04+0.17i
 & -0.04+0.17i &  -0.04+0.17i & -0.04+0.18i \\
 && 0.004i & 0.004i & 0.004i & 0.004i & 0.004i & -0.0004 & -0.003i
 & -0.01i \\
 $\ov B^0_d \to \pi^0\pi^0$ &       II,VI
 & 1 & 1 & 1 & 1 & 1 & 1 & 1 & 1 \\
 && 0.17-0.64i & 0.17-0.66i & 0.17-0.67i & 0.18-0.69i & 0.17-0.64i
 & 1.6-6.3i & -0.31+1.2i & -0.12+0.48i \\
 && 0.15i & 0.14i & 0.14i & 0.14i & 0.15i & 0.05+1.3i & -0.01-0.24i
 & -0.08i \\
 $\ov B^0_d \to \eta \eta$  &       II,VI
 & 1 & 1 & 1 & 1 & 1 & 1 & 1 & 1 \\
 && -0.17+0.84i & -0.18+0.91i & -0.19+0.97i & -0.21+1.1i & -0.17+0.84i
 & -1.6+8.2i & 0.31-1.6i & 0.12-0.63i \\
 && 0.01+0.10i & 0.01+0.10i & 0.01+0.10i & 0.01+0.09i & 0.01+0.10i
 & 0.12+1.0i & -0.02-0.2i & -0.01-0.08i \\
 $\ov B^0_d \to \eta \eta'$ &       II,VI
 & 1 & 1 & 1 & 1 & 1 & 1 & 1 & 1 \\
 && -0.13+0.64i & -0.16+0.85i & -0.17+1.0i & -0.2+1.3i & -0.13+0.64i
 & -1.2+6.4i & 0.23-1.3i & 0.09-0.51i \\
 && 0.012i & 0.012i & 0.012i & 0.012i & 0.012i & 0.04+0.19i &
 -0.01-0.05i  & -0.02i \\
 $\ov B^0_d \to \eta' \eta'$&       II,VI
 & 1 & 1 & 1 & 1 & 1 & 1 & 1 & 1 \\
 && -0.13+0.45i & -0.14+0.79i & -0.15+1.1i & -0.17+1.5i & -0.13+0.45i
 & -1.1+4.7i & 0.21-0.97i & 0.07-0.40i \\
 && -0.01-0.06i & -0.01-0.06i & -0.01-0.06i & -0.01-0.06i & -0.01-0.06i
 & -0.12-0.5i & 0.02+0.09i & 0.01+0.03i \\
 $ B^- \to \pi^- \pi^0$ &          III
 & 1 & 1 & 1 & 1 & 1 & 1 & 1 & 1 \\
 && 0 & 0 & 0 & 0 & 0 & 0 & 0 & 0 \\
 && 0.03i & 0.03i & 0.03i & 0.03i & 0.03i & 0.03i & 0.03i & 0.03i
 \\
 $ B^- \to \pi^- \eta$  &           III
 & 1 & 1 & 1 & 1 & 1 & 1 & 1 & 1 \\
 && -0.08+0.30i & -0.08+0.32i & -0.08+0.34i & -0.09+0.36i & -0.08+0.30i
 & -0.10+0.38i & -0.12+0.45i & -0.18+0.61i \\
 && 0.02i & 0.02i & 0.02i & 0.02i & 0.02i & 0.03i & 0.03i & 0.04i \\
 $ B^- \to \pi^-\eta'$  &           III
 & 1 & 1 & 1 & 1 & 1 & 1 & 1 & 1 \\
 && -0.07+0.23i & -0.09+0.31i & -0.1+0.37i & -0.12+0.46i & -0.07+0.23i
 & -0.1+0.3i & -0.14+0.37i & -0.24+0.52i \\
 && -0.01i & -0.01i & -0.01i & -0.01i & -0.01i & -0.02i & -0.02i
 & -0.03i \\
   $\ov B^0_d \to   K^- \pi^+$&       IV
 & -0.04+0.22i & -0.04+0.22i & -0.04+0.21i & -0.04+0.21i &
 -0.04+0.22i & -0.04+0.22i & -0.04+0.22i & -0.04+0.22i \\
 && 1 & 1 & 1 & 1 & 1 & 1 & 1 & 1 \\
 && 0.02 & 0.02 & 0.02 & 0.02 & 0.02 & 0.002i & -0.02+0.01i
 & -0.03+0.01i \\
 $ B^-\to \ov K^0\pi^-$       &            IV
  & 0 & 0 & 0 & 0 & 0 & 0 & 0 & 0 \\
 && 1 & 1 & 1 & 1 & 1 & 1 & 1 & 1 \\
 && -0.01 & -0.01 & -0.01 & -0.01 & -0.01 & -0.001i & 0.002-0.001i
 &  0.01 \\
 $ B^- \to K^-   K^0$   &                   IV
  & 0 & 0 & 0 & 0 & 0 & 0 & 0 & 0 \\
 && 1 & 1 & 1 & 1 & 1 & 1 & 1 & 1 \\
 && -0.01 & -0.01 & -0.01 & -0.01 & -0.01 & -0.001i & 0.01
 & 0.02-0.01i \\
 \end{tabular}
\end{center}
\end{table}
 }

\newpage
{\squeezetable
\begin{table}[ht]
{\small Table V. (continued)}
\begin{center}
\begin{tabular}{l l c c c c c c c c }
 &  & \multicolumn{4}{c}{$\nc(LL)=2$}
 &   \multicolumn{4}{c}{$\nc(LL)=\nc(LR)$}  \\ \cline{3-6} \cline{7-10}
\raisebox{2.0ex}[0cm][0cm]{Decay} &
\raisebox{2.0ex}[0cm][0cm]{Class} & 2 & 3 & 5 & $\infty$ & 2 & 3 &
5 & $\infty$  \\ \hline
 $\ov B^0_d \to \pi^0 \eta$ &       VI
 & 0.02-0.10i & 0.02-0.09i & 0.02-0.09i & 0.02-0.08i & 0.02-0.10i
 & -0.01i & -0.01+0.05i & -0.03+0.13i \\
 && 1 & 1 & 1 & 1 & 1 & 1 & 1 & 1 \\
 && -0.02+0.01i & -0.02+0.01i & -0.02+0.01i & -0.02+0.004i
 & -0.02+0.01i & -0.01 & -0.001 & 0.01 \\
 $\ov B^0_d \to \pi^0 \eta'$&       VI
 & 0.11-0.27i & 0.09-0.2i& 0.08-0.16i & 0.06-0.13i & 0.11-0.27i &
 0.01-0.03i & -0.06+0.13i & -0.16+0.33i \\
 && 1 & 1 & 1 & 1 & 1 & 1 & 1 & 1 \\
 && -0.17+0.07i & -0.12+0.05i & -0.1+0.04i & -0.08+0.03i &
 -0.17+0.07i & -0.14+0.06i & -0.13+0.05i & -0.11+0.04i  \\
 $\ov B^0_d\to\ov K^0\pi^0$&        VI
 & 0.01-0.05i & 0.01-0.05i & 0.01-0.04i & 0.01-0.04i & 0.01-0.05i
 & -0.01i & 0.03i & -0.01+0.06i \\
 && 1 & 1 & 1 & 1 & 1 & 1 & 1 & 1 \\
 && -0.14+0.04i & -0.14+0.04i & -0.13+0.04i & -0.13+0.04i &
 -0.14+0.04i & -0.13+0.04i & -0.12+0.04i & -0.11+0.03i \\
 $\ov B^0_d \to\ov K^0\eta$&       VI
 & -0.10-0.05i & -0.11-0.05i & -0.11-0.06i & -0.12-0.06i &
  -0.10-0.05i & -0.01-0.01i & 0.06+0.03i & 0.15+0.07i \\
 && 1 & 1 & 1 & 1 & 1 & 1 & 1 & 1 \\
 && -0.19+0.05i & -0.20+0.05i & -0.21+0.05i & -0.22+0.06i &
 -0.19+0.05i & -0.18+0.05i & -0.18+0.05i & -0.17+0.05i \\
 $\ov B^0_d\to\ov K^0\eta'$&       VI
 & 0.08-0.01i & 0.07-0.01i & 0.06-0.01i & 0.06 & 0.08-0.01i
 & 0.01 & -0.04+0.01i & -0.11+0.01i \\
 && 1 & 1 & 1 & 1 & 1 & 1 & 1 & 1 \\
 && -0.03+0.01i & -0.03+0.01i & -0.03+0.01i & -0.02+0.01i &
 -0.03+0.01i & -0.02+0.01i & -0.01+0.004i & -0.004 \\
 $ \ov B^0_d\to K^0\ov K^0$ &             VI
  & 0 & 0 & 0 & 0 & 0 & 0 & 0 & 0 \\
 && 1 & 1 & 1 & 1 & 1 & 1 & 1 & 1 \\
 && -0.01 & -0.01 & -0.01 & -0.01 & -0.01 & -0.001i & 0.01
 & 0.02-0.01i \\
 $ B^-\to K^- \pi^0$    &           VI
 & -0.05+0.27i & -0.05+0.26i & -0.04+0.26i & -0.04+0.25i &
   -0.05+0.27i & -0.04+0.23i & -0.04+0.2i & -0.03+0.15i \\
 && 1 & 1 & 1 & 1 & 1 & 1 & 1 & 1 \\
 && 0.15-0.04i & 0.15-0.04i & 0.14-0.04i & 0.13-0.04i & 0.15-0.04i
 & 0.13-0.04i & 0.11-0.03i & 0.09-0.03i \\
 $ B^- \to K^- \eta$    &           VI
 & -0.03-0.40i & -0.03-0.43i & -0.04-0.45i & -0.04-0.48i & -0.03-0.40i
 & 0.06-0.36i & 0.13-0.33i & 0.22-0.29i \\
 && 1 & 1 & 1 & 1 & 1 & 1 & 1 & 1 \\
 && -0.24+0.05i & -0.25+0.06i & -0.26+0.06i & -0.27+0.07i &
 -0.24+0.05i & -0.19+0.04i & -0.14+0.04i & -0.09+0.02i \\
 $ B^- \to K^- \eta'$   &           VI
 & 0.07+0.07i & 0.06+0.06i & 0.06+0.05i & 0.05+0.05i & 0.07+0.07i
 & 0.08i & -0.05+0.08i & -0.12+0.08i \\
 && 1 & 1 & 1 & 1 & 1 & 1 & 1 & 1 \\
 && -0.02+0.01i & -0.02+0.01i & -0.02+0.01i & -0.02+0.01i &
 -0.02+0.01i & -0.02+0.01i & -0.02+0.01i & -0.02+0.01i \\
\end{tabular}
\end{center}
\end{table}
 }

{\squeezetable
\begin{table}[ht]
{\small Table VI. Same as Table V except for charmless
$B_{u,d}\to VP$ decays.}
\begin{center}
\begin{tabular}{l l c c c c c c c c c}
 &  & \multicolumn{4}{c}{$\nc(LL)=2$}
 &   \multicolumn{4}{c}{$\nc(LL)=\nc(LR)$}  \\ \cline{3-6} \cline{7-10}
\raisebox{2.0ex}[0cm][0cm]{Decay} &
\raisebox{2.0ex}[0cm][0cm]{Class} & 2 & 3 & 5 & $\infty$ & 2 & 3 &
5 & $\infty$  \\ \hline $\ov B^0_d \to \rho^-  \pi^+ $&      I & 1
& 1 & 1 & 1 & 1 & 1 & 1 & 1 \\
&& -0.02+0.08i & -0.02+0.08i & -0.02+0.08i & -0.02+0.08i
 & -0.02+0.08i & -0.02+0.08i & -0.02+0.09i & -0.02+0.09i \\
&& 0.01i & 0.01i & 0.01i & 0.01i & 0.01i & 0.001i & -0.002i &
-0.006i \\
$\ov B^0_d \to \rho^+  \pi^-$&       I
& 1 & 1 & 1 & 1& 1 & 1 & 1 & 1 \\
&& -0.01i & -0.01i & -0.02i & -0.02i & -0.01i &
-0.004i & -0.002i & 0.003i \\
&& 0.01i & 0.01i & 0.01i & 0.01i & 0.01i & 0.001i & -0.002i & -0.006i \\
$\ov B^0_d \to \rho^+K^{-} $&      I,IV
& 1 & 1 & 1 & 1 & 1 & 1 & 1 & 1 \\
&& 0.23+0.7i & 0.24+0.83i & 0.26+0.95i & 0.28+1.1i &
0.23+0.7i & 0.18+0.61i & 0.15+0.55i & 0.11+0.47i \\
&& -0.05-0.1i & -0.05-0.11i & -0.05-0.11i & -0.05-0.11i &
-0.05-0.1i & -0.01-0.03i & 0.01+0.03i & 0.05+0.1i \\
$\ov B^0_d \to \rho^0 \pi^0$&       II
& 1 & 1 & 1 & 1& 1 & 1 & 1 & 1 \\
&& 0.08-0.27i & 0.08-0.27i & 0.08-0.26i & 0.08-0.26i
 & 0.08-0.27i & 0.77-2.7i & -0.15+0.55i & -0.06+0.22i \\
&& 0.15i & 0.15i & 0.15i & 0.15i & 0.15i & 1.3i & -0.24i & -0.09i \\
$\ov B^0_d \to \omega \pi^0$ &      II
& 1 & 1 & 1 & 1 & 1 & 1 & 1 & 1 \\
&& -0.5+1.6i & -0.36+1.0i & -0.29+0.72i & -0.18+0.26i & -0.5+1.6i &
-2.7+7.3i & 0.28-0.54i & -0.02+0.24i \\
&& 0.01+0.26i & 0.01+0.23i & 0.01+0.23i & 0.01+0.23i & 0.01+0.23i &
0.06+2.5i & -0.01-0.52i & -0.01-0.22i \\
$ \ov B^0_d \to \omega\eta $& II
& 1 & 1 & 1 & 1 & 1 & 1 & 1 & 1 \\
&& -0.1+0.29i & -0.08+0.18i & -0.06+0.11i & -0.04-0.01i & -0.1+0.29i &
-0.52+0.98i & 0.04+0.05i & -0.02+0.14i \\
&& 0.06i & 0.06i & 0.06i & 0.06i &
0.06i & 0.03+0.64i & -0.01-0.13i & -0.06i \\
$ \ov B^0_d \to\omega \eta'$& II
& 1 & 1 & 1 & 1 & 1 & 1 & 1 & 1 \\
&& -0.1+0.15i & -0.1+0.16i & -0.1+0.17i & -0.1+0.19i &
-0.1+0.15i & -0.61-0.23i & 0.07+0.26i & 0.21i \\
&& -0.004i & -0.01i & -0.004i & -0.004i & -0.004i & 0.04-0.01i & -0.02i
& -0.01i \\
$\ov B^0_d \to\rho^0 \eta $&      II
& 1 & 1 & 1 & 1 & 1 & 1 & 1 & 1 \\
&& 0.25-0.5i & 0.26-0.52i & 0.26-0.54i &0.26-0.59i &
0.25-0.5i & 2.5-5.5i & -0.48+1.2i & -0.19+0.49i \\
&& -0.03+0.21i & -0.03+0.22i & -0.03+0.22i & -0.03+0.22i & -0.03+0.2i &
-0.27+1.8i & 0.05-0.3i & 0.02-0.10i \\
$ \ov B^0_d \to \rho^0 \eta'$&  II,VI
& 1 & 1 & 1 & 1 & 1 & 1 & 1 & 1 \\
&& 0.26 & 0.51-0.35i & 0.71-0.67i & 1.0-1.1i & 0.26 &
2.8-0.72i & -0.6+0.3i & -0.26+0.20i \\
&& -0.14+0.43i & -0.16+0.45i & -0.16+0.46i & -0.17+0.46i & -0.14+0.43i &
-1.4+4.0i & 0.26-0.73i & 0.1-0.26i \\
$ B^- \to \rho^0 \pi^- $  &  III
& 1 & 1 & 1 & 1 & 1 & 1 & 1 & 1 \\
&& 0.03-0.12i & 0.03-0.12i & 0.03-0.12i & 0.03-0.13i &
0.03-0.12i & 0.04-0.17i & 0.06-0.23i & 0.12-0.48i \\
&& 0.05i & 0.05i & 0.05i & 0.05i & 0.05i & 0.06i & 0.08i & 0.14i \\
$ B^- \to\rho^- \pi^0$   &  III
& 1 & 1 & 1 & 1 & 1 & 1 & 1 & 1 \\
&& -0.02+0.08i & -0.02+0.08i & -0.02+0.08i & -0.02+0.08i &
-0.02+0.08i & -0.03+0.09i & -0.03+0.10i & -0.03+0.11i \\
&& -0.01i & -0.01i & -0.01i & -0.01i & -0.01i & -0.01i &
-0.01i & -0.02i \\
$ B^- \to \omega \pi^- $  &  III
& 1 & 1 & 1 & 1 & 1 & 1 & 1 & 1 \\
&& -0.05+0.17i & -0.04+0.11i & -0.03+0.07i & -0.02+0.01i &
-0.05+0.17i & -0.040.11i & -0.03+0.04i & 0.02-0.21i \\
&& 0.01i & 0.01i & 0.01i & 0.01i & 0.01i & 0.02i & 0.03i & 0.04i \\
$ B^- \to \rho^- \eta $  &   III
& 1 & 1 & 1 & 1 & 1 & 1 & 1 & 1 \\
&& 0.01i & -0.01i & -0.02i & 0.01-0.03i & 0.01i & 0.01i & 0.01i & 0.01i \\
&& 0.001i & 0.0004i & 0.0004i & 0.0004i & 0.001i & -0.002i & -0.005i
& -0.01i \\
$ B^- \to \rho^- \eta'$  &   III
& 1 & 1 & 1 & 1 & 1 & 1 & 1 & 1 \\
&& -0.04+0.07i & -0.06+0.12i & -0.06+0.16i &-0.08+0.23i &
-0.04+0.07i &-0.05+0.09i & -0.06+0.11i & -0.07+0.15i \\
&& 0.02i & 0.02i & 0.02i & 0.02i & 0.02i & 0.02i & 0.02i &
0.02i \\
$ B^- \to \rho^0 K^- $ &     III,VI
& 1 & 1 & 1 & 1 & 1 & 1 & 1 & 1 \\
&& 0.16+0.49i & 0.17+0.59i & 0.18+0.67i & 0.2+0.79i
& 0.16+0.49i & 0.18+0.59i & 0.2+0.71i & 0.27+1.1i \\
&& -0.38-0.9i & -0.38-0.89i & -0.38-0.9i & -0.38-0.9i &
-0.38-0.9i & -0.48-1.1i & -0.62-1.4i & -1.1-2.5i \\
$\ov B^0_d \to K^{*-} \pi^+$ &       IV
& -0.08+0.54i & -0.08+0.54i & -0.08+0.54i & -0.08+0.54i &
-0.08+0.54i & -0.08+0.52i & -0.08+0.51i & -0.08+0.49i \\
&& 1 & 1 & 1 & 1 & 1 & 1 & 1 & 1 \\
&& 0.05-0.01i & 0.05-0.01i & 0.05-0.01i & 0.05-0.01i &
0.05-0.01i & 0.01 & -0.02+0.01i & -0.06+0.02i \\
$ B^- \to\ov K^{*0}\pi^- $&  IV
& 0 & 0 & 0 & 0 & 0 & 0 & 0 & 0 \\
&& 1 & 1 & 1 & 1 & 1 & 1 & 1 & 1 \\
&& -0.03+0.01i & -0.03+0.01i & -0.03+0.01i & -0.03+0.01i &
-0.03+0.01i & -0.004 & 0.01 & 0.03-0.01i \\
$ B^- \to  K^{*0} K^- $&     IV
& 0 & 0 & 0 & 0 & 0 & 0 & 0 & 0 \\
&& 1 & 1 & 1 & 1 & 1 & 1 & 1 & 1 \\
&& -0.03+0.01i & -0.03+0.01i & -0.03+0.01i & -0.03+0.01i &
-0.03+0.01i & -0.004 & 0.012 & 0.03-0.01i \\
$ B^- \to K^{*-} K^0$ &         IV
& 0 & 0 & 0 & 0 & 0 & 0 & 0 & 0 \\
&& 1 & 1 & 1 & 1 & 1 & 1 & 1 & 1 \\
&& 0.09 & 0.08 & 0.07 & 0.06 & 0.09 & 0.03
& -0.03+0.01i & -0.15+0.03i \\
\end{tabular}
\end{center}
\end{table}
}

{\squeezetable
\begin{table}[ht]
{\small Table VI. (continued)}
\begin{center}
\begin{tabular}{l l c c c c c c c c c}
 &  & \multicolumn{4}{c}{$\nc(LL)=2$}
 &   \multicolumn{4}{c}{$\nc(LL)=\nc(LR)$}  \\ \cline{3-6} \cline{7-10}
\raisebox{2.0ex}[0cm][0cm]{Decay} &
\raisebox{2.0ex}[0cm][0cm]{Class} & 2 & 3 & 5 & $\infty$ & 2 & 3 &
5 & $\infty$  \\ \hline $ \ov B^0_d \to \phi \pi^0$  &   V & 0 & 0
& 0 & 0 & 0 & 0 & 0 & 0 \\ && 1 & 1 & 1 & 1 & 1 & 1 & 1 & 1 \\ &&
-0.35+0.12i & -0.97+0.62i & 1.3+0.27i & 0.34-0.04i & -0.35+0.12i &
1.+0.06i & 0.28-0.04i & 0.14-0.03i \\ $ \ov B^0_d \to \phi \eta $
&   V & 0 & 0 & 0 & 0 & 0 & 0 & 0 & 0 \\ && 1 & 1 & 1 & 1 & 1 & 1
& 1 & 1 \\ && -0.35+0.12i & -0.97+0.62i & 1.3+0.27i & 0.34-0.04i &
-0.35+0.12i & 1.+0.06i & 0.28-0.04i & 0.14-0.03i \\ $ \ov B^0_d
\to  \phi \eta'$  &  V & 0 & 0 & 0 & 0 & 0 & 0 & 0 & 0 \\ && 1 & 1
& 1 & 1 & 1 & 1 & 1 & 1 \\ && -0.35+0.12i & -0.97+0.62i &
1.3+0.27i & 0.34-0.04i & -0.35+0.12i & 1.+0.06i & 0.28-0.04i &
0.14-0.03i \\ $ B^- \to \phi\pi^- $  &         V & 0 & 0 & 0 & 0 &
0 & 0 & 0 & 0 \\ && 1 & 1 & 1 & 1 & 1 & 1 & 1 & 1 \\ &&
-0.35+0.12i & -0.97+0.62i & 1.3+0.27i & 0.34-0.04i & -0.35+0.12i &
1.+0.06i & 0.28-0.04i & 0.14-0.03i \\ $\ov B^0_d\to K^{*0}\ov
K^0$&  VI & 0 & 0 & 0 & 0 & 0 & 0 & 0 & 0 \\ && 1 & 1 & 1 & 1 & 1
& 1 & 1 & 1 \\ && 0.09 & 0.08 & 0.07 & 0.06 & 0.09 & 0.03 &
-0.03+0.01i & -0.15+0.03i \\ $\ov B^0_d\to \ov K^{*0} K^0$ &  VI &
0 & 0 & 0 & 0 & 0 & 0 & 0 & 0 \\ && 1 & 1 & 1 & 1 & 1 & 1 & 1 & 1
\\ && -0.03+0.01i & -0.03+0.01i & -0.03+0.01i & -0.03+0.01i &
-0.03+0.01i & -0.004 & 0.012 & 0.03-0.01i \\ $\ov B^0_d\to\ov
K^{*0}\pi^0$&     VI & 0.01-0.07i & 0.01-0.07i & 0.01-0.07i &
0.01-0.07i & 0.01-0.07i & -0.01i & -0.01+0.03i & -0.01+0.08i \\ &&
1 & 1 & 1 & 1 & 1 & 1 & 1 & 1 \\ && -0.21+0.07i & -0.21+0.07i &
-0.21+0.07i & -0.21+0.07i & -0.21+0.07i & -0.18+0.06i &
-0.16+0.05i & -0.13+0.04i \\ $\ov B^0_d\to\rho^0\ov K^0$&      VI
& -0.18+0.54i & -0.14+0.48i & -0.12+0.43i & -0.09+0.37i &
-0.18+0.54i & -0.02+0.07i & 0.12-0.41i & 0.32-1.3i \\ && 1 & 1 & 1
& 1 & 1 & 1 & 1 & 1 \\ && 1.8-0.14i & 1.6-0.18i & 1.4-0.18i &
1.2-0.18i & 1.8-0.14i & 2.1-0.22i & 2.3-0.3i & 2.7-0.46i \\ $\ov
B^0_d \to \omega\ov K^0$&    VI & 0.02+0.28i & 0.92-0.39i &
0.10-0.24i & 0.03-0.11i & 0.02+0.28i & 0.01-0.03i & -0.02+0.06i &
-0.02+0.09i \\ && 1 & 1 & 1 & 1 & 1 & 1 & 1 & 1 \\ && 0.21-0.1i &
-0.6-0.6i & -0.22 & -0.10+0.01i & 0.21-0.1i & -0.27 & -0.12+0.02i
& -0.08+0.02i \\ $\ov B^0_d \to \ov K^{*0} \eta$&  VI & 0.07+0.03i
& 0.06+0.03i & 0.06+0.03i & 0.05+0.02i & 0.07+0.03i & 0.01 &
-0.04-0.02i & -0.10-0.04i \\ && 1 & 1 & 1 & 1 & 1 & 1 & 1 & 1 \\
&& 0.11-0.04i & 0.10-0.03i & 0.09-0.03i & 0.08-0.03i & 0.11-0.04i
& 0.13-0.04i & 0.14-0.04i & 0.16-0.05i \\ $\ov B^0_d\to\ov
K^{*0}\eta'$&    VI & -0.58-0.67i & 0.47-0.84i & 0.46-0.27i &
0.28-0.09i & -0.58-0.67i & -0.06-0.11i & 0.01+0.82i & -1.3+1.5i \\
&& 1 & 1 & 1 & 1 & 1 & 1 & 1 & 1 \\ && 0.43+0.32i & -0.17+0.56i &
-0.24+0.22i & -0.16+0.09i & 0.43+0.32i & 0.28+0.35i & 0.07+0.3i &
0.02-0.03i \\ $\ov B^0_d \to \phi \ov K^0 $& VI & 0 & 0 & 0 & 0 &
0 & 0 & 0 & 0 \\ && 1 & 1 & 1 & 1 & 1 & 1 & 1 & 1 \\ &&
-0.11+0.03i & -0.13+0.04i & -0.16+0.05i & -0.22+0.08i &
-0.11+0.03i & -0.13+0.04i & -0.16+0.05i & -0.32+0.12i \\ $ B^- \to
K^{*- }\pi^0$  &        VI & -0.09+0.59i & -0.09+0.59i &
-0.09+0.59i & -0.09+0.59i & -0.09+0.59i & -0.08+0.52i &
-0.08+0.48i & -0.07+0.43i \\ && 1 & 1 & 1 & 1 & 1 & 1 & 1 & 1 \\
&& 0.19-0.06i & 0.18-0.06i & 0.18-0.06i & 0.18-0.06i & 0.19-0.06i
& 0.13-0.04i & 0.10-0.03i & 0.05-0.02i \\ $ B^- \to \rho^- \ov
K^0$  &     VI & 0 & 0 & 0 & 0 & 0 & 0 & 0 & 0 \\ && 1 & 1 & 1 & 1
& 1 & 1 & 1 & 1 \\ && 0.09 & 0.07 & 0.06 & 0.06 & 0.09 & 0.03 &
-0.03+0.01i & -0.13+0.03i \\ $ B^-\to \phi K^- $  &          VI &
0 & 0 & 0 & 0 & 0 & 0 & 0 & 0 \\ && 1 & 1 & 1 & 1 & 1 & 1 & 1 & 1
\\ && -0.11+0.03i & -0.13+0.04i & -0.16+0.05i & -0.22+0.08i &
-0.11+0.03i & -0.13+0.04i & -0.16+0.05i & -0.32+0.12i \\ $ B^- \to
K^{*-} \eta$  &     VI & 0.44i & 0.4i & 0.37i & 0.33i & 0.44i &
-0.06+0.41i & -0.11+0.39i & -0.17+0.36i \\ && 1 & 1 & 1 & 1 & 1 &
1 & 1 & 1 \\ && 0.19-0.05i & 0.17-0.05i & 0.16-0.04i & 0.14-0.04i
& 0.19-0.05i & 0.14-0.04i & 0.10-0.03i & 0.05-0.02i \\ $ B^- \to
K^{*-} \eta'$  &    VI,III & 0.31-1.2i & 1.8-0.18i & 0.85+0.55i &
0.34+0.4i & 0.31-1.2i & 1.4-0.82i & 2.6+0.48i & 1.1+3.8i \\ && 1 &
1 & 1 & 1 & 1 & 1 & 1 & 1 \\ && 0.17+0.07i & -0.01+0.26i &
-0.10+0.11i & -0.07+0.04i & 0.17+0.07i & 0.25+0.18i & 0.26+0.39i &
-0.13+0.6i \\ $ B^-  \to \omega K^- $  &   VI,III & -0.06+1.0i &
3.0-1.6i & 0.34-0.84i & 0.11-0.39i & -0.06+1.0i & 0.29-0.71i &
0.06-0.24i & 0.02-0.08i \\ && 1 & 1 & 1 & 1 & 1 & 1 & 1 & 1 \\ &&
0.33-0.17i & 0.97-0.76i & -0.33-0.01i & -0.15+0.02i & 0.33-0.17i &
-0.30+0.01i & -0.11+0.02i & -0.05+0.01i \\
\end{tabular}
\end{center}
\end{table}
}

\vskip 0.4cm
{\squeezetable
\begin{table}[ht]
{\small Table VII. Same as Table V except for charmless
$B_{u,d}\to VV$ decays.}
\begin{center}
\begin{tabular}{l l c c c c c c c c l}
 &  & \multicolumn{4}{c}{$\nc(LL)=2$}
 &   \multicolumn{4}{c}{$\nc(LL)=\nc(LR)$}  \\ \cline{3-6} \cline{7-10}
\raisebox{2.0ex}[0cm][0cm]{Decay} &
\raisebox{2.0ex}[0cm][0cm]{Class} & 2 & 3 & 5 & $\infty$ & 2 & 3 &
5 & $\infty$ \\ \hline
 $\ov B^0_d \to \rho^-  \rho^+$   &I
  & 1 & 1 & 1 & 1 & 1 & 1 & 1 & 1 \\
  && 0.17+0.06i & 0.17+0.06i & 0.17+0.06i & 0.17+0.06i
  & 0.17+0.06i & 0.18+0.06i & 0.18+0.06i & 0.19+0.06i \\
  && 0.01 & 0.01 & 0.01 & 0.01 & 0.01 & 0.001 & -0.01 & -0.01 \\
 $\ov B^0_d \to \rho^0 \rho^0$    &II
  & 1 & 1 & 1 & 1 & 1 & 1 & 1 & 1 \\
  && -0.77-0.25i & -0.77-0.25i & -0.77-0.25i & -0.77-0.25i
  & -0.77-0.25i & -7.7-2.4i & 1.5+0.47i & 0.61+0.18i \\
  && 0.32+0.02i & 0.32+0.02i & 0.32+0.02i & 0.32+0.02i
  & 0.32+0.02i & 2.9+0.17i & -0.53-0.03i & -0.19-0.01i \\
 $ \ov B^0_d \to \omega\omega$    &II
   & 1 & 1 & 1 & 1 & 1 & 1 & 1 & 1 \\
  && 1.3+0.46i & 0.91+0.35i & 0.63+0.27i & 0.2+0.15i
  & 1.3+0.46i & 5.9+2.4i & -0.27-0.2i & 0.34+0.06i \\
  && 0.08+0.01i & 0.08+0.01i & 0.08+0.01i & 0.08+0.01i
  & 0.08+0.01i & 0.92+0.05i & -0.2-0.01i & -0.09 \\
 $B^-\to  \rho^- \rho^0$          &III
  & 1 & 1 & 1 & 1 & 1 & 1 & 1 & 1 \\
  && 0 & 0 & 0 & 0 & 0 & 0 & 0 & 0 \\
  && 0.07 & 0.07 & 0.07 & 0.07 & 0.07 & 0.07 & 0.07 & 0.07 \\
 $ B^- \to \rho^-\omega$          &III
  & 1 & 1 & 1 & 1 & 1 & 1 & 1 & 1 \\
  && 0.36+0.12i & 0.3+0.11i & 0.25+0.09i & 0.18+0.07i
  & 0.36+0.12i & 0.3+0.11i & 0.24+0.09i & 0.13+0.06i \\
  && 0.02 & 0.02 & 0.02 & 0.02 & 0.02 & 0.02 & 0.02 & 0.02 \\
 $\ov B^0_d \to K^{*-} \rho^+$    &IV
  & -0.22+0.07i & -0.22+0.07i & -0.22+0.07i & -0.22+0.07i
  & -0.22+0.07i & -0.21+0.07i & -0.21+0.06i & -0.2+0.06i \\
  && 1 & 1 & 1 & 1 & 1 & 1 & 1 & 1 \\
  && 0.05-0.01i & 0.05-0.01i & 0.05-0.01i & 0.05-0.01i
  & 0.05-0.01i & 0.01 & -0.02+0.01i & -0.06+0.02i \\
 $\ov B^0_d \to\ov K^{*0}\rho^0$ &IV
  & 0.05-0.02i & 0.05-0.02i & 0.05-0.02i & 0.05-0.02i
  & 0.05-0.02i & 0.01 & -0.03+0.01i & -0.06+0.02i \\
  && 1 & 1 & 1 & 1 & 1 & 1 & 1 & 1 \\
  && -0.38+0.1i & -0.38+0.1i & -0.38+0.1i & -0.38+0.1i
  & -0.38+0.1i & -0.34+0.09i & -0.32+0.08i & -0.29+0.07i \\
 $\ov B^0_d \to\ov K^{*0}K^{*0}$ &IV
  & 0 & 0 & 0 & 0 & 0 & 0 & 0 & 0 \\
  && 1 & 1 & 1 & 1 & 1 & 1 & 1 & 1 \\
  && -0.03+0.01i & -0.03+0.01i & -0.03+0.01i & -0.03+0.01i
  & -0.03+0.01i & -0.004 & 0.01 & 0.03-0.01i \\
 $ B^- \to K^{*- }\rho^0$         &IV
  & -0.27+0.09i & -0.27+0.09i & -0.27+0.09i & -0.27+0.09i
  & -0.27+0.09i & -0.22+0.07i & -0.18+0.06i & -0.14+0.04i \\
  && 1 & 1 & 1 & 1 & 1 & 1 & 1 & 1 \\
  && 0.4-0.11i & 0.4-0.11i & 0.4-0.11i & 0.41-0.11i
  & 0.4-0.11i & 0.34-0.09i & 0.3-0.07i & 0.26-0.06i \\
 $ B^- \to \ov K^{*0} \rho^-$    &IV
  & 0 & 0 & 0 & 0 & 0 & 0 & 0 & 0 \\
  && 1 & 1 & 1 & 1 & 1 & 1 & 1 & 1 \\
  && -0.03+0.01i & -0.03+0.01i & -0.03+0.01i & -0.03+0.01i
  & -0.03+0.01i & -0.004 & 0.01 & 0.03-0.01i \\
 $ B^- \to K^{*- } K^{*0}$        &IV
  & 0 & 0 & 0 & 0 & 0 & 0 & 0 & 0 \\
  && 1 & 1 & 1 & 1 & 1 & 1 & 1 & 1 \\
  && -0.03+0.01i & -0.03+0.01i & -0.03+0.01i & -0.03+0.01i
  & -0.03+0.01i & -0.004 & 0.01 & 0.03-0.01i \\
 $ \ov B^0_d \to \rho^{0} \phi$   &V
  & 0 & 0 & 0 & 0 & 0 & 0 & 0 & 0 \\
  && 1 & 1 & 1 & 1 & 1 & 1 & 1 & 1 \\
  && -0.35+0.12i & -0.97+0.62i & 1.3+0.27i & 0.34-0.04i
  & -0.35+0.12i & 1.0+0.06i & 0.28-0.04i & 0.14-0.03i \\
 $ \ov B^0_d \to\omega \phi$     &V
  & 0 & 0 & 0 & 0 & 0 & 0 & 0 & 0 \\
  && 1 & 1 & 1 & 1 & 1 & 1 & 1 & 1 \\
  && -0.35+0.12i & -0.97+0.62i & 1.3+0.27i & 0.34-0.04i
  & -0.35+0.12i & 1.0+0.06i & 0.28-0.04i & 0.14-0.03i \\
 $ B^- \to\rho^{-} \phi$          &V
  & 0 & 0 & 0 & 0 & 0 & 0 & 0 & 0 \\
  && 1 & 1 & 1 & 1 & 1 & 1 & 1 & 1 \\
  && -0.35+0.12i & -0.97+0.62i & 1.3+0.27i & 0.34-0.04i
  & -0.35+0.12i & 1.0+0.06i & 0.28-0.04i & 0.14-0.03i \\
 $ \ov B^0_d \to \rho^0 \omega$   & VI
  & -0.04+0.01i & -0.05+0.02i & -0.06+0.02i & -0.08+0.03i
  & -0.04+0.01i & -0.01 & 0.04-0.02i & 0.24-0.1i \\
  && 1 & 1 & 1 & 1 & 1 & 1 & 1 & 1 \\
  && -0.12+0.03i & -0.14+0.04i & -0.17+0.05i & -0.23+0.08i
  & -0.12+0.03i & -0.14+0.04i & -0.17-0.05i & -0.31+0.11i \\
 $\ov B^0_d \to \ov K^{*0}\omega$  &VI
  & -0.03+0.01i & -0.04+0.01i & -0.05+0.02i & -0.11+0.07i
  & -0.03+0.01i & -0.01 & 0.07-0.04i & -0.15+0.02i \\
  && 1 & 1 & 1 & 1 & 1 & 1 & 1 & 1 \\
  && 0.05-0.02i & 0.07-0.02i & 0.09-0.03i & 0.20-0.11i
  & 0.05-0.02i & 0.12-0.04i & 0.36-0.19i & -0.33+0.04i \\
 $ \ov B^0_d \to\ov K^{*0} \phi$ &VI
  & 0 & 0 & 0 & 0 & 0 & 0 & 0 & 0 \\
  && 1 & 1 & 1 & 1 & 1 & 1 & 1 & 1 \\
  && -0.11+0.03i & -0.13+0.04i & -0.16+0.05i & -0.22+0.08i
  & -0.11+0.03i & -0.13+0.04i & -0.16+0.05i & -0.32+0.12i \\
 $ B^- \to K^{*-} \omega $        &VI
  & -0.17+0.06i & -0.22+0.09i & -0.3+0.13i & -0.64+0.41i
  & -0.17+0.06i & -0.26+0.11i & -0.6+0.35i & 0.38-0.06i \\
  && 1 & 1 & 1 & 1 & 1 & 1 & 1 & 1 \\
  && 0.1-0.03i & 0.13-0.04i & 0.18-0.06i & 0.39-0.22i
  & 0.1-0.03i & 0.13-0.04i & 0.25-0.12i & -0.08 \\
 $ B^- \to K^{*-} \phi$           &VI
  & 0 & 0 & 0 & 0 & 0 & 0 & 0 & 0 \\
  && 1 & 1 & 1 & 1 & 1 & 1 & 1 & 1 \\
  && -0.11+0.03i & -0.13+0.04i & -0.16+0.05i & -0.22+0.08i
  & -0.11+0.03i & -0.13+0.04i & -0.16+0.05i & -0.32+0.12i \\
\end{tabular}
\end{center}
\end{table}
}

\vskip 0.4cm
{\squeezetable
\begin{table}[ht]
{\footnotesize Table VIII. Branching ratios (in units of
$10^{-6}$) averaged over CP-conjugate modes for charmless
$B_{u,d}\to PP$ decays. Predictions are made for $k^2=m_b^2/2$,
$\eta=0.370,~\rho=0.175$, and $\nc(LR)=2,3,5,\infty$ with
$\nc(LL)$ being fixed to be 2 in the first case and treated to be
the same as $\nc(LR)$ in the second case. Results using the BSW
model and the light-cone sum rule for heavy-to-light form factors
are shown in the upper and lower entries, respectively.
Experimental values (in units of $10^{-6}$) are taken from
\cite{CLEO,Gao,CLEOomega2,Roy,Behrens1,Behrens2,CLEOpiK}. Our
preferred predictions are those for $\nc(LL)=2$ and $\nc(LR)=5$.}
\begin{center}
\begin{tabular}{l l c c c c c c c c l}
 &  & \multicolumn{4}{c}{$\nc(LL)=2$}
 &   \multicolumn{4}{c}{$\nc(LL)=\nc(LR)$}  \\ \cline{3-6} \cline{7-10}
\raisebox{2.0ex}[0cm][0cm]{Decay} &
\raisebox{2.0ex}[0cm][0cm]{Class} & 2 & 3 & 5 & $\infty$ & 2 & 3 &
5 & $\infty$  & \raisebox{2.0ex}[0cm][0cm]{Expt.}\\ \hline
 $\ov B^0_d \to \pi^+ \pi^-$&        I
  &11.3&11.3&11.4&11.4&11.3&12.8&14.0&15.9& $ < 8.4$ \\
 &&9.49&9.51&9.53&9.55&9.49&10.7&11.7&13.3& \\
 $\ov B^0_d \to \pi^0\pi^0$ &       II,VI
  &0.33&0.33&0.34&0.34&0.33&0.09&0.21&0.90&$<9.3$\\
 &&0.28&0.28&0.28&0.29&0.28&0.08&0.18&0.75& \\
 $\ov B^0_d \to \eta \eta$  &       II,VI
  &0.24&0.26&0.27&0.30&0.24&0.13&0.16&0.43&$<18$\\
 &&0.20&0.21&0.22&0.24&0.20&0.10&0.13&0.35& \\
 $\ov B^0_d \to \eta \eta'$ &       II,VI
  &0.27&0.33&0.40&0.51&0.27&0.10&0.14&0.50& $<27$\\
 &&0.22&0.27&0.32&0.41&0.22&0.08&0.11&0.40& \\
 $\ov B^0_d \to \eta' \eta'$&       II,VI
  &0.08&0.11&0.14&0.21&0.08&0.02&0.03&0.16&$<47$\\
 &&0.06&0.09&0.11&0.17&0.06&0.01&0.02&0.13& \\
  $ B^- \to \pi^- \pi^0$ &          III
  &8.63&8.63&8.63&8.63&8.63&6.82&5.52&3.83&$< 16$\\
 &&7.23&7.23&7.23&7.23&7.23&5.71&4.63&3.21& \\
 $ B^- \to \pi^- \eta$  &           III
  &5.92&6.00&6.06&6.16&5.92&4.70&3.85&2.79& $<15$\\
 &&4.89&4.96&5.01&5.09&4.89&3.88&3.18&2.30& \\
 $ B^- \to \pi^-\eta'$  &           III
  &3.70&3.88&4.07&4.39&3.70&2.74&2.09&1.29& $<12$\\
 &&3.03&3.19&3.34&3.60&3.03&2.26&1.73&1.07& \\
 $\ov B^0_d \to   K^- \pi^+$&       IV
  &14.0&14.9&15.7&16.8&14.0&15.6&16.9&18.9&$14\pm3\pm2$\\
 &&14.0&12.4&13.0&14.0&14.0&12.9&14.0&15.7& \\
 $ B^-\to \ov K^0\pi^-$       &            IV
  &16.0&17.1&17.9&19.3&16.0&18.9&21.4&25.4&$14\pm5\pm2 $\\
 &&13.3&14.2&14.9&16.0&13.3&15.7&17.8&21.1& \\
 $ B^- \to K^-   K^0$   &                   IV
  &0.85&0.91&0.95&1.03&0.85&1.00&1.14&1.35&$ <9.3$\\
 &&0.68&0.73&0.77&0.82&0.68&0.81&0.91&1.08& \\
 $\ov B^0_d\to\ov K^0\pi^0$&        VI
  &5.92&6.37&6.74&7.32&5.92&6.75&7.47&8.64&$<41$\\
 &&4.93&5.30&5.61&6.10&4.93&5.62&6.23&7.21& \\
 $ \ov B^0_d\to K^0\ov K^0$ &             VI
  &0.80&0.85&0.90&0.96&0.80&0.94&1.07&1.27&$ <17$\\
 &&0.64&0.69&0.72&0.77&0.64&0.76&0.86&1.02& \\
 $\ov B^0_d \to \pi^0 \eta$ &       VI
  &0.22&0.25&0.27&0.30&0.22&0.24&0.26&0.29& $<8$\\
 &&0.18&0.20&0.22&0.25&0.18&0.20&0.21&0.24& \\
 $\ov B^0_d \to \pi^0 \eta'$&       VI
  &0.08&0.15&0.22&0.34&0.08&0.07&0.07&0.07& $<11$\\
 &&0.06&0.12&0.18&0.28&0.06&0.06&0.06&0.06& \\
 $\ov B^0_d \to\ov K^0\eta$&       VI
  &0.95&0.84&0.75&0.63&0.95&1.32&1.67&2.30&$<33 $\\
 &&0.73&0.64&0.57&0.48&0.73&1.02&1.29&1.78& \\
 $\ov B^0_d\to\ov K^0\eta'$&       VI
  &25.5&35.1&43.8&58.8&25.5&27.2&28.6&30.7&$ 59^{+18}_{-16}\pm9 $\\
 &&20.4&28.0&34.9&46.8&20.4&21.7&22.8&24.6& \\
 $ B^-\to K^- \pi^0$    &           VI
  &9.45&9.98&10.4&11.1&9.45&10.7&11.8&13.5&$15\pm4\pm3 $\\
 &&7.83&8.26&8.62&9.18&7.83&8.85&9.73&11.1& \\
 $ B^- \to K^- \eta$    &           VI
  &1.57&1.44&1.33&1.19&1.57&2.17&2.75&3.81&$<14$\\
 &&1.23&1.12&1.04&0.92&1.23&1.70&2.15&2.99& \\
 $ B^- \to K^- \eta'$   &           VI
  &26.3&36.3&45.5&61.1&26.3&27.4&28.3&29.7&$74^{+~8}_{-13}\pm10$\\
 &&21.0&28.9&36.2&48.7&21.0&21.9&22.6&23.7& \\
\end{tabular}
\end{center}
\end{table}
\vskip 0.4cm
 }

\begin{table}[ht]
{\footnotesize Table IX. Same as Table VIII except for charmless
$B_{u,d}\to VP$ decays.}
\begin{center}
\begin{tabular}{l l c c c c c c c c l}
 &  & \multicolumn{4}{c}{$\nc(LL)=2$}
 &   \multicolumn{4}{c}{$\nc(LL)=\nc(LR)$}  \\ \cline{3-6} \cline{7-10}
\raisebox{2.0ex}[0cm][0cm]{Decay} &
\raisebox{2.0ex}[0cm][0cm]{Class} & 2 & 3 & 5 & $\infty$ & 2 & 3 &
5 & $\infty$  & \raisebox{2.0ex}[0cm][0cm]{Expt.}\\ \hline
\parbox[c]{2.8cm}{$\ov B^0_d \to \rho^-  \pi^+ $ \\
\\$\ov B^0_d \to \rho^+  \pi^- $\\} & \parbox{1cm}{I \\
\\${\rm I}$\\} &
\parbox[c]{1cm}{\centering 29.6\\ 24.1 \\ 7.39\\ 12.8 }&
\parbox[c]{1cm}{\centering 29.6 \\24.1\\7.15\\12.5 }
 & \parbox[c]{1cm}{\centering 29.6 \\ 24.1\\7.14\\12.5}
& \parbox[c]{1cm}{\centering 29.6 \\ 24.1\\7.14\\12.5} &
\parbox[c]{1cm}{\centering 29.6 \\ 24.1\\7.39\\12.8} &
\parbox[c]{1cm}{\centering 33.4 \\ 27.2\\8.06\\14.1} &
\parbox[c]{1cm}{\centering 36.5 \\ 29.8\\8.83\\15.4} &
\parbox[c]{1cm}{\centering 41.6 \\ 33.9\\10.0\\17.6} & $
\Bigg\}35^{+11}_{-10}\pm 5$ \vspace{0.8mm}
\\
%$\ov B^0_d \to \rho^-  \pi^+ $&      I
%&29.6&29.6&29.6&29.6&29.6&33.4&36.5&41.6&$<88$\\
%&&24.1&24.1&24.1&24.1&24.1&27.2&29.8&33.9& \\
%$\ov B^0_d \to \rho^+  \pi^-$&       I
%&7.39&7.15&7.14&7.14&7.39&8.06&8.83&10.0&$<88$\\
%&&12.8&12.5&12.5&12.5&12.8&14.1&15.4&17.6& \\
$\ov B^0_d \to\rho^+K^{-} $&      I,IV
&1.04&1.20&1.34&1.58&1.04&1.16&1.26&1.42&$ <25$\\
&&1.80&2.08&2.33&2.75&1.80&2.01&2.18&2.46& \\ $\ov B^0_d \to
\rho^0 \pi^0$&       II
&0.82&0.81&0.81&0.81&0.81&0.03&0.33&2.31&$<18$\\
&&0.89&0.88&0.88&0.88&0.89&0.03&0.35&2.49& \\ $\ov B^0_d \to
\omega \pi^0$ &      II
&0.24&0.17&0.12&0.08&0.24&0.08&0.05&0.17&$<14$\\
&&0.18&0.11&0.07&0.03&0.18&0.08&0.04&0.03& \\ $ \ov B^0_d
\to\omega \eta $& II &0.51&0.48&0.46&0.45&0.51&0.02&0.16&1.25&$<12
$\\ &&0.53&0.51&0.50&0.49&0.53&0.02&0.17&1.37& \\ $ \ov B^0_d
\to\omega \eta'$& II
&0.33&0.35&0.36&0.37&0.33&0.01&0.15&1.04&$<60$
\\ &&0.33&0.35&0.36&0.37&0.33&0.01&0.15&1.04& \\
$\ov B^0_d\to\rho^0 \eta $&      II
&0.05&0.05&0.05&0.05&0.05&0.01&0.03&0.12&$<13$ \\
&&0.01&0.01&0.01&0.01&0.01&0.01&0.01&0.02& \\ $ \ov B^0_d
\to\rho^0 \eta'$&      II,VI
&0.02&0.02&0.03&0.05&0.02&0.003&0.01&0.06&$<23$ \\
&&0.004&0.004&0.01&0.05&0.004&0.01&0.01&0.02& \\ $ B^- \to
\rho^-\pi^0$   &  III
&18.5&18.5&18.5&18.5&18.5&18.3&18.1&17.8&$<77$ \\
&&16.1&16.1&16.1&16.1&16.1&15.0&14.1&12.9& \\ $ B^- \to
\rho^0\pi^- $  &  III
&8.09&8.08&8.09&8.09&8.09&4.78&2.77&0.81&$15\pm5\pm4$\\
&&11.5&11.5&11.5&11.5&11.5&8.10&5.81&3.11& \\ $ B^- \to
\omega\pi^- $  &  III &7.97&7.82&7.72&7.62&7.97&4.84&2.92&1.01& $<
23$ \\ &&11.3&11.2&11.1&11.0&11.3&8.16&6.04&3.53& \\ $ B^- \to
\rho^-\eta $  &   III &11.6&11.3&11.3&11.3&11.6&10.3&9.59&8.51&
$<32$ \\ &&9.76&9.72&9.72&9.73&9.76&8.07&6.87&5.25& \\ $ B^- \to
\rho^-\eta'$  &   III &7.41&7.56&7.71&7.99&7.41&6.63&6.04&5.20&
$<47$ \\ &&6.43&6.63&6.84&7.26&6.43&5.21&4.33&3.17& \\ $ B^- \to
\rho^0 K^-$ &     III,VI
&0.48&0.47&0.48&0.51&0.48&0.30&0.21&0.15&$ <22$\\
&&0.73&0.78&0.83&0.93&0.73&0.51&0.37&0.21& \\ $\ov B^0_d \to
K^{*-} \pi^+$ &       IV
&4.83&4.83&4.83&4.83&4.83&5.38
&5.85&6.59&$22^{+8+4}_{-6-5}$\\
&&3.89&3.89&3.89&3.89&3.89&4.34&4.72&5.32& \\ $ B^- \to\ov
K^{*0}\pi^- $&  IV
&5.35&5.35&5.35&5.35&5.35&6.84&8.17&10.4&$<27$\\
&&4.32&4.32&4.32&4.32&4.32&5.52&6.59&8.38& \\ $ B^-\to  K^{*0} K^-
$&     IV &0.27&0.28&0.28&0.28&0.27&0.35&0.42&0.54&$<12$\\
&&0.22&0.23&0.23&0.23&0.22&0.29&0.35&0.44& \\ $ B^- \to K^{*-}K^0$
&         IV &0.03&0.02&0.03&0.04&0.03&0.01&0.01&0.01&$-$\\
&&0.01&0.04&0.06&0.08&0.01&0.02&0.02&0.01& \\
\end{tabular}
\end{center}
\end{table}

\begin{table}[ht]
{\footnotesize Table IX. (Continued)}
\begin{center}
\begin{tabular}{l l c c c c c c c c l}
 &  & \multicolumn{4}{c}{$\nc(LL)=2$}
 &   \multicolumn{4}{c}{$\nc(LL)=\nc(LR)$}  \\ \cline{3-6} \cline{7-10}
\raisebox{2.0ex}[0cm][0cm]{Decay} &
\raisebox{2.0ex}[0cm][0cm]{Class} & 2 & 3 & 5 & $\infty$ & 2 & 3 &
5 & $\infty$  & \raisebox{2.0ex}[0cm][0cm]{Expt.}\\ \hline
$ \ov B^0_d \to \phi\pi^0$  &   V
&0.01&0.001&0.01&0.03&0.01&0.01&0.05&0.17&$<5.4$\\
&&0.01&0.0004&0.004&0.02&0.01&0.01&0.04&0.13& \\
$ \ov B^0_d \to\phi \eta $ &   V
&0.003&0.0003&0.003&0.02&0.003&0.01&0.03&0.09&$<9$\\
&&0.002&0.0002&0.002&0.01&0.002&0.003&0.02&0.07& \\
$ \ov B^0_d\to  \phi \eta'$  &  V
&0.002&0.0002&0.002&0.01&0.002&0.003&0.02&0.06&$<31$\\
&&0.002&0.0001&0.001&0.01&0.002&0.002&0.01&0.04& \\
$ B^- \to\phi\pi^- $  &   V
&0.01&0.001&0.01&0.06&0.01&0.02&0.10&0.36&$<4$\\
&&0.01&0.001&0.01&0.05&0.01&0.01&0.08&0.29& \\
$\ov B^0_d\to K^{*0}\ov K^0$&  VI
&0.01&0.02&0.02&0.03&0.01&0.01&0.01&0.004&$-$\\
&&0.03&0.04&0.05&0.07&0.03&0.02&0.02&0.01& \\
$\ov B^0_d\to \ov K^{*0} K^0$ &  VI
&0.26&0.26&0.26&0.26&0.26&0.33&0.39&0.50& $-$\\
&&0.21&0.21&0.21&0.21&0.21&0.27&0.32&0.41& \\
$\ov B^0_d\to\ov K^{*0}\pi^0$&     VI
&1.76&1.76&1.77&1.77&1.76&2.16&2.53&3.15&$<28$\\
&&1.11&1.12&1.13&1.14&1.1&1.32&1.52&1.88& \\
$\ov B^0_d\to\rho^0\ov K^0$&      VI
&0.95&1.05&1.18&1.38&0.95&1.02&1.16&1.45&$ <27$\\
&&0.95&1.25&1.43&1.72&0.95&1.12&1.21&1.42& \\
$\ov B^0_d \to \omega\ov K^0$&    VI
&0.62&0.05&0.50&2.70&0.62&0.28&2.44&9.73&$<57 $\\
&&0.34&0.10&0.72&3.03&0.34&0.39&2.38&8.66& \\
$\ov B^0_d \to \ov K^{*0} \eta$&  VI
&3.57&4.26&4.85&5.81&3.57&3.83&4.05&4.40&$<30$\\
&&4.32&5.44&6.42&8.04&4.32&4.27&4.23&4.22& \\
$\ov B^0_d\to\ov K^{*0}\eta'$&    VI
&0.08&0.16&0.52&1.55&0.08&0.13&0.16&0.24&$<20$\\
&&0.61&0.14&0.32&1.64&0.61&0.99&1.28&1.80& \\
$\ov B^0_d \to \phi\ov K^0 $& VI
&10.7&7.01&4.60&1.96&10.7&5.63&2.73&0.34&$<28$\\
&&8.81&5.75&3.78&1.61&8.81&4.62&2.24&0.28& \\
$ B^- \to K^{*-}\pi^0$  &        VI
&3.27&3.27&3.26&3.26&3.27&3.63&3.93&4.42&$<99$\\
&&3.01&3.01&3.00&2.99&3.01&3.34&3.63&4.11& \\
$ B^- \to\rho^- \ov K^0$  &     VI
&0.32&0.45&0.57&0.77&0.32&0.24&0.19&0.12&  $<48$\\
&&0.56&0.78&0.99&1.34&0.56&0.42&0.32&0.20& \\
$ B^-\to \phi K^- $&          VI
&10.9&7.55&4.96&2.11&10.9&6.07&2.94&0.36& $<5.9$\\
&&9.08&6.20&4.07&1.73&9.08&4.98&2.42&0.30& \\
$ B^- \to K^{*-}\eta$  &     VI
&3.74&4.41&5.00&5.95&3.74&3.45&3.24&2.94&$<30$\\
&&4.48&5.60&6.59&8.24&4.48&3.89&3.44&2.84& \\
$ B^- \to K^{*-}\eta'$  &    VI,III
&0.54&0.43&0.65&1.53&0.54&0.69&0.85&1.16&$<87$\\
&&1.41&0.53&0.49&1.54&1.41&2.00&2.58&3.65& \\
$ B^-  \to \omega K^- $  &   VI,III
&0.88&0.52&1.21&3.90&0.88&0.93&3.91&13.2&$15^{+7}_{-6}\pm 2$\\
&&0.82&0.88&1.81&4.70&0.82&1.44&4.43&13.1&
\end{tabular}
\end{center}
\end{table}

\vskip 0.4cm
\begin{table}[ht]
{\footnotesize Table X. Same as Table VIII except for charmless
$B_{u,d}\to VV$ decays.}
\begin{center}
\begin{tabular}{l l c c c c c c c c l}
 &  & \multicolumn{4}{c}{$\nc(LL)=2$}
 &   \multicolumn{4}{c}{$\nc(LL)=\nc(LR)$}  \\ \cline{3-6} \cline{7-10}
\raisebox{2.0ex}[0cm][0cm]{Decay} &
\raisebox{2.0ex}[0cm][0cm]{Class} & 2 & 3 & 5 & $\infty$ & 2 & 3 &
5 & $\infty$  & \raisebox{2.0ex}[0cm][0cm]{Expt.}\\ \hline
 $\ov B^0_d \to \rho^-  \rho^+$   &I
  &21.9&21.9&21.9&21.9&21.9&24.7&27.0&30.7&  $ <2200$\\
 &&35.8&35.8&35.8&35.8&35.8&40.3&44.2&50.3& \\
 $\ov B^0_d \to \rho^0 \rho^0$    &II
  &0.55&0.55&0.55&0.55&0.55&0.05&0.25&1.57&$<40$\\
 &&0.90&0.90&0.90&0.90&0.90&0.07&0.41&2.57& \\
 $ \ov B^0_d \to \omega\omega$    &II
  &0.65&0.55&0.50&0.45&0.65&0.07&0.16&1.24& $ <19$\\
 &&1.04&0.89&0.81&0.72&1.04&0.11&0.26&2.00& \\
 $B^-\to  \rho^- \rho^0$          &III
  &17.2&17.2&17.2&17.2&17.2&13.6&11.0&7.64& $<120$\\
 &&28.1&28.1&28.1&28.1&28.1&22.2&18.0&12.5& \\
 $ B^- \to \rho^-\omega$          &III
  &17.2&17.1&17.0&16.8&17.2&13.9&11.5&8.37& $<61$\\
 &&27.9&27.7&27.5&27.2&27.9&22.6&18.6&13.5& \\
 $\ov B^0_d \to K^{*-} \rho^+$    &IV
  &3.65&3.65&3.65&3.65&3.65&4.08&4.43&4.99&$ -$\\
 &&5.86&5.86&5.86&5.86&5.86&6.54&7.11&8.01& \\
 $\ov B^0_d \to\ov K^{*0}\rho^0$ &IV
  &0.88&0.87&0.86&0.86&0.88&0.99&1.12&1.38&$ <460 $ \\
 &&1.26&1.25&1.24&1.22&1.26&1.38&1.55&1.92& \\
 $\ov B^0_d \to\ov K^{*0}K^{*0}$ &IV
  &0.20&0.20&0.20&0.20&0.20&0.25&0.30&0.38& $-$\\
 &&0.40&0.40&0.40&0.40&0.40&0.51&0.60&0.77& \\
 $ B^- \to K^{*- }\rho^0$         &IV
  &3.53&3.58&3.59&3.61&3.53&4.00&4.40&5.11&$ <900$\\
 &&6.10&6.14&6.16&6.19&6.10&6.87&7.60&8.89& \\
 $ B^- \to \ov K^{*0} \rho^-$    &IV
  &4.00&4.00&4.00&4.00&4.00&5.11&6.11&7.76&$-$\\
 &&6.42&6.42&6.42&6.42&6.42&8.21&9.80&12.5& \\
 $ B^- \to K^{*- } K^{*0}$        &IV
  &0.21&0.21&0.21&0.21&0.21&0.27&0.32&0.41& $-$\\
 &&0.42&0.42&0.42&0.42&0.42&0.54&0.64&0.82& \\
 $ \ov B^0_d \to \rho^{0} \phi$   &V
  &0.005&0.0004&0.004&0.02&0.005&0.006&0.04&0.13&  $ <13$\\
 &&0.01&0.0006&0.007&0.04&0.01&0.01&0.06&0.20& \\
 $ \ov B^0_d \to\omega \phi$     &V
  &0.005&0.0004&0.004&0.02&0.005&0.006&0.04&0.13& $ <21$\\
 &&0.01&0.0006&0.007&0.04&0.01&0.01&0.06&0.20& \\
 $ B^- \to\rho^{-} \phi$          &V
  &0.01&0.001&0.01&0.05&0.01&0.01&0.08&0.28&  $ <16$\\
 &&0.02&0.0014&0.015&0.08&0.02&0.02&0.12&0.43& \\
 $ \ov B^0_d \to \rho^0 \omega$   & VI
  &0.18&0.12&0.08&0.04&0.18&0.10&0.05&0.02& $ <11$\\
 &&0.30&0.20&0.13&0.06&0.30&0.16&0.08&0.02& \\
$\ov B^0_d \to \ov K^{*0}\omega$&VI
  &5.57&3.06&1.56&0.27&5.57&1.93&0.37&0.41&$  <23$\\
 &&9.97&5.14&2.43&0.29&9.97&2.97&0.35&1.44& \\
 $ \ov B^0_d \to\ov K^{*0} \phi$ &VI
  &8.75&5.58&3.66&1.56&8.75&4.48&2.17&0.27&$<21$\\
 &&16.8&11.0&7.20&3.06&16.8&8.81&4.27&0.53& \\
 $ B^- \to K^{*-} \omega $        &VI
  &5.65&3.26&1.90&0.88&5.65&1.82&0.56&1.78&$ <87$\\
 &&10.1&5.50&3.03&1.38&10.1&2.82&0.81&4.21& \\
 $ B^- \to K^{*-} \phi$           &VI
  &9.31&5.93&3.90&1.66&9.31&4.77&2.31&0.29&$<41$\\
 &&17.9&11.7&7.66&3.25&17.9&9.37&4.54&0.56&
\end{tabular}
\end{center}
\end{table}

\end{document}